%% file: main.tex
\documentclass{vgtc}                          % final (conference style)
\graphicspath{{figures/}{pictures/}{images/}{./}} % where to search for the images

\usepackage{times}                     % we use Times as the main font
\usepackage{tabu}                      % only used for the table example
\usepackage{booktabs}                  % only used for the table example
\usepackage{lipsum}                    % used to generate placeholder text
\usepackage{mwe}                       % used to generate placeholder figures

\usepackage{mathptmx}                  % use matching math font
\usepackage{amsmath}
\usepackage{amsfonts}
\usepackage{subcaption}
\usepackage{soul}

\newcommand\blfootnote[1]{%
  \begingroup
  \renewcommand\thefootnote{}\footnote{#1}%
  \addtocounter{footnote}{-1}%
  \endgroup
}

\onlineid{0}

\vgtccategory{Research}

\vgtcinsertpkg

\title{Estimation and Visualization of Isosurface Uncertainty from Linear and High-Order Interpolation Methods}

\author{%
 Timbwaoga A. J. Ouermi \thanks{e-mail: touermi@sci.utah.edu} \\
 \scriptsize SCI Institute, University of Utah
 \and Jixian Li \thanks{e-mail: jixianli@sci.utah.edu} \\
 \scriptsize SCI Institute, University of Utah
 \and Tushar Athawale \thanks{e-mail: athawaletm@ornl.gov} \\
 \scriptsize Oak Ridge National Laboratory
 \and Chris R. Johnson \thanks{e-mail: crj@sci.utah.edu} \\ 
 \scriptsize SCI Institute, University of Utah
}
\teaser{
  \centering
    \begin{subfigure}[b]{0.19 \textwidth}
        \centering
        \includegraphics[width=0.99 \columnwidth]{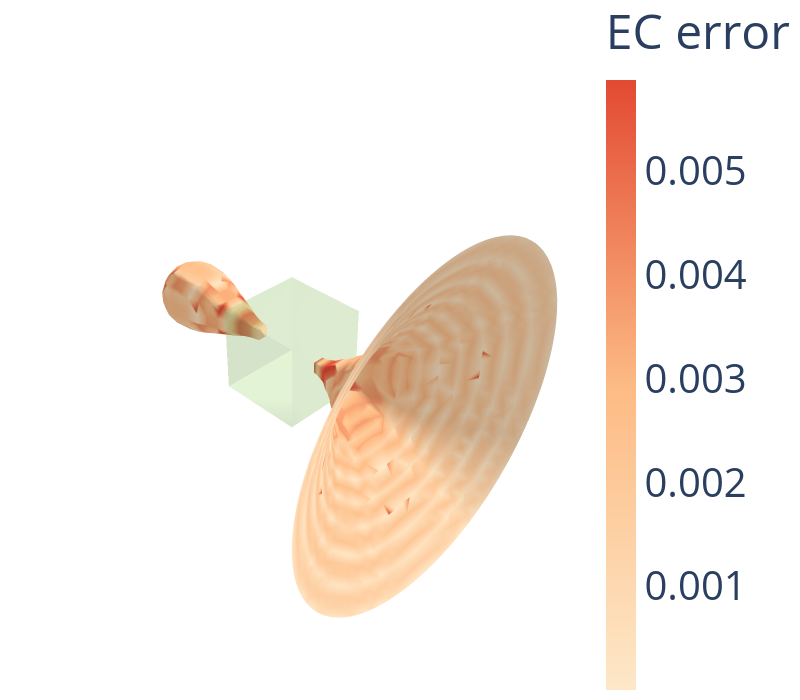}
        \caption{Teardrop local selection}
        \label{subfig:teardrop_box_32x32x32}
    \end{subfigure}
    \begin{subfigure}[b]{0.26 \textwidth}
        \centering
        \includegraphics[width=0.99 \columnwidth]{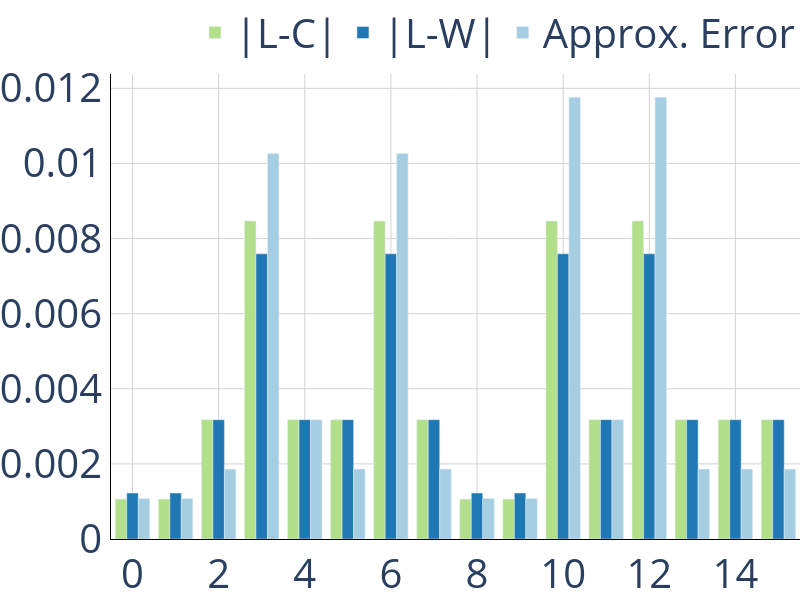}
        \caption{Vertex-by-vertex comparison}
        \label{subfig:teardrop_box_LC_LW_32x32x32}
    \end{subfigure}
    \begin{subfigure}[b]{0.19 \textwidth}
        \centering
        \includegraphics[width=0.99 \columnwidth]{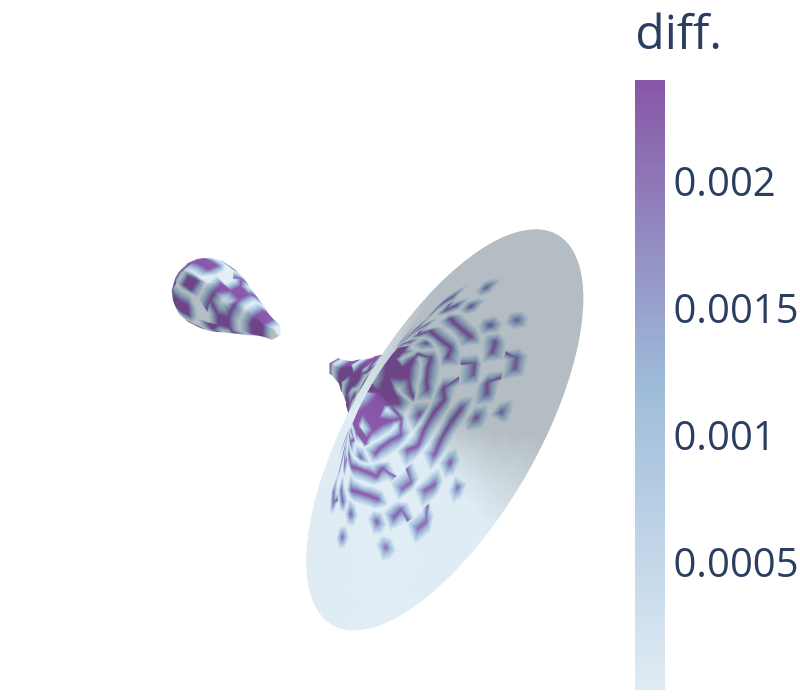}
        \caption{ L vs W}
        \label{subfig:teardrop_LW_32x32x32}
    \end{subfigure}
    \begin{subfigure}[b]{0.19 \textwidth}
        \centering
        \includegraphics[width=0.99 \columnwidth]{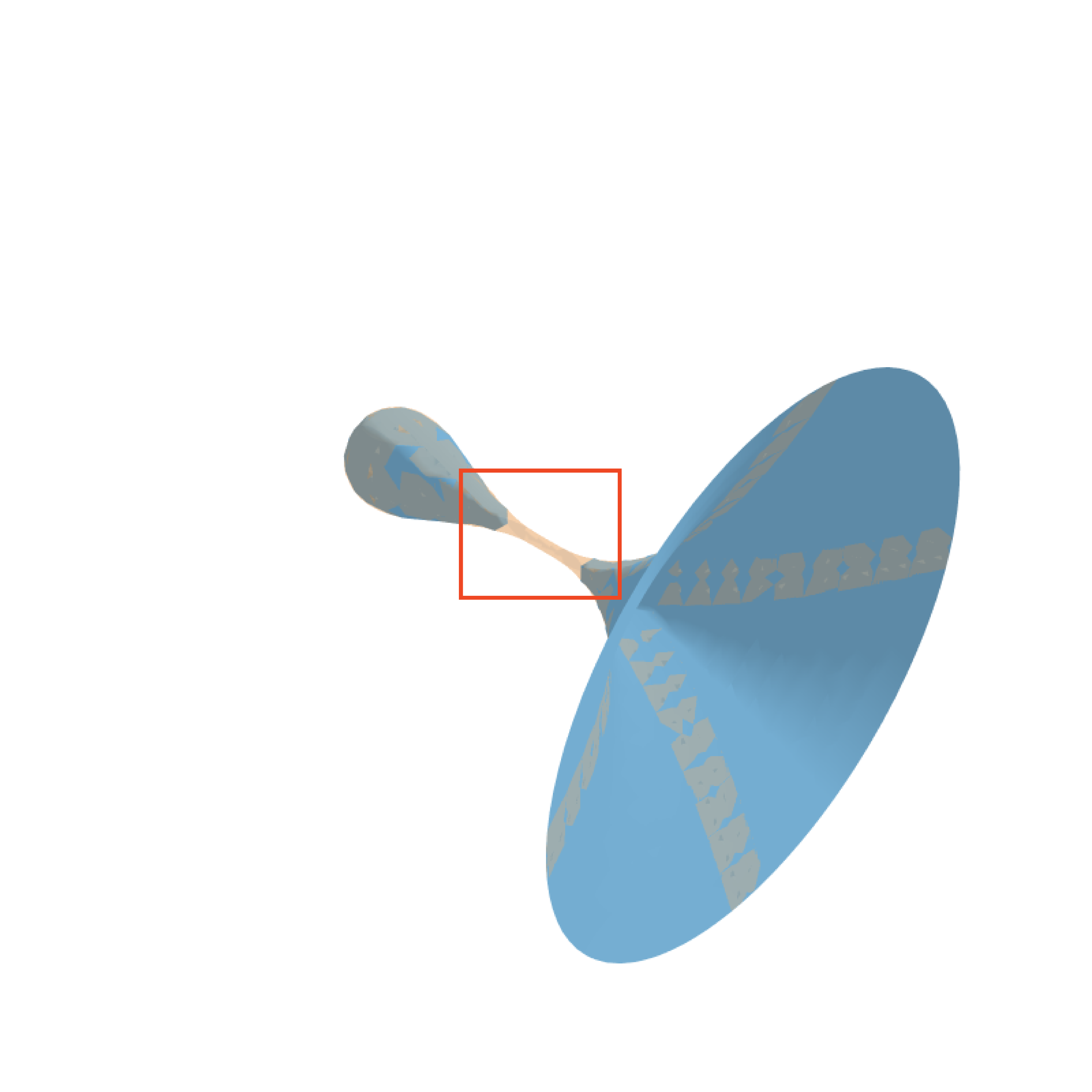}
        \caption{Possible hidden feature}
        \label{subfig:teardrop_global_hidden_features_32x32x32}
    \end{subfigure}
    \begin{subfigure}[b]{0.14 \textwidth}
        \centering
        \includegraphics[width=0.99 \columnwidth]{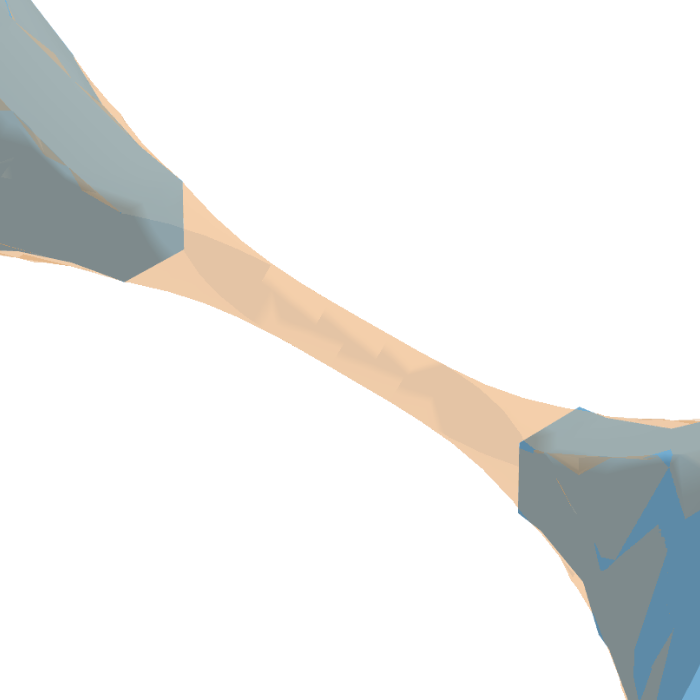}
        \caption{Zoomed red box}
        \label{subfig:teardrop_zoom1_global_hidden_features_32x32x3}
    \end{subfigure}
    \begin{subfigure}[b]{0.19 \textwidth}
        \centering
        \includegraphics[width=0.99 \columnwidth]{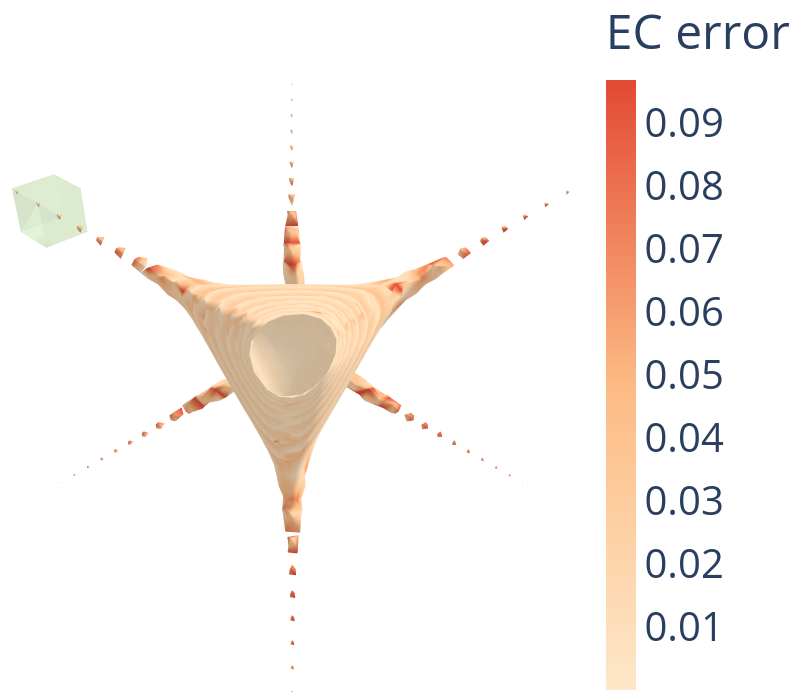}
        \caption{Tubey local selection}
        \label{subfig:tubey_box_32x32x32}
    \end{subfigure}
    \begin{subfigure}[b]{0.26 \textwidth}
        \centering
        \includegraphics[width=0.99 \columnwidth]{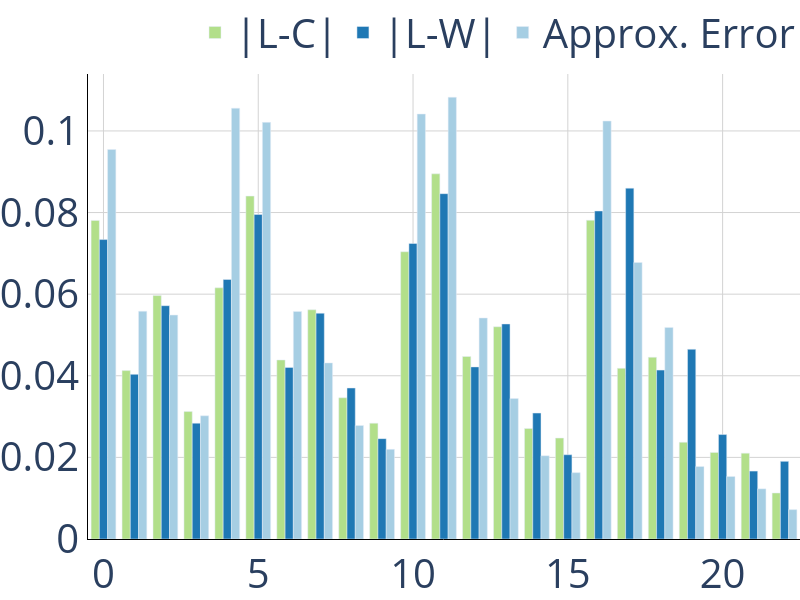}
        \caption{Vertex-by-vertex comparison}
        \label{subfig:tubey_box_LC_LW_32x32x32}
    \end{subfigure}
    \begin{subfigure}[b]{0.19 \textwidth}
        \centering
        \includegraphics[width=0.99 \columnwidth]{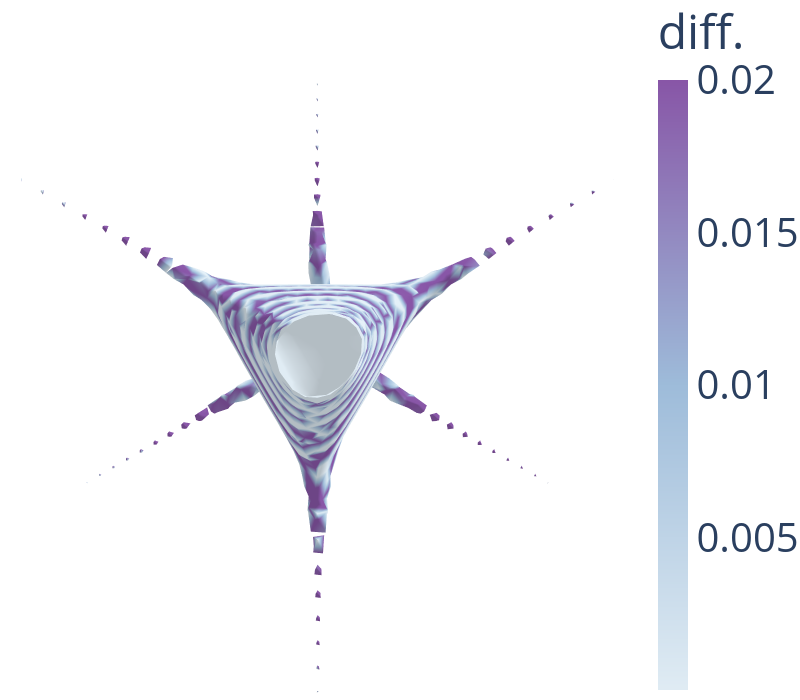}
        \caption{ L vs W}
        \label{subfig:tubey_LW_32x32x32}
    \end{subfigure}
    \begin{subfigure}[b]{0.19 \textwidth}
        \centering
        \includegraphics[width=0.99 \columnwidth]{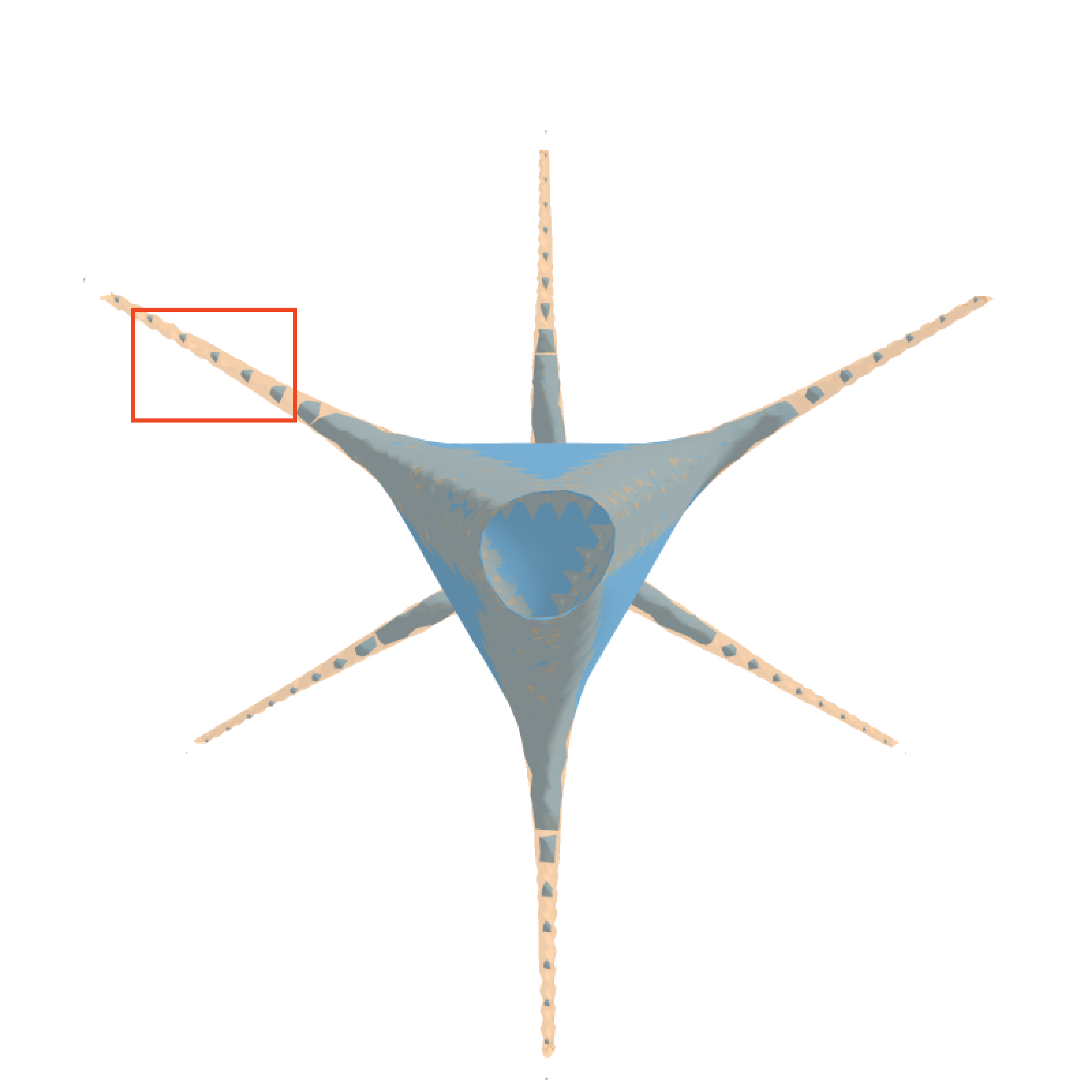}
        \caption{Possible hidden feature }
        \label{subfig:tubey_global_hidden_features_32x32x32}
    \end{subfigure}
    \begin{subfigure}[b]{0.14 \textwidth}
        \centering
        \includegraphics[width=0.99 \columnwidth]{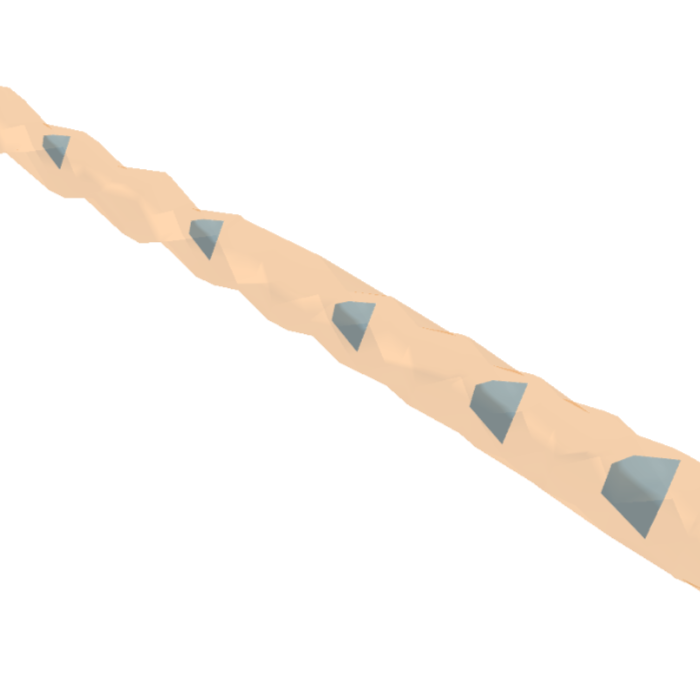}
        \caption{Zoomed red box}
        \label{subfig:tubey_zoom1_global_hidden_features_32x32x32}
    \end{subfigure}
    \vspace{-3mm}
    \caption{Our proposed visualization system highlights the geometric and topological errors introduced by linear interpolation methods and allows users to query local vertex differences between interpolation methods. The first column (\cref{subfig:teardrop_box_32x32x32} and \cref{subfig:tubey_box_32x32x32}) shows the approximated isosurface uncertainty and local selection using the colormap and transparent box, respectively. The second column (\cref{subfig:teardrop_box_LC_LW_32x32x32} and \cref{subfig:tubey_box_LC_LW_32x32x32}) shows the differences between linear and cubic, linear and WENO methods and the approximated error for each vertex inside the transparent boxes. The third column shows a global comparison between linear and WENO interpolation methods. The fourth and fifth columns (\cref{subfig:teardrop_global_hidden_features_32x32x32}, \cref{subfig:teardrop_zoom1_global_hidden_features_32x32x3}, \cref{subfig:tubey_global_hidden_features_32x32x32}, and \cref{subfig:tubey_zoom1_global_hidden_features_32x32x32}) show a comparison between isosurfaces with (transparent orange) and without (opaque blue) possible hidden features that indicates isosurface feature uncertainty.}
    \label{fig:figs_local_plus_hidden_futeres}
}

\abstract{
   Isosurface visualization is fundamental for exploring and analyzing 3D volumetric data. Marching cubes (MC) algorithms with linear interpolation are commonly used for isosurface extraction and visualization.  Although linear interpolation is easy to implement, it has limitations when the underlying data is complex and high-order, which is the case for most real-world data. Linear interpolation can output vertices at the wrong location. Its inability to deal with sharp features and features smaller than grid cells 
   can lead to an incorrect isosurface with holes and broken pieces. 
   Despite these limitations, isosurface visualizations typically do not include insight into the spatial location and the magnitude of these errors. We utilize high-order interpolation methods with MC algorithms and interactive visualization to highlight these uncertainties. Our visualization tool helps identify the regions of high interpolation errors. It also allows users to query local areas for details and compare the differences between isosurfaces from different interpolation methods. In addition, we employ high-order methods to identify and reconstruct possible features that linear methods cannot detect.  We showcase how our visualization tool helps explore and understand the extracted isosurface errors through synthetic and real-world data.
} % end of abstract

\keywords{Marching cubes, linear interpolation, high-order interpolation isosurface uncertainty, weighted essentially non-oscillatory method}

\begin{document}


\firstsection{Introduction}

\maketitle
\input{introduction}

%
\input{related_work}
%
\input{background}
%
\input{methods}

% %
\input{results}
% %
\input{discussion_conclusion}
%%
\input{appendix}
\acknowledgments{
This work was partially supported by the Intel OneAPI CoE, the Intel Graphics and Visualization Institutes of XeLLENCE, the DOE Ab-initio Visualization for Innovative Science (AIVIS) grant 2428225, and the U.S. Department of Energy (DOE) RAPIDS-2 SciDAC project under contract number DE-AC0500OR22725.}

% \newpage
%\bibliographystyle{abbrv}
\bibliographystyle{abbrv-doi}

\bibliography{references}
\end{document}

%% file: introduction.tex
% \ssection{Introduction}
%% importance of surface vis
Isosurface visualization is fundamental for the exploration and analysis of 3D volumetric data. Several domains, including medical imaging \cite{LEE2001, WOLF2005, RISTOVSKI2014, Gillmann2021}, dynamic simulations \cite{Muller2009, EDMUNDS2012}, and crystallography \cite{Spackman2021} rely on isosurface visualization to explore and analyze important features that provide key insight for decision-making. The marching cubes (MC) ~\cite{Willam1987} algorithm is widely used for extracting and visualizing these isosurfaces. The MC algorithms are commonly coupled with linear interpolation. Although local and fast, linear interpolation can introduce significant errors that lead to undesirable visual artifacts and degrade the isosurface approximation ~\cite{Fuhrmann2015, Balazs2019}. In addition, the standard MC algorithms with linear interpolation fail to detect and reconstruct hidden and sharp features not captured by the mesh resolution \cite{Kobbelt2001, Ju2002, Ho2004, Ho2005, NEWMAN2006, YU2008, Dietrich2009}. \blfootnote{This manuscript has been authored by UT-Battelle, LLC under Contract No. DE-AC05-00OR22725 with the U.S. Department of Energy. The publisher, by accepting the article for publication, acknowledges that the U.S. Government retains a non-exclusive, paid up, irrevocable, world-wide license to publish or reproduce the published form of the manuscript, or allow others to do so, for U.S. Government purposes. The DOE will provide public access to these results in accordance with the DOE Public Access Plan (\url{http://energy.gov/downloads/doe-public-access-plan}).}Sharp features correspond to edges and corners, while hidden features are isosurface patches ignored by MC topological cases and linear interpolation. Despite the MC algorithms' well-known errors and limitations, most visualizations don't offer insights into how those errors could impact the extracted isosurface. Instead of treating the extracted isosurface as a complete piece, we observe that some regions are more (or less) trustworthy than others. Characterizing the quality of extracted isosurfaces and the type of errors can help better understand the reliability and limitations of the extracted isosurface and its features. However, obtaining the true surface specification is often impossible due to insufficient information. Scalar fields are commonly stored on a sampled uniform grid. Linear interpolation directly computes the surface's geometry using those values, ignoring some of the important information, such as gradients or high-order coefficients. 
Therefore, it is important to highlight the regions with high errors and variations and allow the users to determine the correct surfaces based on their expertise and domain-specific knowledge.

To investigate the isosurface uncertainty, one considers the errors caused by noisy input data (data uncertainty) and/or the isosurface extraction procedure (model uncertainty). Here, we investigate the latter in the context of interpolation methods. Current approaches have focused on characterizing the uncertainty from noisy data applied to MC with linear interpolation~\cite{Athawale2013, Schlegel2012, Pothkow2013, Athawale2016, Athawale2021, Han2022}. These approaches apply probabilistic methods to the noisy data to estimate the isosurface uncertainty. Other approaches compare the extracted isosurface against a high-resolution target isosurface~\cite{Cignoni1998}. 
%The probabilistic and isosurface comparison methods rely on additional information and may require computationally expensive sampling algorithms. 
Many real-world examples lack finer resolution datasets or target isosurface to compare against. Several studies propose quadratic and cubic interpolation methods to improve the accuracy of level-crossing at each edge, and therefore the isosurface accuracy \cite{Fuhrmann2015, Balazs2019, Cohen-Or2000, Marschner1994, Mitchell1988, CATMULL1974, Theisel2002}. However, these methods don't provide sufficient insight into the accuracy improvement from linear to higher-order interpolation.

To address the issue of feature recovery, several studies extend the original MC algorithm to recover sharp features by inserting additional points or refining the cells containing the sharp features \cite{Kobbelt2001, Ju2002, Ho2004, Ho2005, NEWMAN2006, YU2008, Dietrich2009}. These methods require access to normals or finer-resolution data to identify and recover sharp edges. In practice, finer resolution data are often unavailable and normals at the interior of the cell edges are approximated using linear interpolation, which has large errors as previously indicated. Moreover, these methods do not identify features in cells where all node values are above or below the provided isovalue.

In this paper, we provide insights into uncertainties caused by interpolation methods through interactive visualization and address the aforementioned limitations. 
%We first approximate linear interpolation errors at vertices of extracted surfaces. We provide an overview for users to identify high-error regions based on a user-selected error tolerance. We then provide comparative visualization to allow users to see the difference between linear and other high-order methods. Those are the positional differences between vertices or potential features missed by the linear methods.
%We also allow the users to select a local region and inspect details within the region. 
We summarize our contribution as follows:
% \begin{enumerate}
    (1) We construct an analytical error approximation of the edge-crossing vertex in the MC algorithms for visualizing the reconstructed isosurface error. This method effectively approximates the edge-crossing error and is computationally more efficient than the isosurface comparison methods. In addition, the method is applicable to any volumetric data as it directly computes the estimated error from the volume data and doesn't require sampling or additional information such as ensemble or high-resolution data.
    (2) We introduce a method for detecting and reconstructing possible hidden features missed with the MC algorithm with linear interpolation. We use the divided differences (slopes) of each cell's edges and its neighbors to identify the cell with possible hidden features. The target cells are fitted with local cubic Lagrange polynomials that are then used to divide the cell and reconstruct the hidden features. The current approaches for sharp feature reconstruction don't recover hidden features because they don't consider cells where all node values are above or below the provided isovalue. 
    %Our method can successfully identify and reconstruct these hidden features.
    %
    (3) We present a visualization tool
    % \footnote{We plan to make our visualization tool publicly available upon publication.} 
    that employs error approximation, isosurface positional variation for different interpolation methods, and possible feature reconstruction with other techniques to provide a platform for the exploration and analysis of isosurface uncertainty. The framework can effectively highlight edge-crossing errors and isosurface feature uncertainty. 
    %
% \end{enumerate}

%% file: related_work.tex
\section{Related Work}
\label{sec:related_work}
% Since the introduction of the MC~\cite{Willam1987} algorithm, a significant effort has been invested to evaluate and improve its performance~\cite{NEWMAN2006,Nielson1991,Claudia1994,Neilson2003,shu1995adaptive,Nielson2004,Chen2021,Liao_2018_CVPR,Lewiner2003,SCI:She95a,Michael2007,Bloomenthal19883,Gokul2004}. Here, we discuss several methods for quantifying and visualizing the uncertainty from MC algorithms. 
%First, we highlight techniques for estimating the uncertainty of the crossing location on each edge inside a given cell. Then, we discuss methods for feature preservation that can be used to visualize feature uncertainty in MC.

\subsection{Edge-Crossing Uncertainty}
\label{subsec:isosurface_uncertainty}
%Uncertainty vis for isosurface
The approximation and visualization of isosurface uncertainty is a challenging problem~\cite{Brodlie2012, Johnson2004}. Statistical methods for parametric~\cite{Athawale2013} and nonparametric~\cite{Pothkow2013, Athawale2016} models provide measure metrics that can be used to visualize the most probable isosurface and its uncertainty. For instance, Athawale et al. provided a closed form for computing the expected position and variance of the level-crossing in the MC algorithm for parametric~\cite{Athawale2013} and nonparametric~\cite{Athawale2016} distributions. Topology case count and entropy-based methods can resolve ambiguity in MC algorithm and visualize isosurface uncertainty~\cite{Athawale2021}. The statistical approaches may require solving the level-set crossing problem for each cell many times or sampling methods such as Monte Carlo sampling algorithms that are computationally expensive~\cite{Han2022, Wang2023}. The closed forms in~\cite{Athawale2013, Athawale2016, Athawale2021} improve the computational performance for independent noise models, however, no closed forms are available for more complex noise models such as multivariate Gaussian noise models. The isosurface uncertainty characterized by the statistical methods relies on ensemble data and doesn't explicitly account for the uncertainty from the interpolation method (model uncertainty) which is the focus of this work. 
When the target isosurface is accessible, the uncertainty can be derived by computing the error between the target and approximated isosurfaces~\cite{Cignoni1998, Aspert2002}. 
However, the error computation is computationally expensive as it relies on isosurface sampling, and the target isosurface is often unavailable. 

\subsection{Feature Uncertainty}
\label{subsec:feature_uncertainty}
We consider the isosurface variation from MC with and without feature-preserving methods. These feature uncertainties impact the overall isosurface structure. 
% The standard MC algorithms fail to recover sharp features (edges and corners) that are not captured by mesh resolution. 
Several studies have extended the MC algorithms to incorporate feature-preserving techniques that can recover these sharp features~\cite{Kobbelt2001, Ju2002, Ho2004, Ho2005, NEWMAN2006, YU2008, Dietrich2009, BAGLEY2016}. These methods use information about the cell derivatives to better represent the underlying sharp features. Recently, machine learning-based approaches have been proposed for more accurate MC with feature preservation~\cite{Chen2021, FENG2023, Chen2022, Remelli2020, Gao2020}.

Kobelt et al.~\cite{Kobbelt2001} propose a surface extraction method from directed distance field and surface normals of a geometric object that preserves sharp features. The normals are used to detect the sharp features and new sample points are added inside the cell to recover the hidden features. The Dual contouring algorithm proposed in~\cite{Ju2002} uses the edges intersection and the normals at those intersections to process the cells with sharp features. This method doesn't explicitly require identifying the cell with sharp features because it uses a quadratic error function to automatically place the additional points. Ho et al.~\cite{Ho2005} propose sampling the edges normals to detect the cell with sharp features in volumetric data. These cells are then subdivided to represent the sharp features. The adaptive refinement of the cell requires access to a finer-resolution version of the data.

% These techniques for sharp feature recovery do not recover hidden features and may require access to a finer resolution of the volume data to both detect and subdivide the target cells. In many practical real-world examples, finer-resolution data are often not available. We propose to employ the local neighboring cells to identify the target cell with possible hidden features and construct a polynomial approximation that can then be used to subdivide the cell.
%

%% file: background.tex
\section{Technical Background}
\label{sec:background}
% Here, we briefly describe the MC algorithm ~\cite{Willam1987} and the interpolation methods considered in this work.

\subsection{Marching Cubes Algorithm} 
\label{sec:mcAlgorithm}
The MC algorithm~\cite{Willam1987} extracts the isosurface as it steps through each cubical cell of the uniform grid. For a single cell, the algorithm first determines the topological configuration based on the relationship between the value on each vertex and the isovalue $k$. The values on the vertices could be either larger or smaller than the isovalue. On each edge, the isosurface crosses the edge if one of the vertex has a value larger than the isovalue while the other is smaller. We connect the edge-crossing points to form surfaces contributing to the final output surface. Comparing vertex values only shows whether there is an edge-crossing point. To determine the exact location of the edge-crossing point, we need to identify the point on the edge where the value is equal to our isovalue. Therefore, we need a method for interpolation between the two vertices of an edge. We will introduce different methods of interpolation in the rest of this section.

\subsection{Linear Interpolation} 
\label{subsec:linear}

Estimating edge-crossing points using linear approximation has advantages in terms of speed and simplicity of the mathematical model. Let $f(x_{0})$ and $f(x_{1})$ be the scalar values sampled at vertex positions $x_{0}$ and $x_{1}$ denoting ends of a cell edge, respectively. The crossing position for the isovalue $k$ on this cell edge is determined by finding $x$ such that $k = f(x_{0}) + \frac{f(x_{1})-f(x_{0})}{x_{1}-x_{0}} (x-x_{0})$. The solution is $x= x_{0} + \frac{x_{1}-x_{0}}{f(x_{1})-f(x_{0})}(k-f(x_{0}))$. To take advantage of vector arithmetic the solution can written as follows: $x = \alpha * x_1 + (1-\alpha)*x_0$, where  $\alpha = \frac{k - f(x_{0})}{f(x_{1})-f(x_{0})}$. 

\subsection{Cubic Interpolation}  
\label{sec:cubic}

Although linear interpolation is efficient, it may lead to significant approximation errors that degrade the quality of the extracted isosurface. Many studies have proposed higher order interpolation methods to improve the accuracy of the level crossing at each edge, and therefore the isosurface accuracy~\cite{Fuhrmann2015, Balazs2019, Cohen-Or2000,Marschner1994, Mitchell1988, CATMULL1974, Theisel2002}. 

For the same edge (or interval) considered in \cref{subsec:linear} the cubic interpolant is $q(x) = c_{0} + c_{1}x + c_{2}x^2  + c_{3}x^{3}$ with $x \in [ x_{0}, x_{1} ]$. The cubic polynomial has four degrees of freedom and requires solving a $4\times4$ system of linear equations to compute the coefficients $c_{i}$, $i=0, \cdots, 3$. A common approach is to use the sampled data values and derivatives at the edge endpoints to build the system of linear equations and find the coefficients. Let 
% \color{blue}
$(f(x_{0}), f^{'}(x_{0}))$ and $(f(x_{1}), f^{'}(x_{1}))$ 
% \color{black}  
be the data values at $x_{0}$ and $x_{1}$, respectively. The coefficients are obtained by solving
\begin{equation*} 
% \begin{aligned}
    q(x_{0}) =  f(x_{0}) \quad  q(x_{1}) =  f(x_{1}) \quad q^{'}(x_{0}) =  f^{'}(x_{0})\quad   q^{'}(x_{1}) = f^{'}(x_{1}).
% \end{aligned}
\end{equation*}
The crossing position for isovalue $k$ on this edge is obtained by finding the roots to $q(x)=k$. We note that the derivatives $f^{'}(x_{0})$ and $f^{'}(x_{1})$ are often not available and therefore approximated using finite difference methods.

\subsection{WENO Interpolation}
\label{subsec:weno}
The weighted essentially non-oscillatory (WENO) method~\cite{ZHANG2016103} is a high-order polynomial reconstruction method developed for solving hyperbolic and convection-diffusion equations. 
WENO achieves high-order accuracy in smooth regions and provides a better representation of regions with sharp gradients compared to standard Lagrange interpolation~\cite{hildebrand1987introduction}.

Let's consider the  1D mesh $\mathcal{M} = \{ \cdots, x_{i-2}, x_{i-1}, x_{i}, x_{i+1}, \cdots \}$ where $i \in \mathbb{N} \cup \{ 0 \}$. For each edge $E_{i}$ where the isovalue $k$ lies between $f_{i}$ and $f_{i+1}$, a high-order polynomial $p_{i}(x)$ is used to approximate the function inside the interval defined by the edge boundaries. The edge-crossing is obtained by finding the roots of the implicit equation $p_{i}(x)=k$. The final interpolant $p_{i}(x)$ is a convex combination of the third-order polynomials 
% \color{blue} 
$p^{(1)}(x)$, $p^{(2)}(x)$, and $p^{(3)}(x)$.
% \color{black}
\begin{equation}\label{eq:p_i}
    p_{i}(x) = w_{1}p^{(1)}(x) + w_{2}p^{(2)}(x) + w_{3}p^{(1)}(x), 
\end{equation}
where $w_{1}, w_{2}$, and $w_{3}$ are nonlinear weights such that $w_{1}+w_{2}+w_{3} =1$. The nonlinear weights are obtained using the ``smoothness" indicator $\beta_{j}$ in~\cite{Jiang1996} that can be approximated as follows:
\begin{equation}
\begin{aligned}\label{eq:smoothness_indicators}
    \beta_{1} = & 13/12 \big( f_{i-2} - 2f_{i-1} + f_{i}\big)^{2} + 1/4 \big( f_{i-2}-4f_{i-1}+ 3u_{i}\big)^{2}, \\
    \beta_{2} = & 13/12 \big( f_{i-1} - 2f_{i} + f_{i+1}\big)^{2} + 1/4 \big( f_{i-1}-f_{i+1}\big)^{2}, \\
    \beta_{3} = & 13/12 \big( f_{i} - 2f_{i+1} + f_{i+2}\big)^{2} + 1/4 \big( 3f_{i}-4f_{i+1}+ f_{i+2}\big)^{2}. \\
\end{aligned}
\end{equation}
The nonlinear weights are dependent on the constants $\gamma_{1} = 1/10$, $\gamma_{2} = 3/5$, $\gamma_{3} = 3/10$.
% \begin{equation} \label{eq:linear_weigts}
%     \gamma_{1} = 1/10, \quad \gamma_{2} = 3/5, \quad \gamma_{3} = 3/10.
% \end{equation}
%
Using the ``smoothness" indicator in \cref{eq:smoothness_indicators} and the constants, the nonlinear weight can be expressed as follows:
\begin{equation}
    w_{j} = \alpha_{j}/(\alpha_{1}+\alpha_{2}+\alpha_{3}), \quad \alpha_{j} = \gamma_{j}/(\epsilon+\beta_{j})^{2}, \quad j=1,2,3.
\end{equation}
The parameter $\epsilon$, typically set to $10^{-6}$, is introduced to avoid division by zero. Jiang and Shu~\cite{Jiang1996} proved that in smooth regions the approximation $p_{i}(x)$ in \cref{eq:p_i} is fifth-order accurate.  
Solving $p_{i}(x)=k$ is equivalent to a root-finding problem for a cubic polynomial. The roots for the WENO polynomial can be found using the cubic formula in~\cite{weisstein2002}. Similar to~\cite{Fuhrmann2015}, the median solution is selected in the cases where multiple valid roots are found. 
% Overall, the WENO method uses a convex combination of third-order (cubic) polynomials with nonlinear weights to achieve fifth-order accuracy. 

%% file: methods.tex
%%%
\begin{figure*}[!htb]
    \centering
    \begin{subfigure}[b]{0.195 \textwidth}
        \centering
        \includegraphics[width= 0.80 \columnwidth]{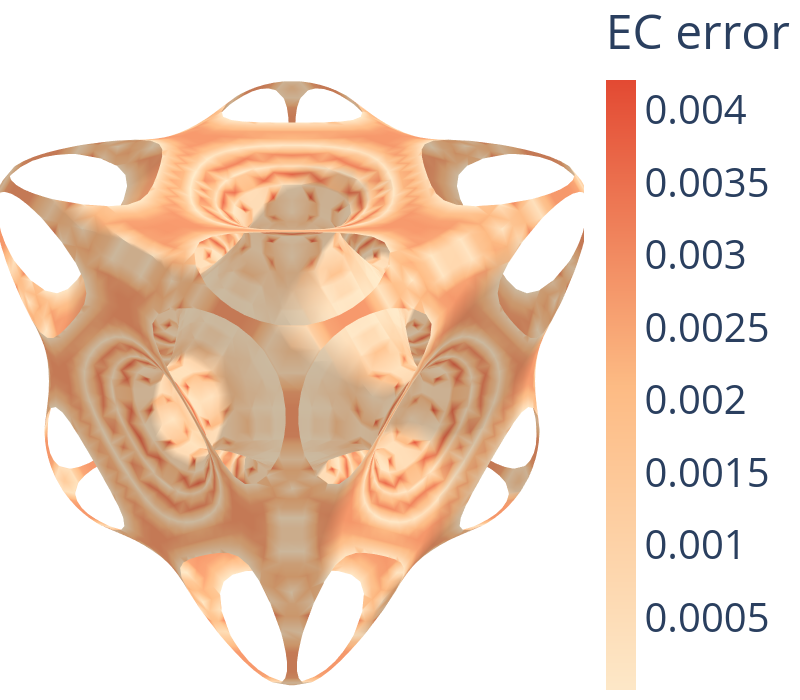}
        \vspace{-2.5mm}
        \caption{Tangle Error ($32^{3}$)}
        \label{subfig:tangle_512x512x512_32x32x32}
    \end{subfigure}
    \begin{subfigure}[b]{0.195 \textwidth}
        \centering
        \includegraphics[width= 0.80 \columnwidth]{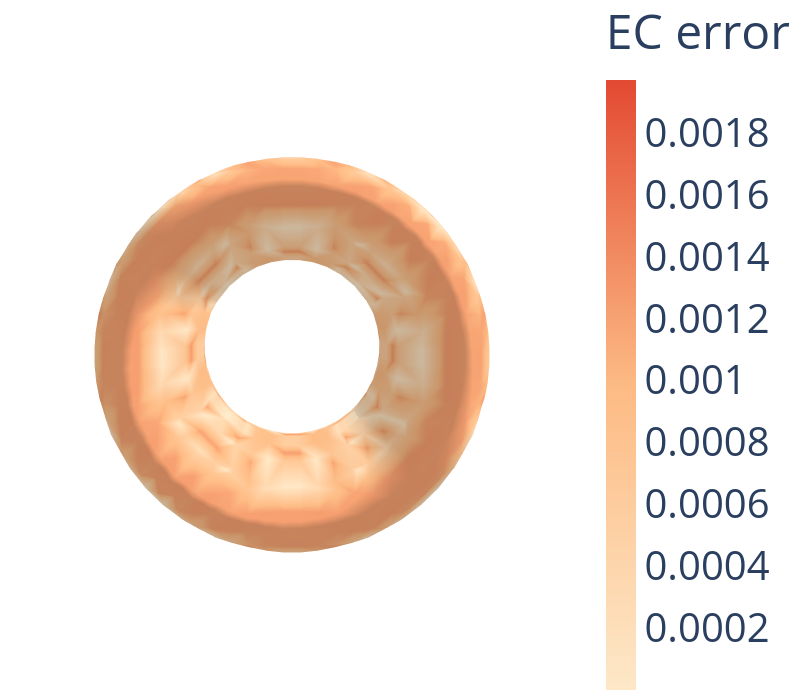}
        \vspace{-2.5mm}
        \caption{Torus Error ($64^{3}$)}
        \label{subfig:torus_512x512x512_64x64x64}
    \end{subfigure}
    \begin{subfigure}[b]{0.195 \textwidth}
        \centering
        \includegraphics[width= 0.80 \columnwidth]{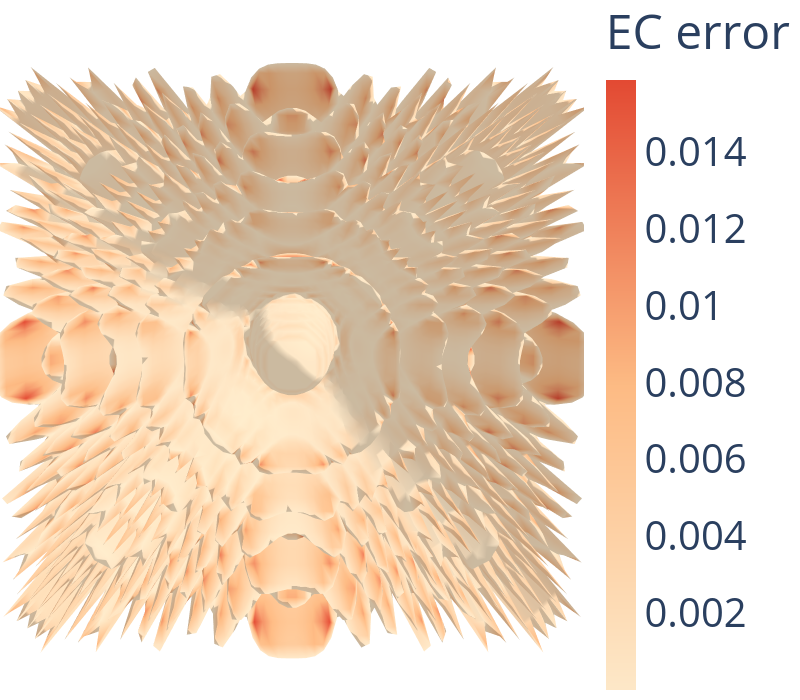}
        \vspace{-2.5mm}
        \caption{M. and L. Error ($64^{3}$)}
        \label{subfig:marschnerlobb_512x512x512_32x32x32}
    \end{subfigure}
    \begin{subfigure}[b]{0.195 \textwidth}
        \centering
        \includegraphics[width= 0.80 \columnwidth]{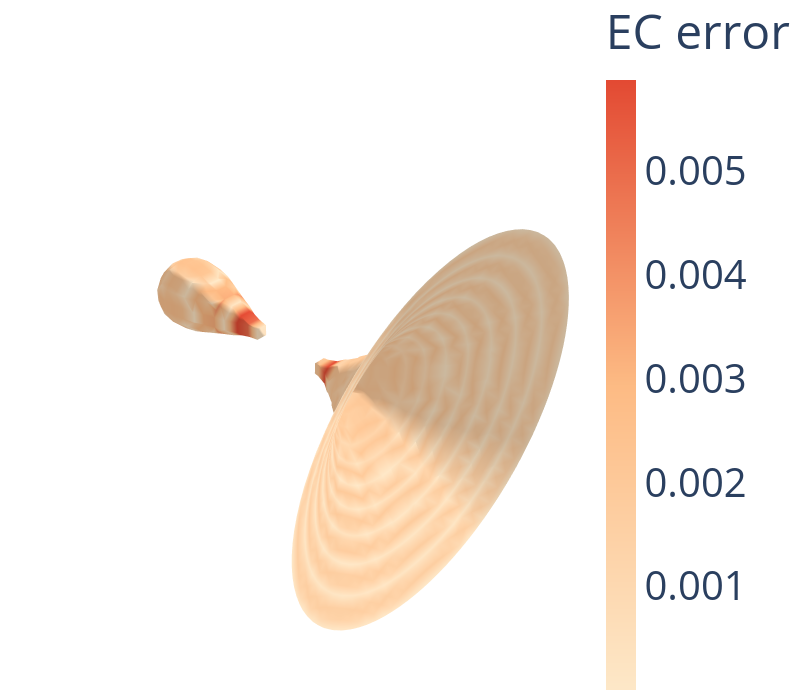}
        \vspace{-2.5mm}
        \caption{Teardrop Error ($32^{3}$)}
        \label{subfig:teardrop_512x512x512_32x32x32}
    \end{subfigure}
    %% %%
    \begin{subfigure}[b]{0.195 \textwidth }
        \centering
        \includegraphics[width= 0.80 \columnwidth]{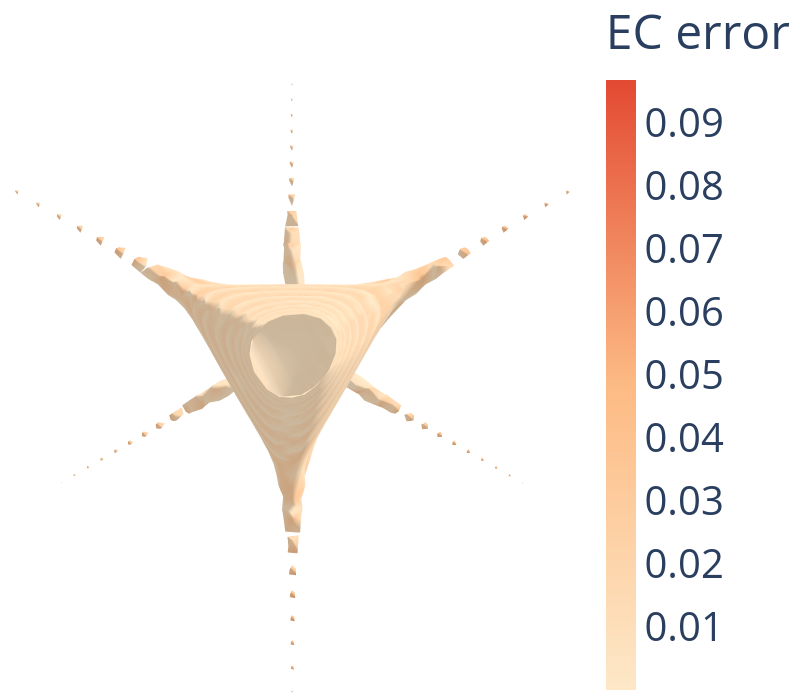}
        \vspace{-2.5mm}
        \caption{Tubey Error ($32^{3}$)}
        \label{subfig:tubey_512x512x512_32x32x32}
    \end{subfigure}
    \begin{subfigure}[b]{0.195 \textwidth}
        \centering
        \includegraphics[width= 0.80 \columnwidth]{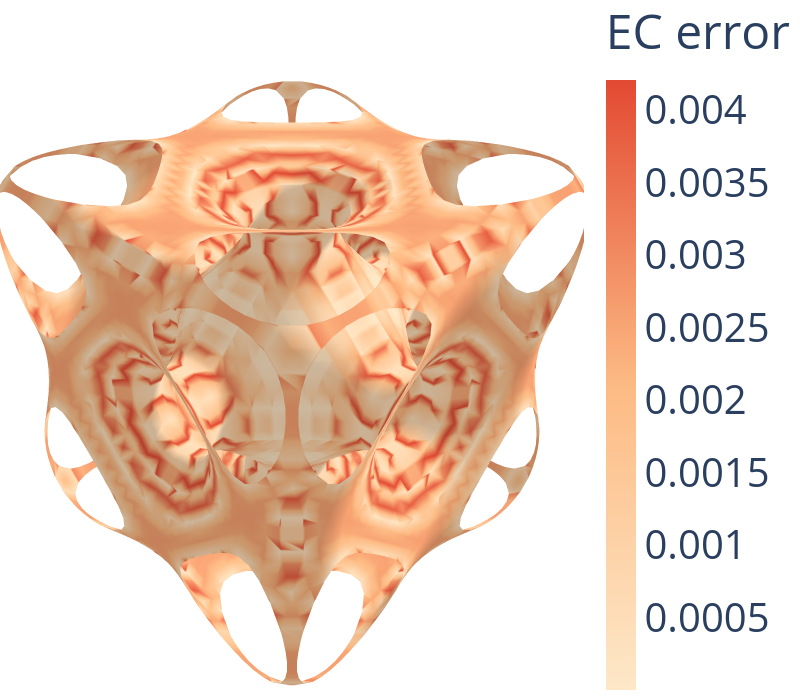}
        \vspace{-2.5mm}
        \caption{Tangle Approx. Error ($32^{3}$)}
        \label{subfig:tangle_Approx_32x32x32}
    \end{subfigure}
    \begin{subfigure}[b]{0.195 \textwidth}
        \centering
        \includegraphics[width= 0.80 \columnwidth]{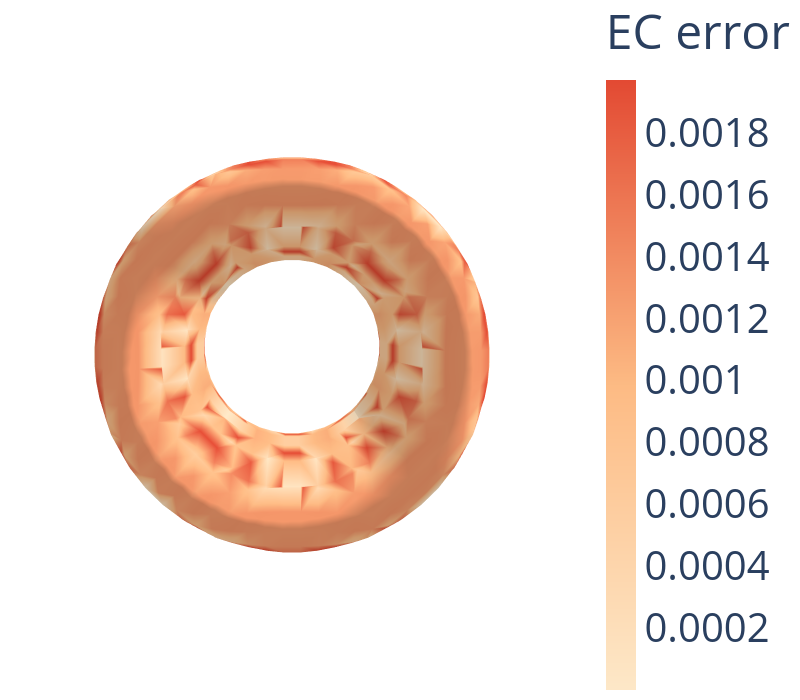}
        \vspace{-2.5mm}
        \caption{Torus Approx. Error ($64^{3}$)}
        \label{subfig:torus_Approx_64x64x64}
    \end{subfigure}
    \begin{subfigure}[b]{0.195 \textwidth}
        \centering
        \includegraphics[width= 0.80 \columnwidth]{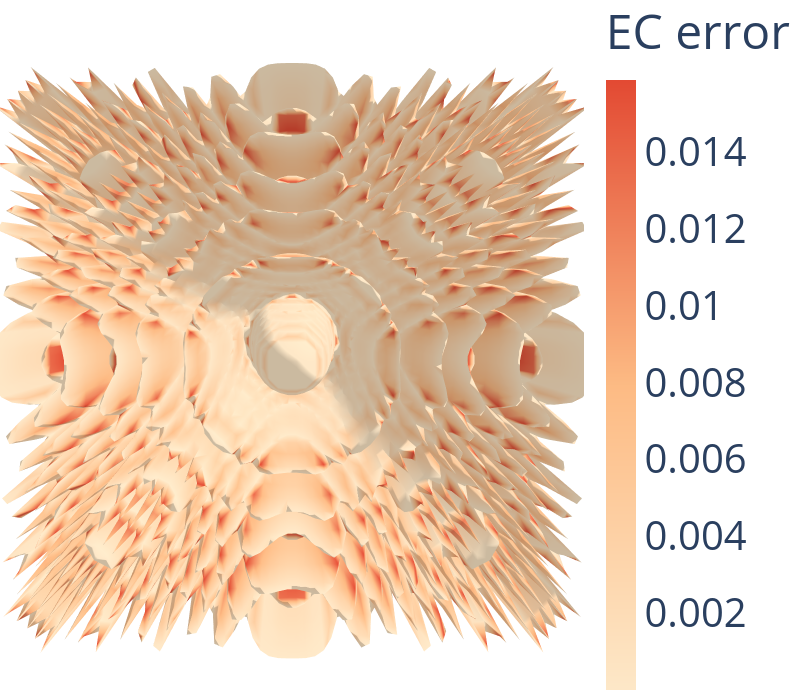}
        \vspace{-2.5mm}
        \caption{M. and L. Approx. Error ($64^{3}$)}
        \label{subfig:marschnerlobb_Approx_32x32x32}
    \end{subfigure}
    \begin{subfigure}[b]{0.195 \textwidth}
        \centering
        \includegraphics[width= 0.80 \columnwidth]{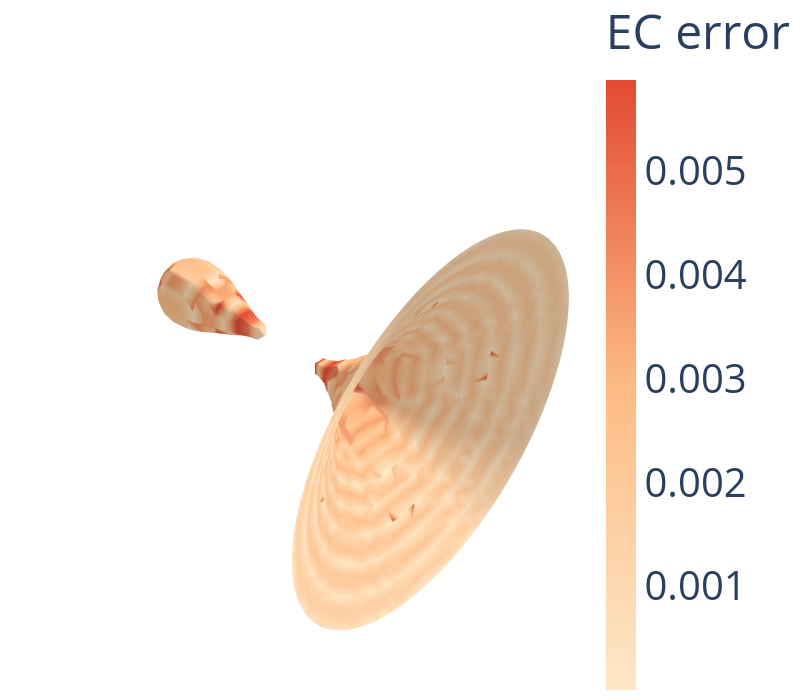}
        \vspace{-2.5mm}
        \caption{Teardrop Approx. Error ($32^{3}$)}
        \label{subfig:teardrop_Approx_32x32x32}
    \end{subfigure}
    \begin{subfigure}[b]{0.195 \textwidth}
        \centering
        \includegraphics[width= 0.80 \columnwidth]{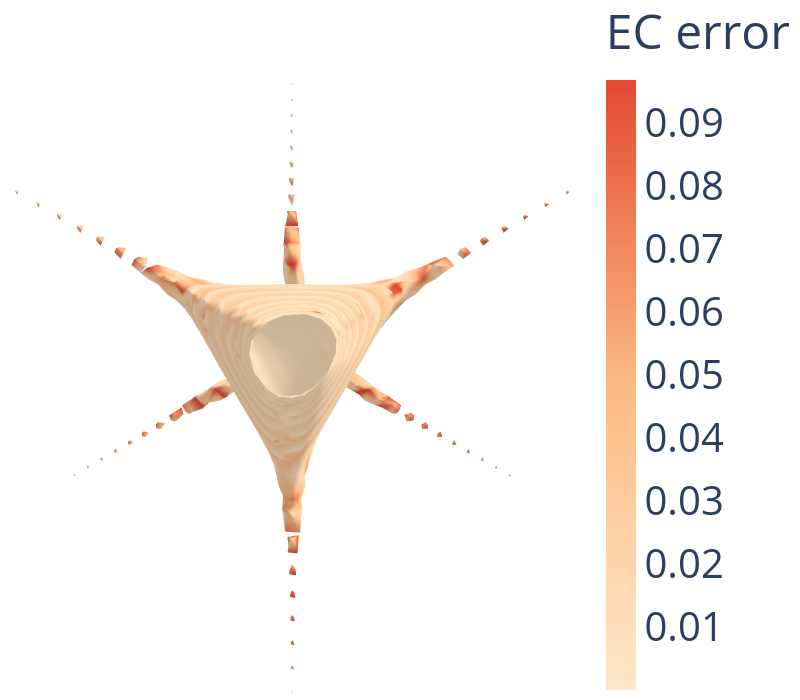}
        \vspace{-2.5mm}
        \caption{Tubey Approx. Error ($32^{3}$)}
        \label{subfig:tubey_Approx_32x32x32}
    \end{subfigure}
    \vspace{-3mm}
    \caption{Comparison between measured and approximate error. The first row corresponds to the measured error and the second to the approximated error. Each  column from left to right corresponds the \textbf{Tangle} ( \cref{subfig:tangle_512x512x512_32x32x32} and \cref{subfig:tangle_Approx_32x32x32} with $k=0.1$), \textbf{Torus} ( \cref{subfig:torus_512x512x512_64x64x64} and \cref{subfig:torus_Approx_64x64x64} with $k=0.0$), \textbf{Marschner and Lobb} (\cref{subfig:marschnerlobb_512x512x512_32x32x32} and \cref{subfig:marschnerlobb_Approx_32x32x32} with $k=0.5$), and \textbf{Teardrop} (\cref{subfig:teardrop_512x512x512_32x32x32} and \cref{subfig:teardrop_Approx_32x32x32} with $k=-0.001$) and \textbf{Tubey} (\cref{subfig:tubey_512x512x512_32x32x32} and \cref{subfig:tubey_Approx_32x32x32} with $k=0.0  $) examples. Our approximated errors show similar patterns to the measured errors. In most cases, it slightly overestimates the errors.}
    \label{fig:edge_crossing_error}
\end{figure*}
%%%
%%%
\begin{figure*}[!htb]
    \centering
    \begin{subfigure}[b]{0.195 \textwidth}
        \centering
        \includegraphics[width= 0.99 \columnwidth]{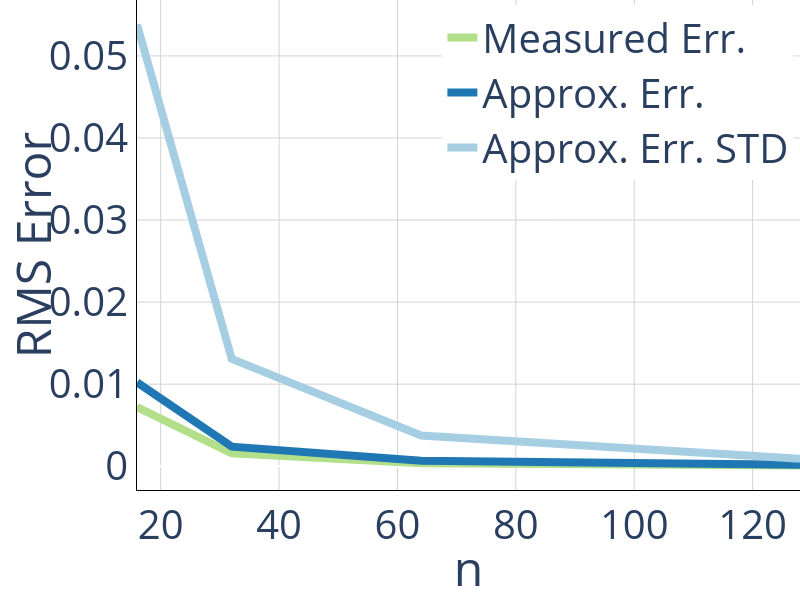}
        \vspace{-5mm}
        \caption{Tangle}
        \label{subfig:tangle_rms_error}
    \end{subfigure}
    \begin{subfigure}[b]{0.195 \textwidth}
        \centering
        \includegraphics[width= 0.99 \columnwidth]{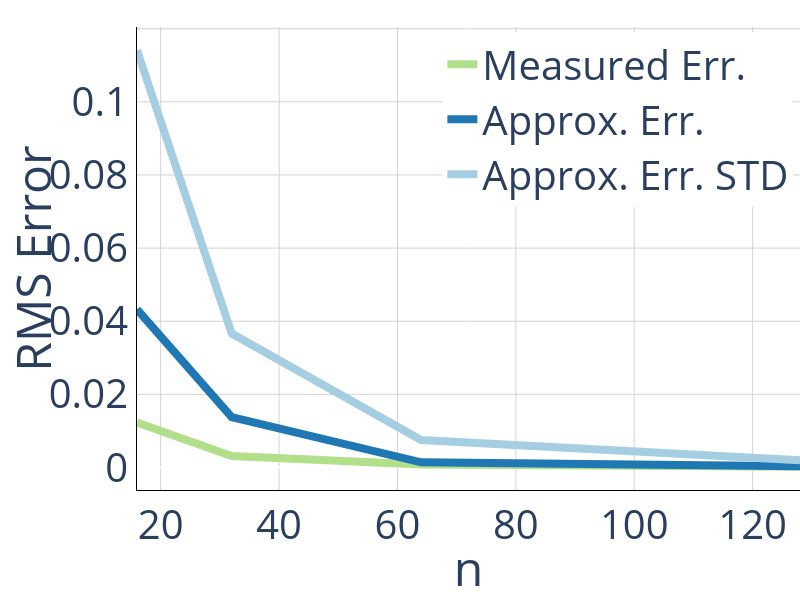}
        \vspace{-5mm}
        \caption{Torus}
        \label{subfig:torus_rms_error}
    \end{subfigure}
    \begin{subfigure}[b]{0.195 \textwidth}
        \centering
        \includegraphics[width= 0.99 \columnwidth]{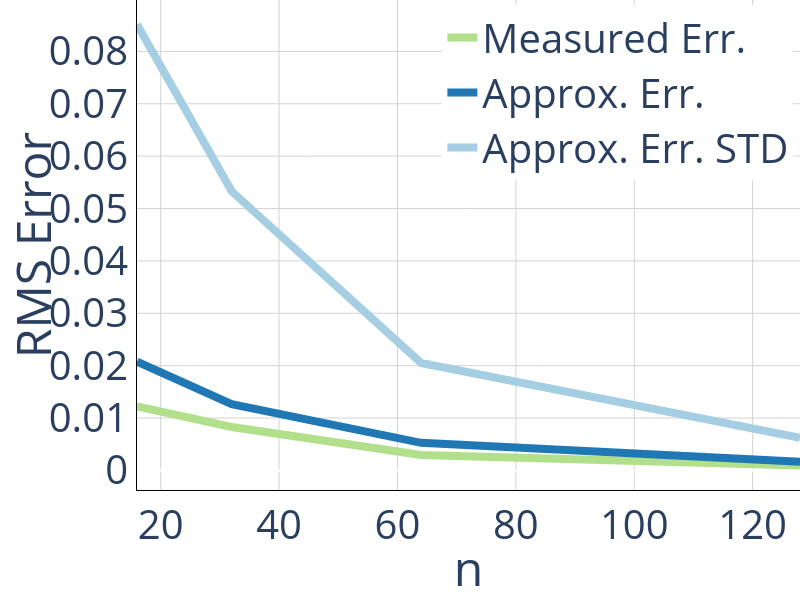}
        \vspace{-5mm}
        \caption{Marschner and Lobb}
        \label{subfig:marschnerlobb_rms_error}
    \end{subfigure}
    \begin{subfigure}[b]{0.195 \textwidth}
        \centering
        \includegraphics[width=0.99 \columnwidth]{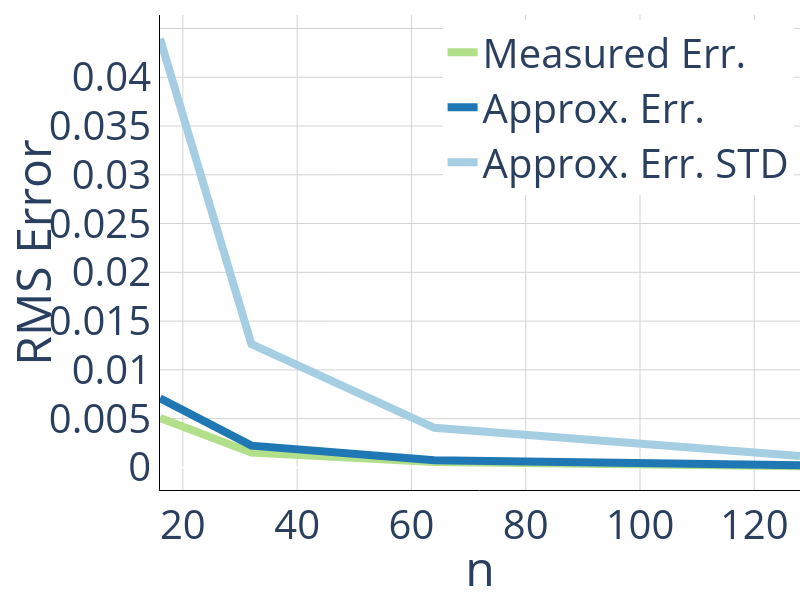}
        \vspace{-5mm}
        \caption{Teardrop}
        \label{subfig:teardrop_rms_error}
    \end{subfigure}
    \begin{subfigure}[b]{0.195 \textwidth}
        \centering
        \includegraphics[width=0.99 \columnwidth]{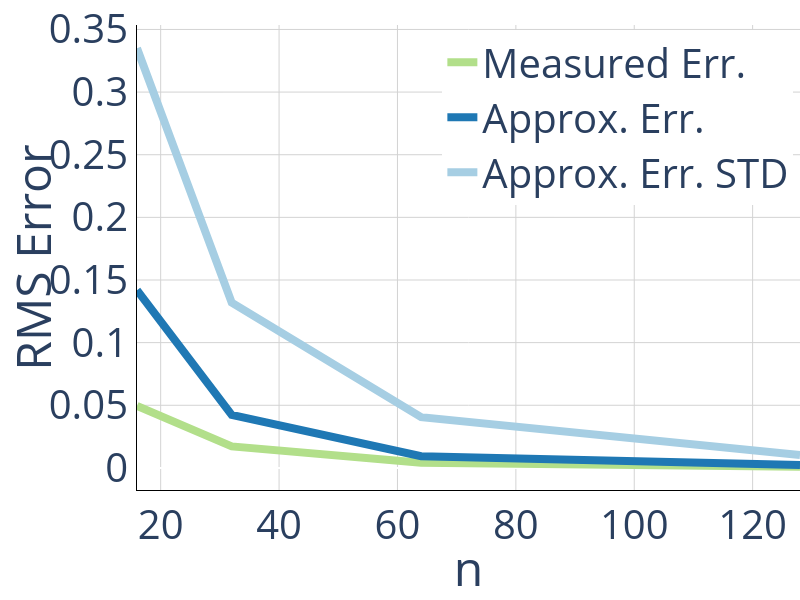}
        \vspace{-5mm}
        \caption{Tubey}
        \label{subfig:tubery_rms_error}
    \end{subfigure}
    %%%%%%%%%%%%%%%%%%%%%%%%%%%%%%%%%%%%%%%%%%%%%%%%%%%%%%%%%%%%%%%%%%%%%%%%%%%%%%%%%%%%%%%%
    \vspace{-3mm}
    \caption{Comparison of measured (in green), our estimated (in blue), and standard approach for error approximation (light blue). The columns from left to right show errors for the \textbf{Tangle} (\cref{subfig:tangle_rms_error}), \textbf{Torus} (\cref{subfig:torus_rms_error}), \textbf{Marschner and Lobb} (\cref{subfig:marschnerlobb_rms_error}), \textbf{Teardrop} (\cref{subfig:teardrop_rms_error}), and Tubey (\cref{subfig:tubery_rms_error}). Our error estimation in \cref{eq:approx_linear_error} is much closer to the measured error compared to the standard approach in \cref{eq:error_bound_approx}.}
    \label{fig:figs_max_mean_rms}
    % \vspace{-7mm}
\end{figure*}

\section{Method}
\label{sec:method}

% This section introduces the different methods for visualizing and analyzing isosurfaces obtained from the MC algorithms. 

\subsection{Edge-Crossing Error Approximation}
\label{subsec:error_approximation}
\begin{figure}[h]%[!htb]
        \centering
        \includegraphics[width=0.35\columnwidth]{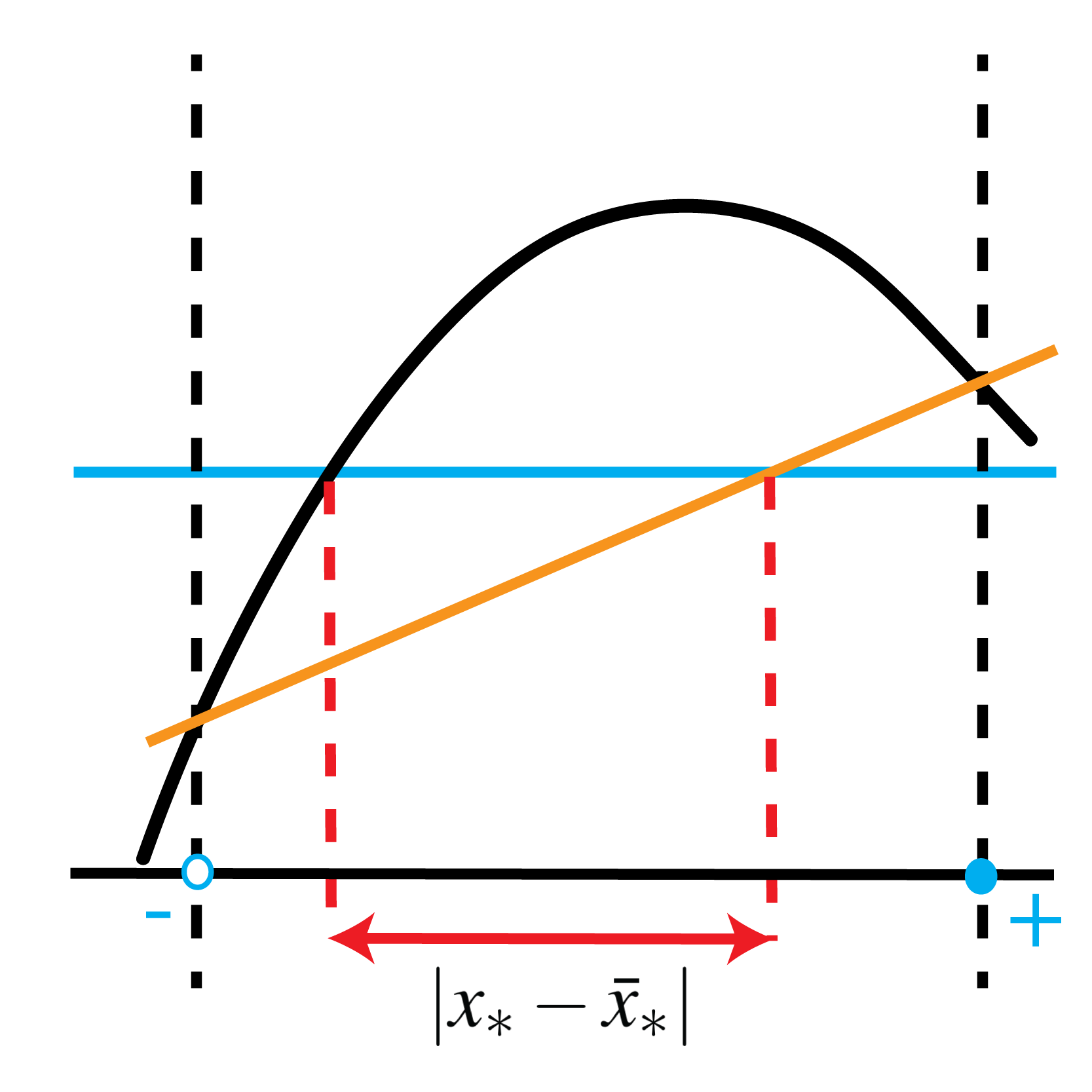}
        \vspace{-3mm}
        \caption{Edge-crossing uncertainty. The underlying function and the linear interpolation are shown in black and orange, respectively. The black line segment with the positive and negative nodes is the edge and the blue line indicates the target isovalue. The red double arrow indicates the approximation error. 
        % \color{blue} 
        The isovalue is indicated by the blue horizontal line. 
        % \color{black}
        }
        \label{fig:vertex-error-approximation}
    \end{figure}

%
    % \color{blue}
    We propose an edge-crossing error approximation for MC algorithms derived from polynomial interpolation error, which is used to visualize isosurface discrepancies and highlight regions with significant errors. 
    % \color{black}
    Taylor series expansion is widely used for local function and error approximation function. For instance, in the context of visualization, Moller et al.~\cite{Moller1998, Moller1997, Moller1996} use it to develop smooth filters for volume data approximation and estimate their local error. We distinguish our approach by using the Taylor series expansion to estimate the edge-crossing error, which has the advantage of providing insight into the isosurface reconstruction error without the need to solve a linear system. The interpolation error from the linear approximation $\ell(x)$ of the function $f(x)$ is
    \begin{align}\label{eq:error}
        e(x) =  \ell(x) - f(x) = \frac{f^{''}(\xi)}{2}(x-x_{i})(x-x_{i+1}),
    \end{align}
    where $\xi \in (x_{i-1}, x_{i+2})$. The MC algorithms solve the implicit problem $\ell(x)=k$ where $k \in \mathbb{R}$ is the isovalue. 
    % \color{blue} 
    Let $x_{*}$ and $\bar{x}_{*}$ the solutions to $f(x) =k$ and $\ell(x)=k$. Substituting the solutions $x_{*}$ and $\bar{x}_{*}$ into $f(x)$ and $\ell(x)$ gives. 
    % \color{black} 
    %
    \begin{align}\label{eq:l_of_x}
        \ell(\bar{x}_{*}) = f_{i}+U[i, i+1] (\bar{x}_{*}-x_{i}) =k, \textrm{ and}
    \end{align}
    \begin{align}\label{eq:f_of_x}
        f(x_{*}) = f_{i}+U[i, i+1] (x_{*}-x_{i}) + \frac{f^{''}(\xi)}{2}(x_{*}-x_{i})(x_{*}-x_{i+1}) = k.
    \end{align}
    The divided difference $U[i, i+1]$ is recursively defined as
    \begin{equation}
        \begin{gathered}
        U[i] = f(x_{i})=f_{i}, \quad U[i,i+1] = \frac{U[i+1] -U[i]}{x_{i+1} - x_{i}}, \\
        U[i, i+j] = \frac{U[i+1, \cdots, i+j]- U[i, \cdots, i+j-1]}{x_{i+j}-x_{i}},    
        \end{gathered}
    \end{equation}
    with $j$ being an integer. Subtracting \cref{eq:l_of_x} from \cref{eq:f_of_x} gives the edge-crossing error 
    \begin{align}
        |x_{*}-\bar{x}_{*}| = \Big|\frac{1}{U[i, i+1]}\frac{f^{''}(\xi)}{2} (x_{*}-x_{i})(x_{*}-x_{i+1}) \Big|.
    \end{align}
    The edge-crossing approximation error can be bounded by 
    \begin{align}\label{eq:error_bound}
        |x_{*}-\bar{x}_{*}| \leq \frac{\max_{\xi \in[x_{i}, x_{i+1}]} |f^{''}(\xi)|}{2U[i, i+1]}(x_{*}-x_{i})(x_{*}-x_{i+1})
    \end{align}
    In practice, $\xi$, $f^{''}$, and $x_{*}$ are not available. Typically the term $\max_{\xi \in[x_{i}, x_{i+1}]} |f^{''}(\xi)|$ is approximation using finite difference $max\big( |U[i-1, i+1]|, |U[i, i+2]| \big)$. The product $(x_{*}-x_{i})(x_{*}-x_{i+1})$ is approximate with the interval size $(x_{i}-x_{i+1})^{2}$. The error bound on the right side of \cref{eq:error_bound} is approximated as follows:
    \begin{equation}\label{eq:error_bound_approx}
        \bar{e}_{b} = \frac{max\Big( |U[i-1, i+1]|, |U[i, i+2]| \Big)}{U[i, i+1]}(x_{i}-x_{i+1})^{2} 
    \end{equation}
    The approximation $\bar{e}_b$ in \cref{eq:error_bound_approx} tends to overestimate the edge-crossing errors $|x_{*}-\bar{x}_{*}|$.
    We estimate the vertex approximation error $\bar{e} \approx |x_{*}-\bar{x}_{*}|$ as follows:
    \begin{equation}\label{eq:approx_linear_error}
        \bar{e} = \frac{max\big( |U[i-1, i+1]|, |U[i, i+2]|  \Big)}{U[i, i+1]}(\bar{x}_{*}-x_{i})* (\bar{x}_{*}-x_{i+1}) 
    \end{equation}
    \cref{eq:approx_linear_error} provides a much tighter approximation of the error compare to \cref{eq:error_bound_approx}. \cref{fig:vertex-error-approximation} shows the difference between the underlying function (black curve) and its linear approximation (orange line). The black horizontal line indicates the edge considered and the blue line shows the isovalue position of the underlying function (black curve) and the linear interpolation (orange line).

We use several datasets to evaluate the approximated edge-crossing error introduced in \cref{eq:approx_linear_error}. The volume datasets are sampled from a \textbf{Tangle}, \textbf{Torus}\cite{Knoll2007}, \textbf{Marschner and Lobb}\cite{Marschner1994}, \textbf{Teardrop} \cite{Knoll2007}, and \textbf{Tubey}~\cite{Bourke2003} functions. Their equations are provided in the appendix.
    
The functions are sampled on a $512^{3}$ uniform mesh to construct the high-resolution data from which the target isosurfaces are extracted. The isosurface error is obtained by calculating the difference between the high- (target) and coarse-resolution isosurfaces using METRO \cite{Cignoni1998}. The results in \cref{fig:edge_crossing_error} and  \cref{fig:figs_max_mean_rms} evaluate and validate the edge-crossing error introduced in \cref{eq:approx_linear_error}. The first and second rows in \cref{fig:edge_crossing_error} show the measured and approximated errors, respectively. The label ``EC error" represents the edge-crossing error. In each example, the measured and approximated errors exhibit similar patterns, demonstrating that the edge-crossing error approximation offers a fast and reliable estimate of the isosurface error arising from linear interpolation. This method is computationally more efficient because the approximated error is directly computed using \cref{eq:approx_linear_error} and doesn't require additional data, computation, or sampling algorithm as in the case of statistical and isosurface comparison methods. For instance, the isosurface extraction, and error estimation for the tangle example in \cref{subfig:tangle_Approx_32x32x32} takes less than a millisecond ($1 ms$) whereas the measured error in \cref{subfig:tangle_512x512x512_32x32x32} takes a few seconds. This cost is significantly magnified with the increase in grid resolution which can hinder the interactivity of visualization. Our approach provides a much more efficient way of visualizing linear interpolation uncertainty with the proposed approximation. The estimated isosurface uncertainty provides quick insight into the quality of the isosurface that can guide decisions about using higher-order interpolation, higher-resolution data, and/or other methods to improve the isosurface quality. %
    
The results in \cref{fig:figs_max_mean_rms} show the root mean squares (RMS) errors. The green line corresponds to the measured errors, the blue to our introduced method, and the light blue to the standard approach for polynomial error approximation. The standard approach (in light blue) ~\cite{hildebrand1987introduction}
significantly overestimates the edge-crossing errors. Our method provides a better approximation of the measured error that can be utilized for visual analysis of isosurface error.

\subsection{Hidden Features Detection and Reconstruction}
\label{subsec:hidden_features}
    The MC algorithms with and without sharp feature recovery fail to identify and reconstruct hidden features not captured by the mesh resolution. These hidden features are isosurface patches not detected by linear interpolation and the topological cases considered in MC algorithms. The hidden features can alter the isosurface connectivity and its overall structure. We propose a method for hidden feature detection and reconstruction that relies on the slopes (divided differences) on cell edges and high-order interpolation. We utilize this method to offer the user two possible isosurface reconstructions: one with hidden feature recovery and one without. Both isosurfaces are visualized together to highlight the differences and provide insight into the feature differences. 
    
    A cell might have a hidden feature if for any of its edges two of the three slopes $U[i-1, i]$, $U[i, i+1]$, and $U[i, i-1]$ of neighboring edges have opposite signs. The detected cell is divided into smaller subcells using tri-cubic Lagrange polynomial interpolation. We note that using linear interpolation instead for the cell refinement does not recover the missing hidden features. The MC algorithm is then applied to each subcell to reconstruct the hidden features.
   
\begin{figure}[!h]
        \centering
        \begin{subfigure}[b]{0.60 \columnwidth}
            \centering
            \includegraphics[width=0.95\columnwidth]{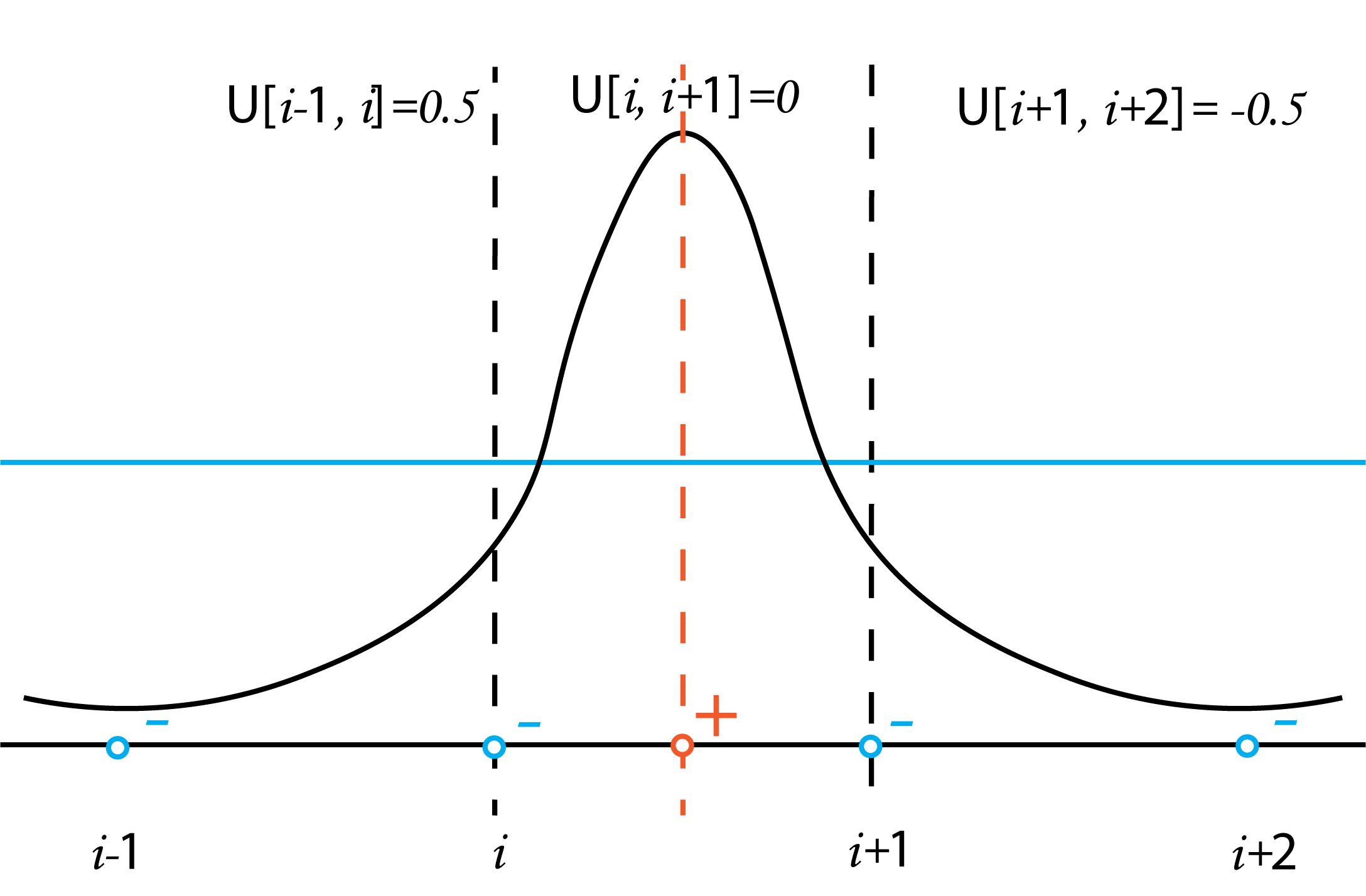}
            \caption{1D hidden feature recovery}
            \label{subfig:hidden_feature_1D}    
        \end{subfigure}
        \begin{subfigure}[b]{0.39 \columnwidth}
            \centering
            \includegraphics[width=0.95 \columnwidth]{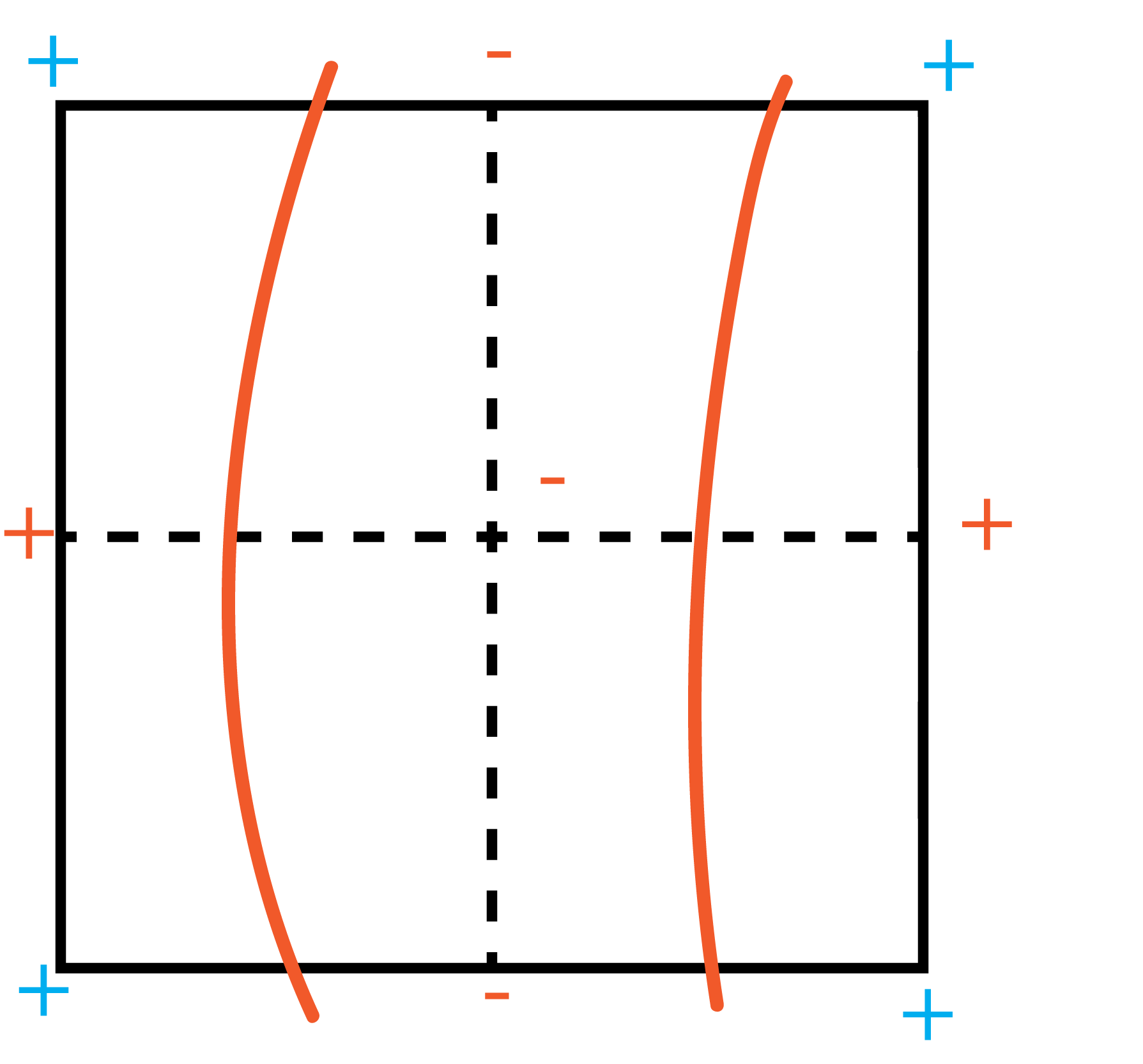}
            \caption{2D hidden feature recovery}
            \label{subfig:hidden_feature_2D}    
        \end{subfigure}
        \vspace{-3mm}
        \caption{Hidden feature recovery in 1D and 2D.\cref{subfig:hidden_feature_1D} shows an example of a hidden feature between $i$ and $i+1$ that can be detected by noting that  $U[i-1, i+1]*U[i+1, i+2] < 0$, meaning the slopes surrounding the hidden features have a different sign. Our method subdivides the cell at the orange dotted line to recover the hidden feature. 
        % \color{blue} 
        The isovalue is indicated by the blue horizontal line. 
        % \color{black}
        \cref{subfig:hidden_feature_2D} shows a 2D example with hidden features that can be recovered by refining the cell. The orange curves are the isocontour inside the cell. MC will miss the contour because all four corners have the same sign. Our method identifies the hidden feature and subdivides the cell at the dotted black lines.}
        \label{fig:hidden_feature_illustrations}
    \end{figure}

    \cref{subfig:hidden_feature_1D} shows a 1D example where the vertex crossings on the middle edge are detected using the slopes (divided differences) of the neighboring edges. The middle edge is split into two new edges that can then be used to detect and approximate the edge-crossing. The scalar value at the split location is obtained using a cubic Lagrange interpolation. \cref{subfig:hidden_feature_2D} provides an illustration using a marching square in the 2D case. The MC algorithms with and without sharp feature-preserving methods don't reconstruct the hidden feature in orange. This cell is considered above the desired isoline because all its node values are larger than the target isovalue. The cell outlined with black lines is divided into smaller cells, indicated by the dashed lines, using cubic Lagrange interpolation. These new cells reveal new edge-crossings that are used to represent the hidden feature and better approximate the overall isosurface.
      \begin{figure}[!h]
        \centering
        \begin{subfigure}[b]{0.45 \columnwidth}
            \centering
            \includegraphics[width=0.90\columnwidth]{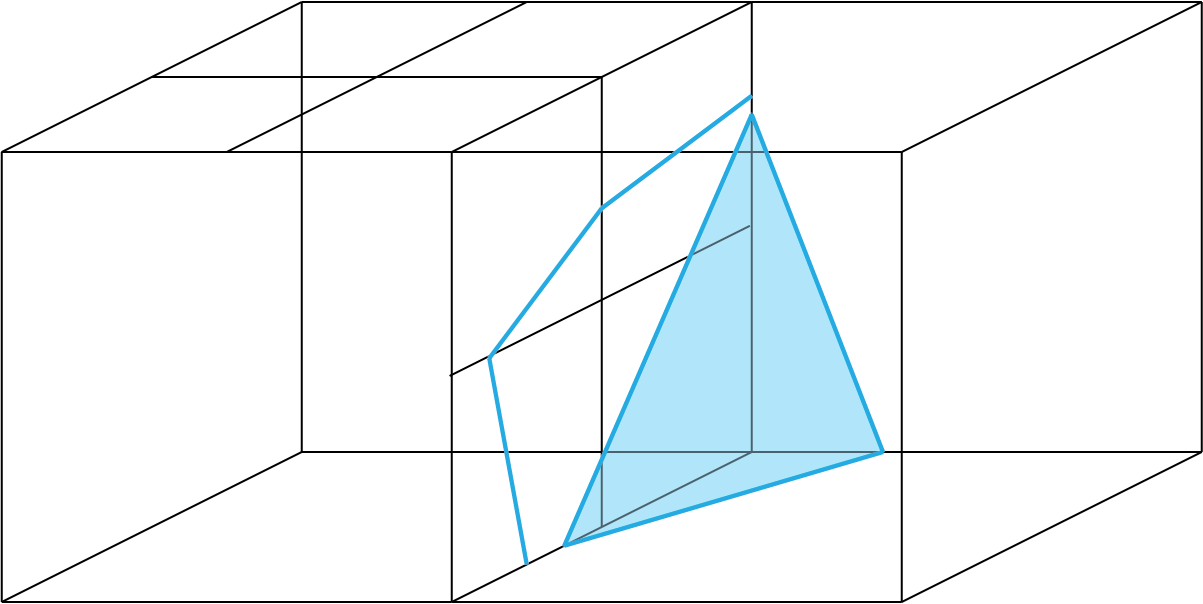}
            \caption{}
            \label{subfig:crack}    
        \end{subfigure}
        \begin{subfigure}[b]{0.45 \columnwidth}
            \centering
            \includegraphics[width=0.90 \columnwidth]{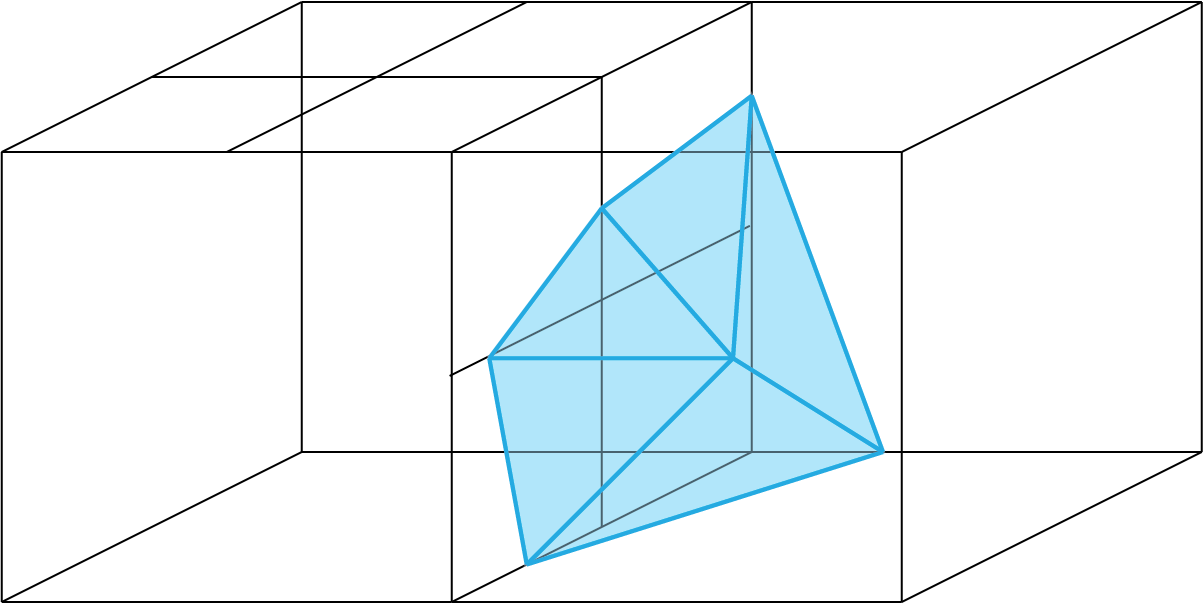}
            \caption{}
            \label{subfig:crack_free}    
        \end{subfigure}
        \vspace{-3mm}
        \caption{Crack patching. \cref{subfig:crack} shows an isosurface crack caused by the refined cell. The cell on the left is divided according to our algorithm, while the cell on the right is from the original Marching Cubes. The extracted vertices on the interface of two cells are mismatched. In \cref{subfig:crack_free} the crack is fixed by (1) matching the boundaries of the blue triangle in \cref{subfig:crack} to the boundary of the blue line in \cref{subfig:crack}, (2) connecting the edges of the new polygon to the center of the triangle in \cref{subfig:crack} to form the crack-free triangulated patch..}
        \label{fig:crack_patching}
    \end{figure}
    \begin{figure}[!h]
        \centering
        \begin{subfigure}[b]{0.24 \columnwidth}
            \centering
            \includegraphics[width=0.99\columnwidth]{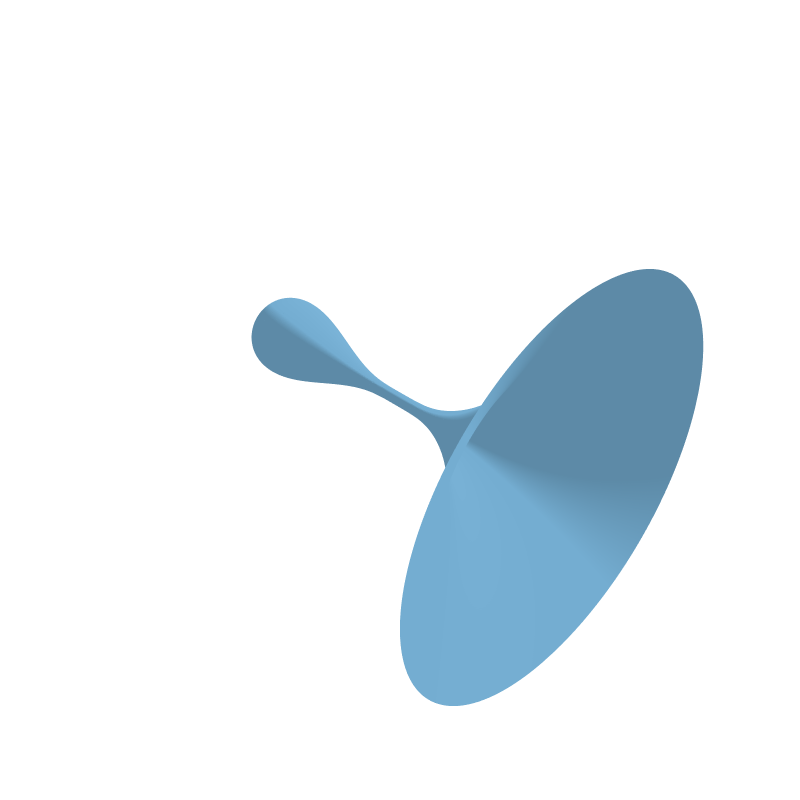}
            \vspace{-3mm}
            \caption{target}
            \label{subfig:teardrop_target}    
        \end{subfigure}
        \begin{subfigure}[b]{0.24 \columnwidth}
            \centering
            \includegraphics[width=0.99\columnwidth]{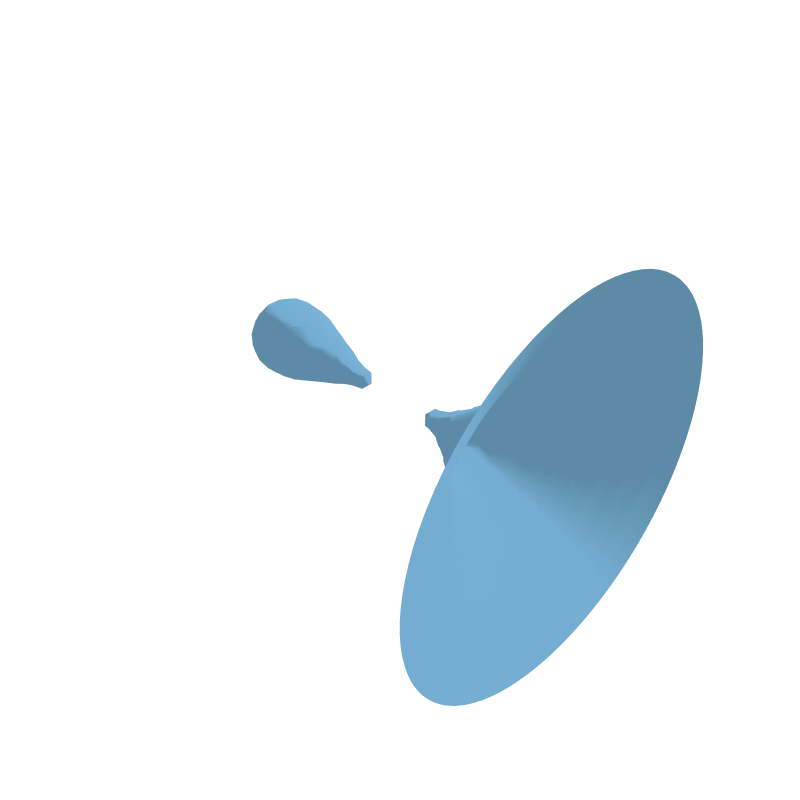}
            \vspace{-3mm}
            \caption{standard MC}
            \label{subfig:teardrop_MC}    
        \end{subfigure}
        \begin{subfigure}[b]{0.24 \columnwidth}
            \centering
            \includegraphics[width=0.99\columnwidth]{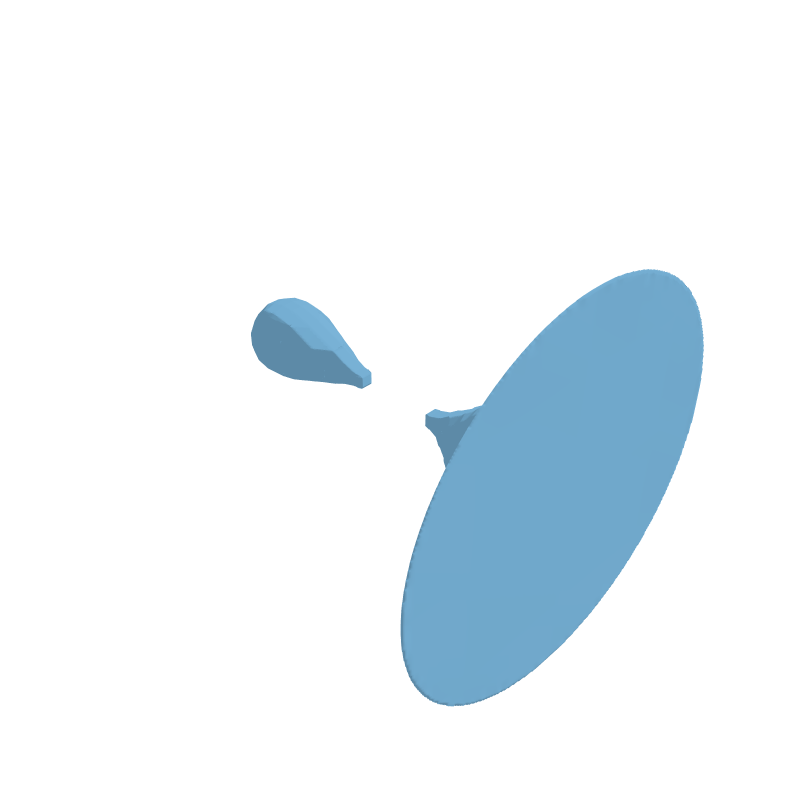}
            \vspace{-3mm}
            \caption{dual contouring}
            \label{subfig:teardrop_dc}    
        \end{subfigure}
        \begin{subfigure}[b]{0.24 \columnwidth}
            \centering
            \includegraphics[width=0.99\columnwidth]{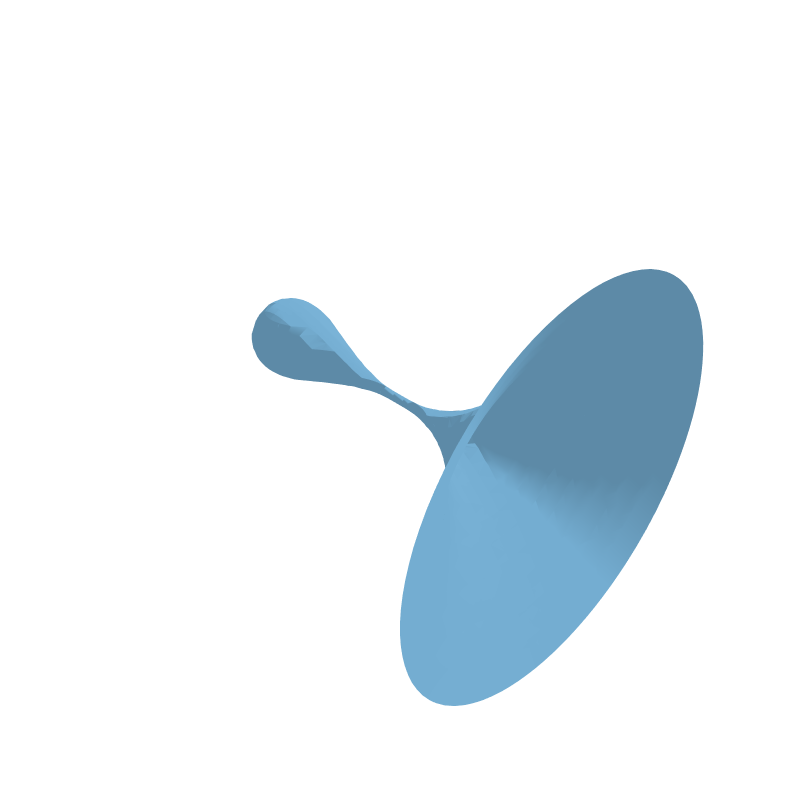}
            \vspace{-3mm}
            \caption{hidden feature}
            \label{subfig:teardrop_hf}    
        \end{subfigure}
        \vspace{-3mm}
        \caption{Comparison between the target, MC, dual contouring, and hidden feature reconstruction method. This example is based on the teardrop example with a resolution of $32^{3}$ and an isovalue $k=-0.001$. The hidden feature recovery method can detect and reconstruct the missing feature that connects the two broken pieces.}
        \label{fig:teadrop-mc-dc-hf}
    \end{figure}
    \begin{figure*}[!h]
        \centering
        \includegraphics[width=0.70 \textwidth]{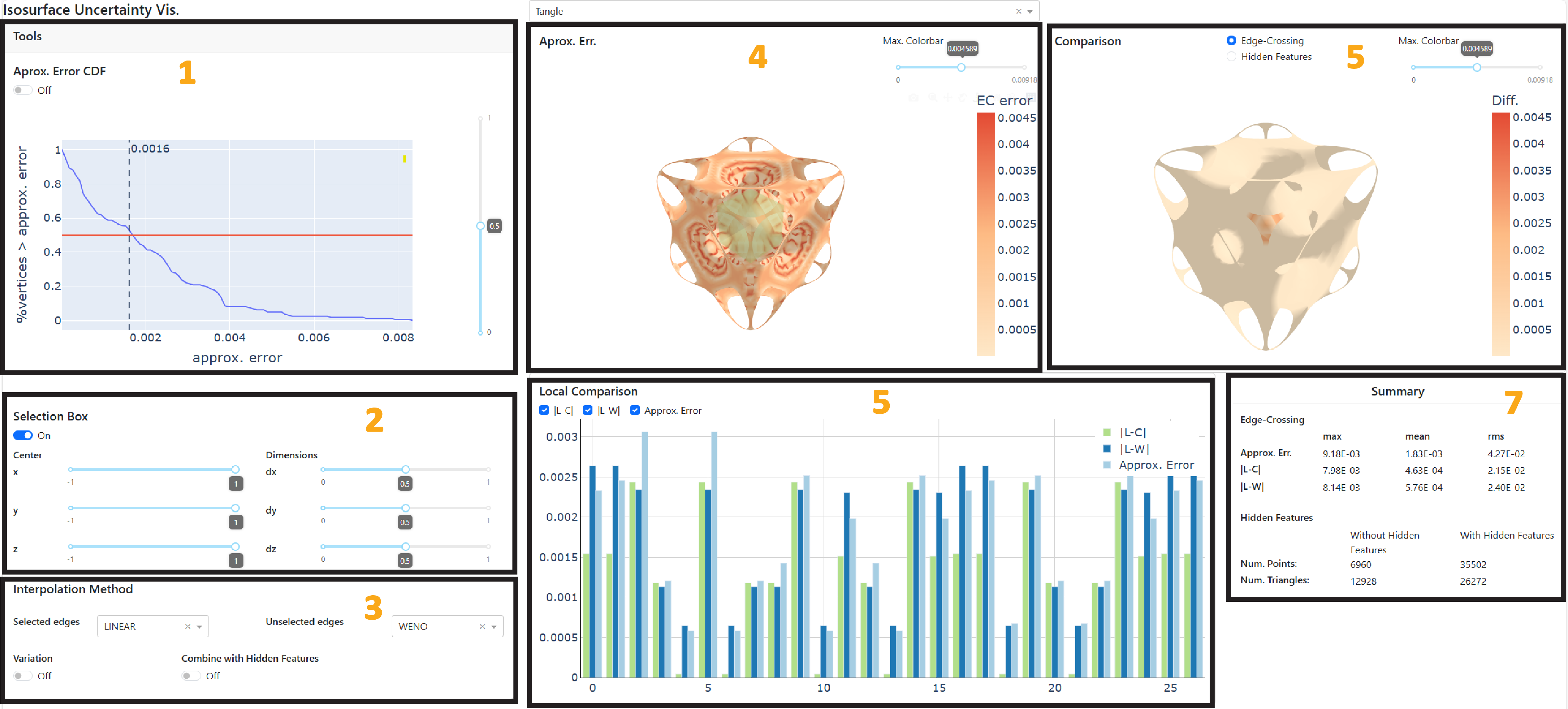}
        \vspace{-3mm}
        \caption{Our visualization framework has seven components. It shows the approximated error cumulative distribution function(CDF), the local selection box specifier, the interpolation method specifier, the approximated error overview, the local vertices comparison, the surface comparison, and the summary.}
        \label{fig:frame-work}
        % \vspace{-8mm}
    \end{figure*}

    These cell refinements lead to cracks in the reconstructed isosurface.
    % \color{blue}
    Crack-free isosurface extraction techniques have been introduced in the context of adaptive mesh refinement~\cite{weber2001, weber2012, Wald2021}
    % \color{black}
    Here, we resolve this issue by connecting the edges at hidden feature face boundaries with edges at the boundaries of the reaming isosurface. 
    % \color{blue}
    Additional edges indicated by the interior blue lines shown in \cref{subfig:crack_free} are introduced to reduce discontinuities at the cell interface.
    % \color{black}
    The cells at the boundaries of the isosurface with no hidden features are modified during the polygon extraction to ensure a crack-free isosurface, as shown in \cref{fig:crack_patching}.

  The interface is designed to highlight high errors and isosurface feature differences based on queries. Several filter tools (sliders, switches, checklists, radio buttons) are introduced for flexibility and to facilitate exploration and analysis. The different views are used to enable simultaneous visualization of different isosurface error metrics. The interface design provides insight into the confidence of the extracted isosurface.
    \cref{fig:teadrop-mc-dc-hf} shows the ground truth, standard MC, dual contouring, and our hidden feature recovery method isosurfaces for the \textbf{Teardrop} dataset. The standard MC and dual contouring do not detect and reconstruct the missing piece.
    % that connects the two components. 
    The dual contouring method inserts additional triangles to enforce closed surfaces. Our method successfully recovers the missing piece and yields an isosurface similar to the target solution. 
    % We utilize the hidden feature recovery method to highlight the limitations of linear interpolation and show the isosurface feature uncertainty. 

\subsection{Isosurface Uncertainty Visual Analysis Tool}
\label{subsec:framework}
    %% small intro
    Here, we introduce a framework for visualizing and analyzing MC isosurface uncertainty using C, Python, and Dash \footnote{\url{https://dash.plotly.com/}}. 
    % As previously highlighted, most isosurface visualizations do not provide insight into the uncertainty of the isosurface reconstruction error including the interpolation model uncertainty. 
    Our framework employs the error approximation in \cref{subsec:error_approximation}, the hidden feature-preserving method in  \cref{subsec:hidden_features}, and several other techniques to enable visual analysis of isosurfaces uncertainty obtained from using different interpolation methods along with MC algorithms.
    % \color{blue}
    The interface is designed to highlight high errors and isosurface feature differences based on queries. Several filter tools (sliders, switches, checklists, radio buttons) are introduced for flexibility and to facilitate exploration and analysis. The different views are used to enable simultaneous visualization of different isosurface error metrics. The interface design provides insight into the confidence of the extracted isosurface. 
    % \color{black}
    Our framework has seven components indicated by the outlined rectangles and the corresponding component number shown in \cref{fig:frame-work} 
    
    % \begin{enumerate}
        % \item %\ta{Maybe name components as $C_1..C_7$. Later you can use these references when describing the results.}
        The first component $C_{1}$ shows a plot of the cumulative percentage of vertices (y-axis) with respect to the approximated error (x-axis) introduced in \cref{subsec:error_approximation}. The vertical slider to the right of the plot in the first component is used to select a cumulative percentage. The switch below ``Approx. Error CDF" turns on and off a binary color map on the isosurface shown in the fifth component. This component provides insight into the percentage of isosurface vertices below and above a selected threshold. 
         
        The second component $C_{2}$ is used to insert a transparent box inside the domain of the fifth component to show a selected local region. The switch below ``Selection Box" must be on to activate the box feature. The position and size of the box are adjusted using ``Center" and ``Dimension", respectively. This component facilitates detailed inspection of local isosurface uncertainty based on a selected region of interest. 
        
        The third component $C_{3}$ is used to select different interpolation methods for selected and unselected edges based on the error threshold in $C_{1}$ or local box selection in $C_{2}$. This feature enables the comparison between different interpolation methods.
        
        The fourth component $C_{4}$ visualizes the estimated isosurface uncertainty obtained from \cref{subsec:error_approximation} and the local selection based on the parameters selected in $C_{2}$. This component highlights regions with high and low edge-crossing errors.
        
        The fifth component $C_{5}$ shows a bar plot that provides a local comparison of the different interpolation methods for selected local regions. This enables a vertex-by-vertex comparison of the edge-crossing error and the difference between interpolation methods. 
        
        The sixth component $C_{6}$ visualizes the comparison between two selected interpolation methods using the ``Edge-Crossing" radio button. The "Hidden Feature" radio button enables the simultaneous visualization of both the isosurface with and without hidden features reconstruction. This component provides insight into the difference and accuracy gain between linear and higher interpolation methods. It also highlights the feature uncertainty between standard MC and hidden feature recovery. 
        
        The seventh component $C_{7}$ provides a summary of the isosurface uncertainties.
    % \end{enumerate}

%% file: results.tex
%
\begin{figure*}[!ht]
    \centering
    \begin{subfigure}[b]{0.16 \textwidth}
        \centering
        \includegraphics[width=0.80 \columnwidth]{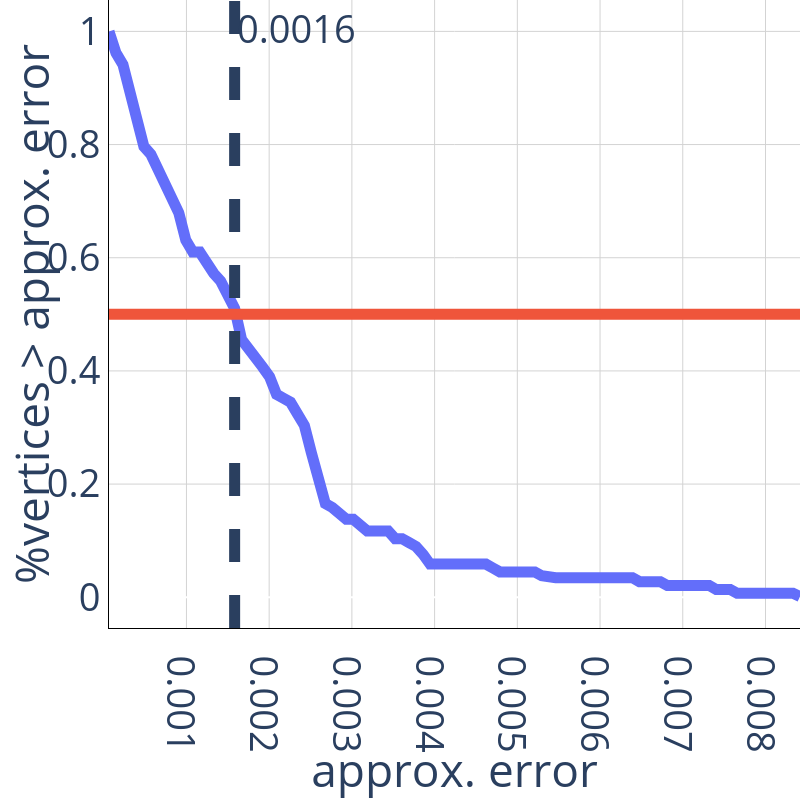}
        \vspace{-2.5mm}
        \caption{}
        \label{subfig:tangle_error_distribution_32x32x32}
    \end{subfigure}
    %%%
    \begin{subfigure}[b]{0.16 \textwidth}
        \centering
        \includegraphics[width=0.99 \columnwidth]{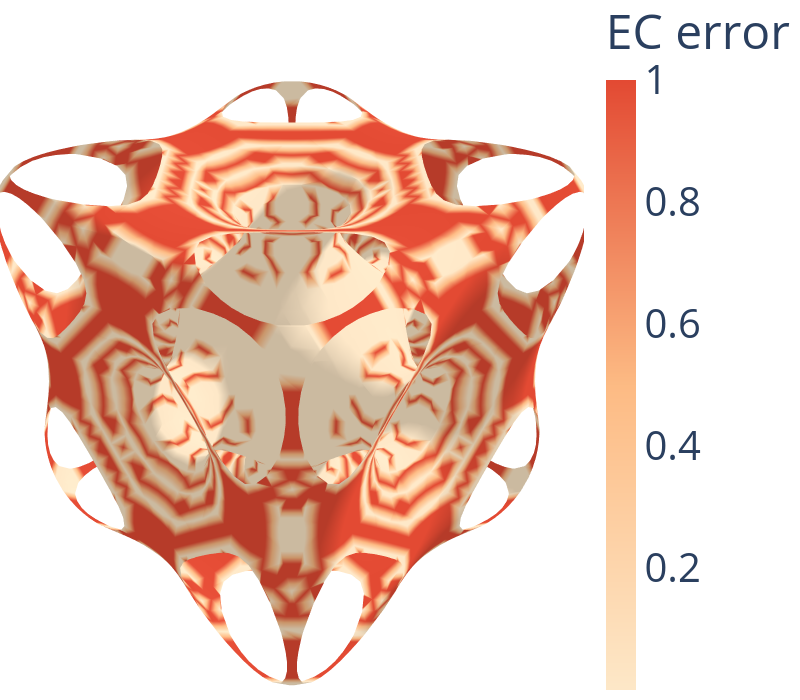}
        \vspace{-2.5mm}
        \caption{ Threshold }
        \label{subfig:tangle_true_error_threshold_32x32x32}
    \end{subfigure}
    %%%
    \begin{subfigure}[b]{0.16 \textwidth}
        \centering
        \includegraphics[width=0.99 \columnwidth]{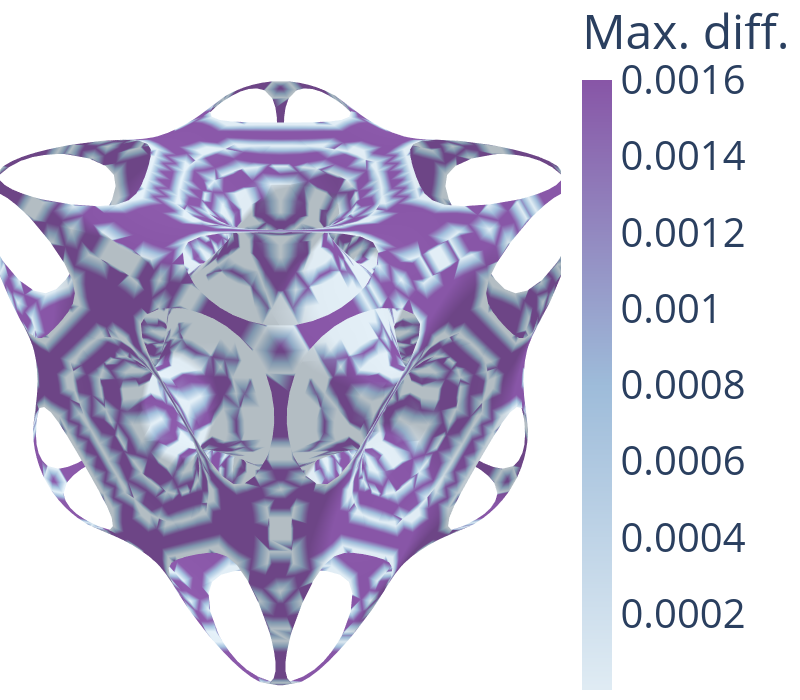}
        \vspace{-2.5mm}
        \caption{Max. Variation }
        \label{subfig:tangle_ALL_32x32x32}
    \end{subfigure}
    \begin{subfigure}[b]{0.16 \textwidth}
        \centering
        \includegraphics[width=0.99 \columnwidth]{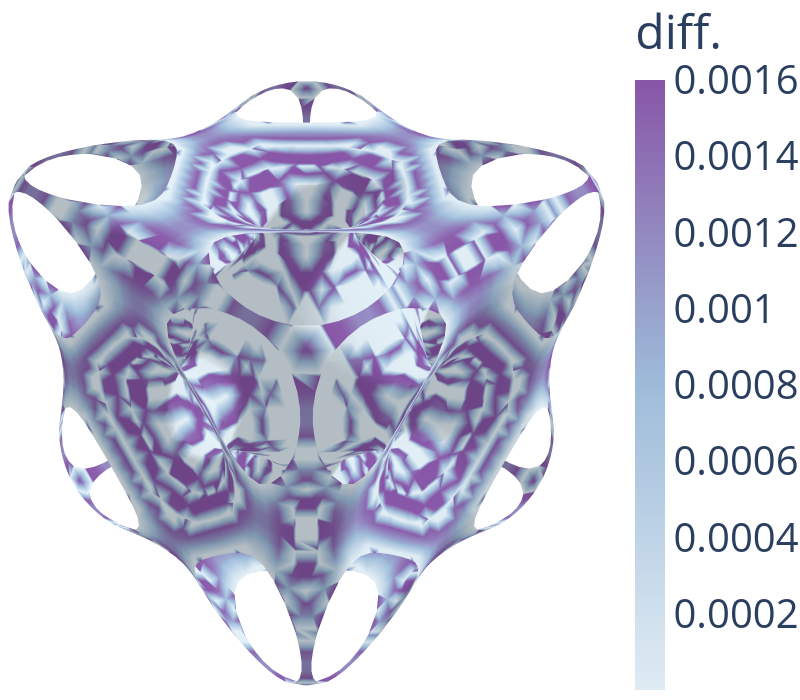}
        \vspace{-2.5mm}
        \caption{L vs C}
        \label{subfig:tangle_threshold_LC_32x32x32}
    \end{subfigure}
     %%%
    \begin{subfigure}[b]{0.16 \textwidth}
        \centering
        \includegraphics[width=0.99 \columnwidth]{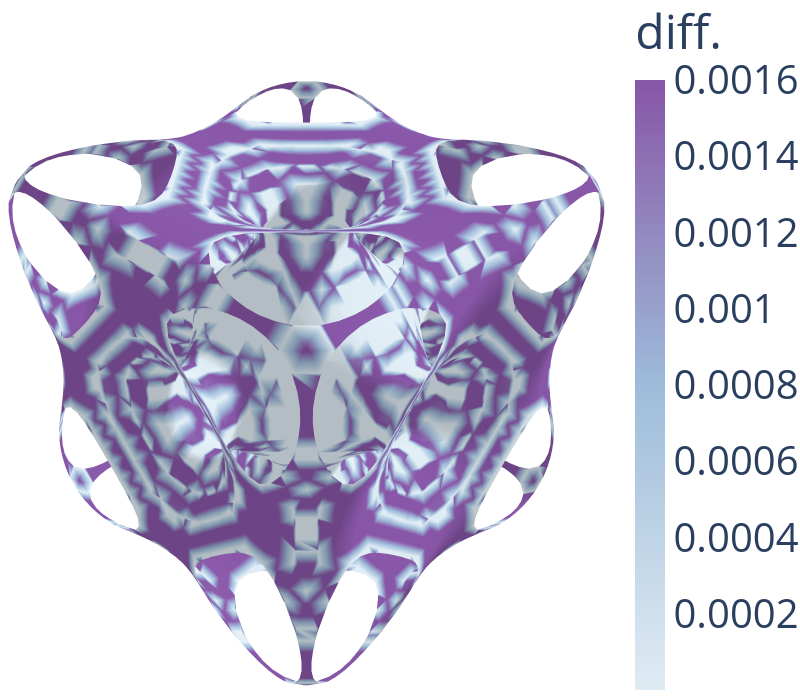}
        \vspace{-2.5mm}
        \caption{L vs W}
        \label{subfig:tangle_threshold_LW_32x32x32}
    \end{subfigure}
    %%%
     \begin{subfigure}[b]{0.16 \textwidth}
        \centering
        \includegraphics[width=0.99 \columnwidth]{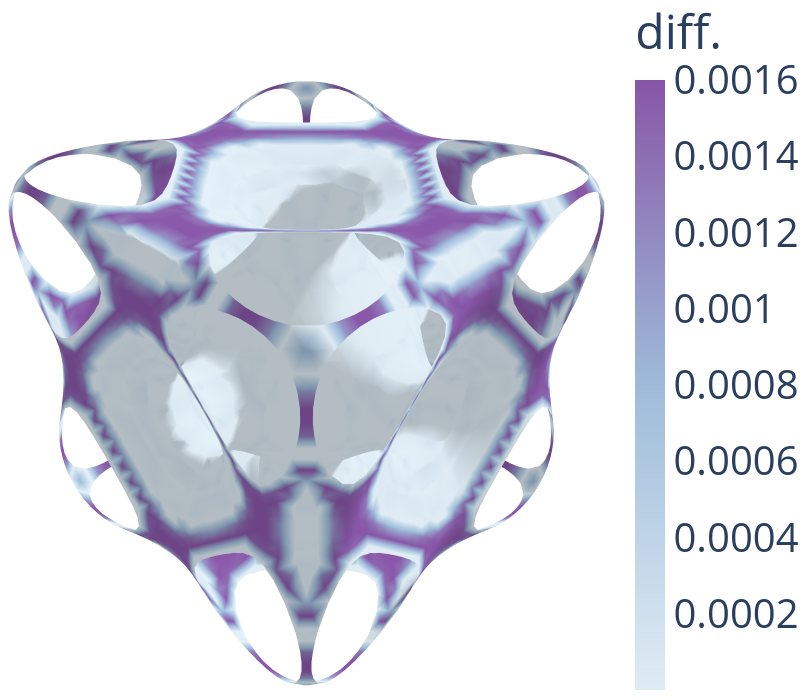}
        \vspace{-2.5mm}
        \caption{C vs W}
        \label{subfig:tangle_threshold_CW_32x32x32}
    \end{subfigure}
    %%%%%%%%%%%%%%%%%%%%%%%%%%%%%%%%%%%%%%%%%%%%%%%%%%%%%%%%%%%%%%%%%%%%%%%%%%%%
    %
    \begin{subfigure}[b]{0.16 \textwidth}
        \centering
        \includegraphics[width=0.80 \columnwidth]{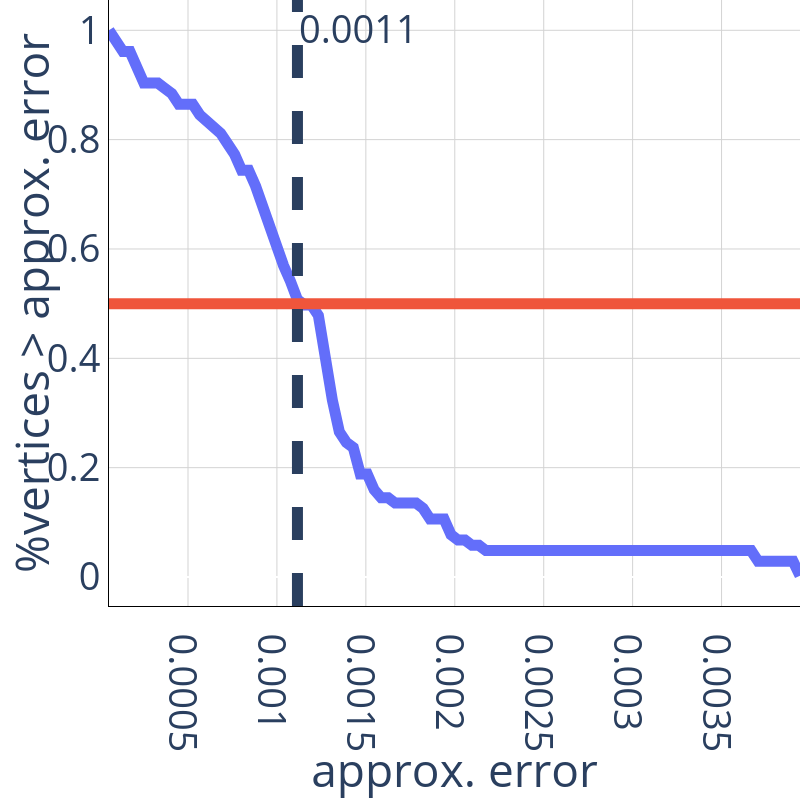}
        \caption{}
        \label{subfig:torus_error_distribution_64x64x64}
    \end{subfigure}
    %%%
    \begin{subfigure}[b]{0.16 \textwidth}
        \centering
        \includegraphics[width=0.99 \columnwidth]{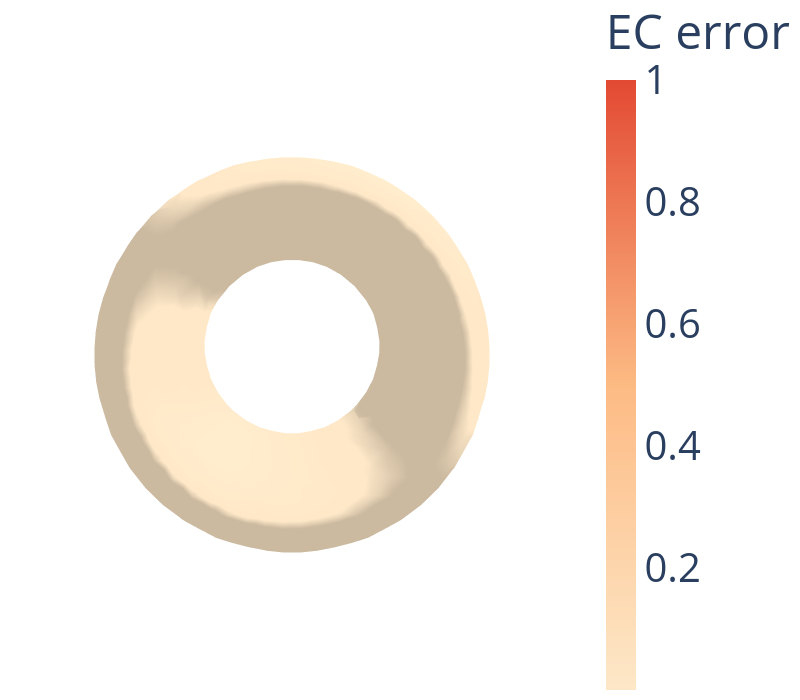}
        \caption{ Threshold }
        \label{subfig:torus_true_error_threshold_64x64x64}
    \end{subfigure}
    %%%
    \begin{subfigure}[b]{0.16 \textwidth}
        \centering
        \includegraphics[width=0.99 \columnwidth]{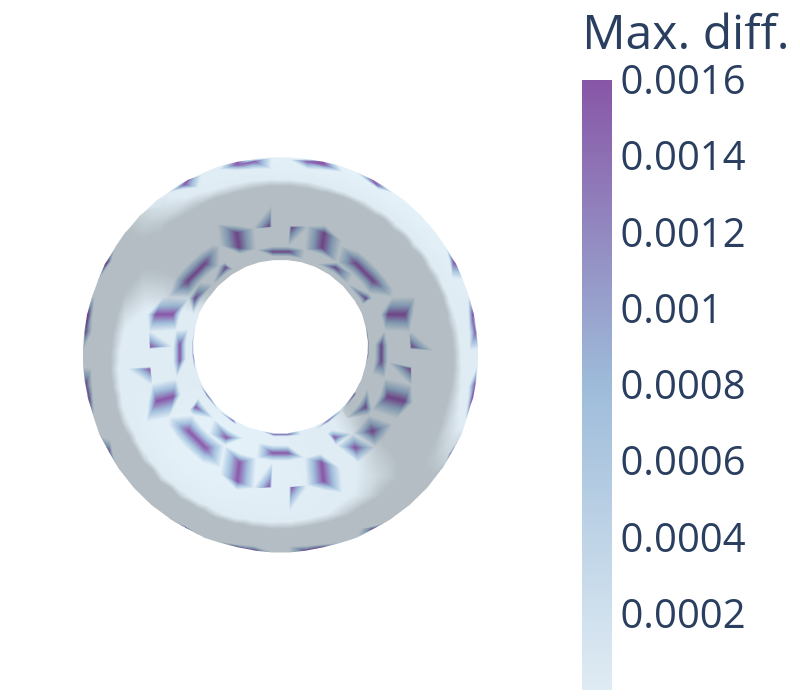}
        \caption{ Max. Variation }
        \label{subfig:torus_ALL_64x64x64}
    \end{subfigure}
    %%%
    \begin{subfigure}[b]{0.16 \textwidth}
        \centering
        \includegraphics[width=0.99 \columnwidth]{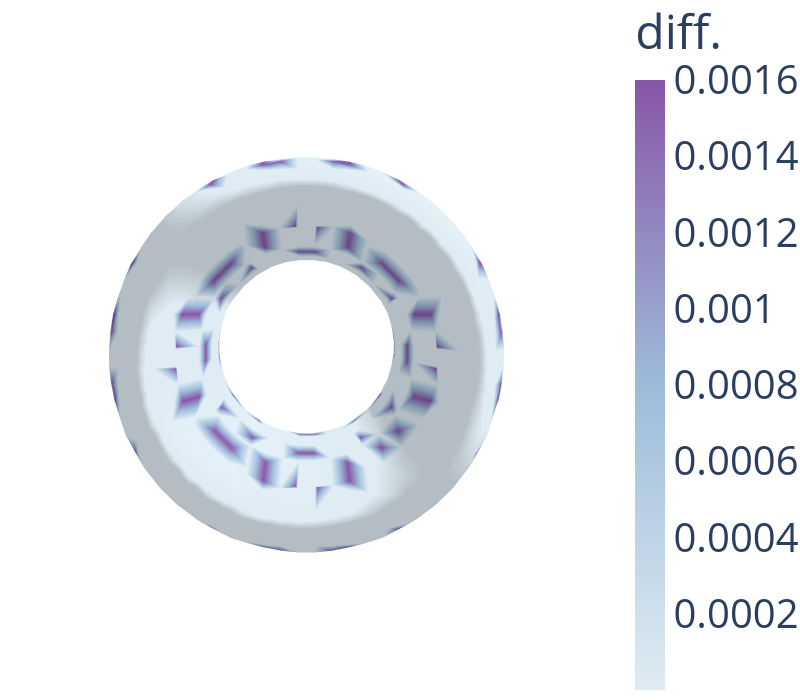}
        \caption{L vs C}
        \label{subfig:torus_threshold_LC_64x64x64}
    \end{subfigure}
     %%%
    \begin{subfigure}[b]{0.16 \textwidth}
        \centering
        \includegraphics[width=0.99 \columnwidth]{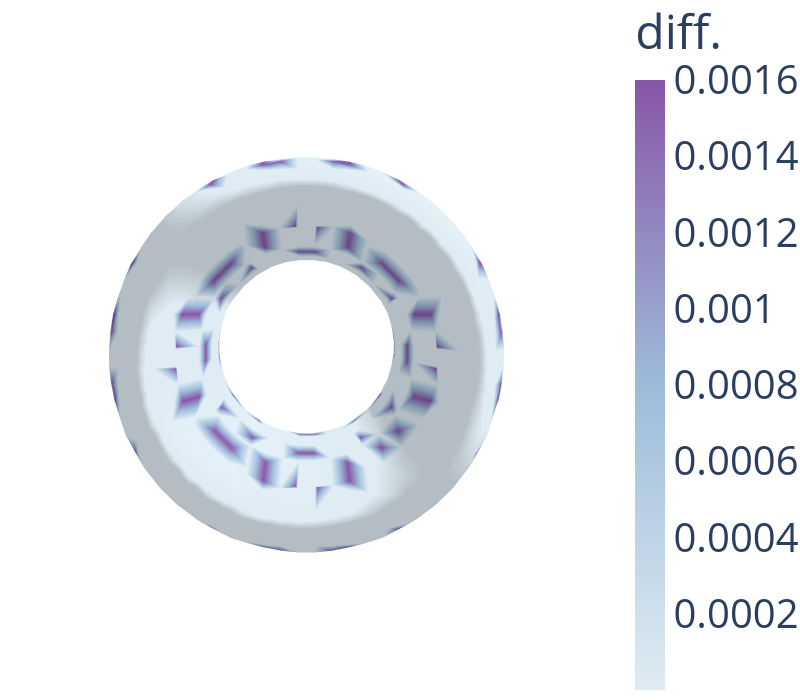}
        \caption{L vs W}
        \label{subfig:torus_threshold_LW_64x64x64}
    \end{subfigure}
    %%%
     \begin{subfigure}[b]{0.16 \textwidth}
        \centering
        \includegraphics[width=0.99 \columnwidth]{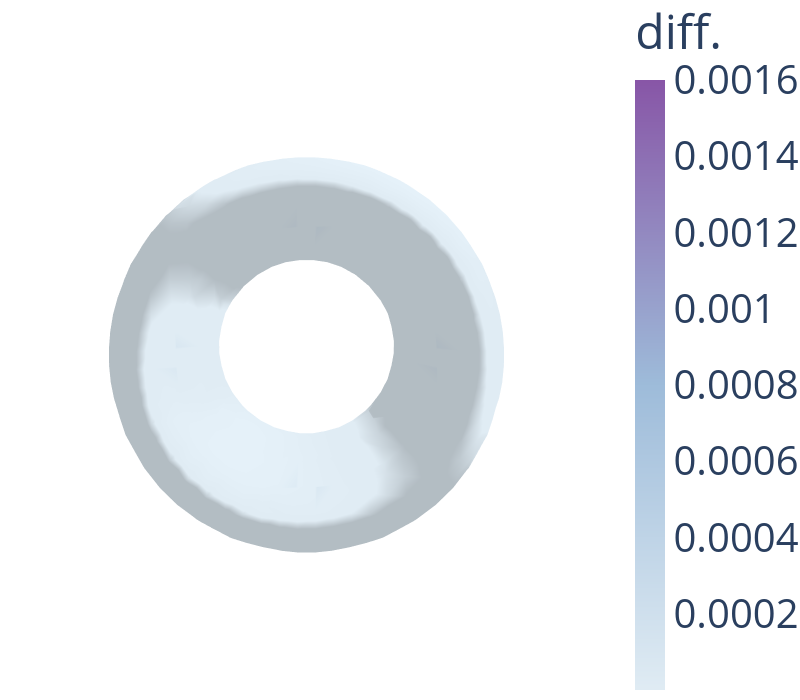}
        \caption{C vs W}
        \label{subfig:torus_threshold_CW_64x64x64}
    \end{subfigure}
    %%%%%%%%%%%%%%%%%%%%%%%%%%%%%%%%%%%%%%%%%%%%%%%%%%%%%%%%%%%%%%%%%%%%%%%%%%%%
    \begin{subfigure}[b]{0.16 \textwidth}
        \centering
        \includegraphics[width=0.80 \columnwidth]{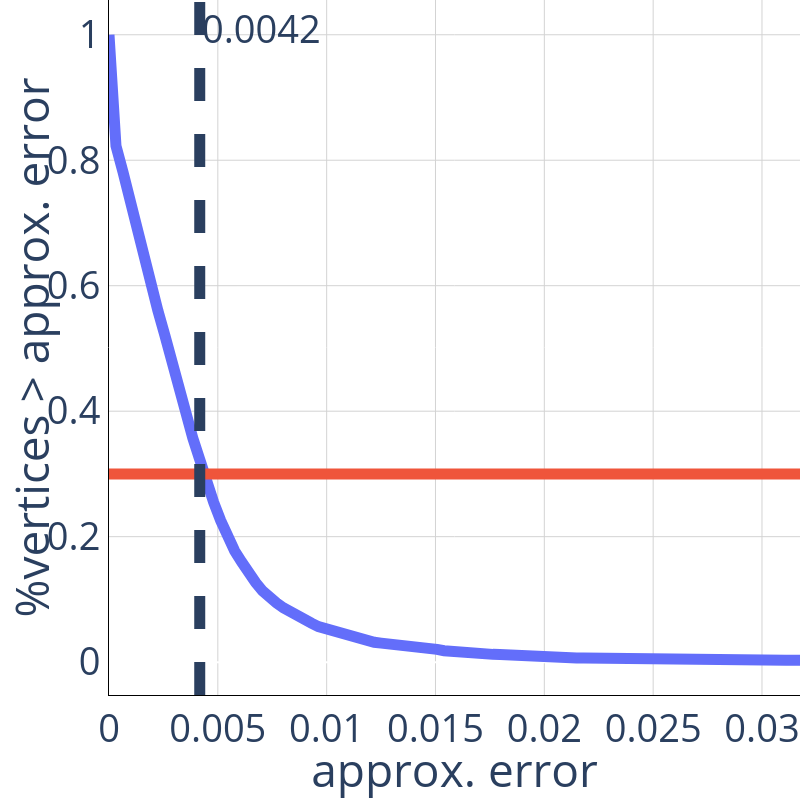}
        \caption{}
        \label{subfig:marschnerlobb_error_distribution_64x64x64}
    \end{subfigure}
    %%%
    \begin{subfigure}[b]{0.16 \textwidth}
        \centering
        \includegraphics[width=0.99 \columnwidth]{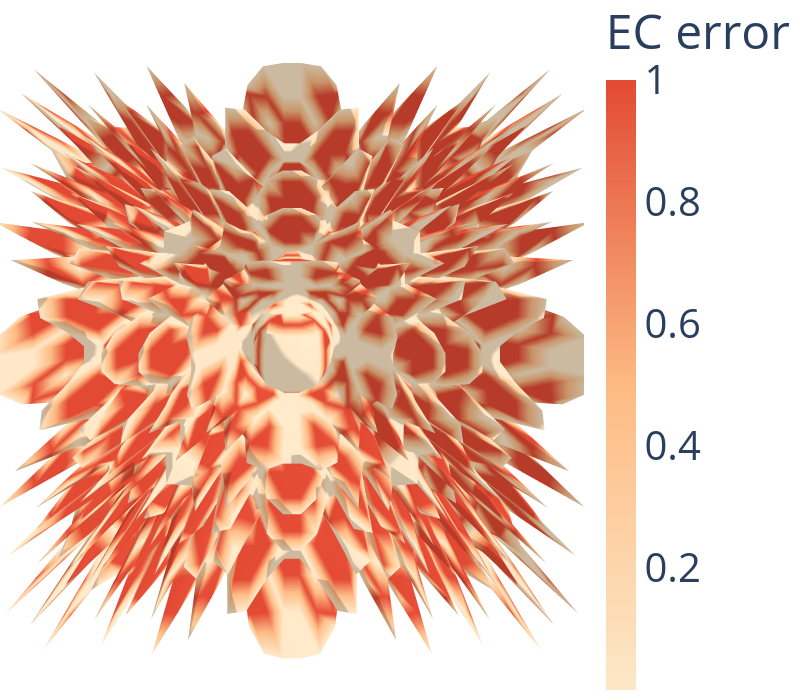}
        \caption{ Threshold }
        \label{subfig:marschnerlobb_true_error_threshold_64x64x64}
    \end{subfigure}
    %%%
    \begin{subfigure}[b]{0.16 \textwidth}
        \centering
        \includegraphics[width=0.99 \columnwidth]{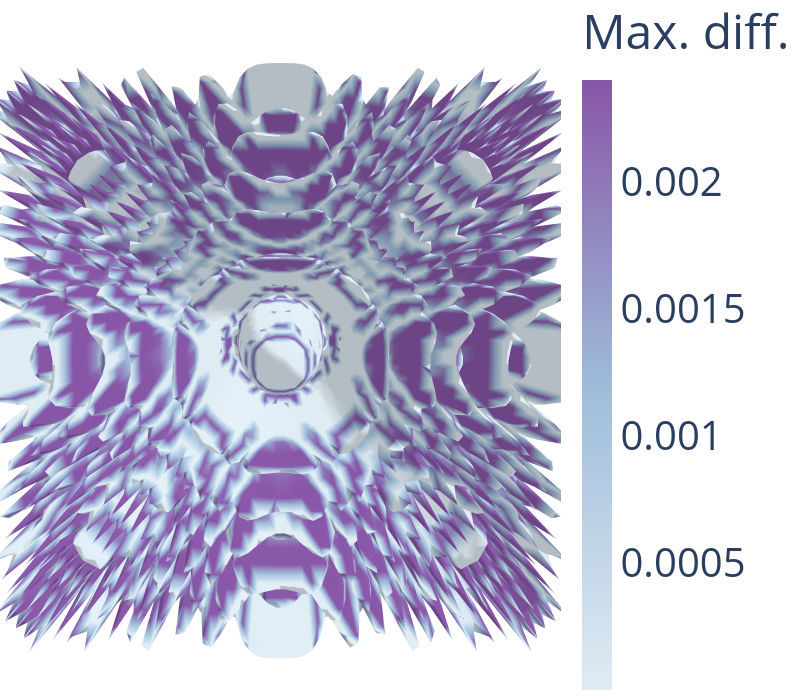}
        \caption{ Max. Variation}
        \label{subfig:marschnerlobb_ALL_64x64x64}
    \end{subfigure}
    %%%
    \begin{subfigure}[b]{0.16 \textwidth}
        \centering
        \includegraphics[width=0.99 \columnwidth]{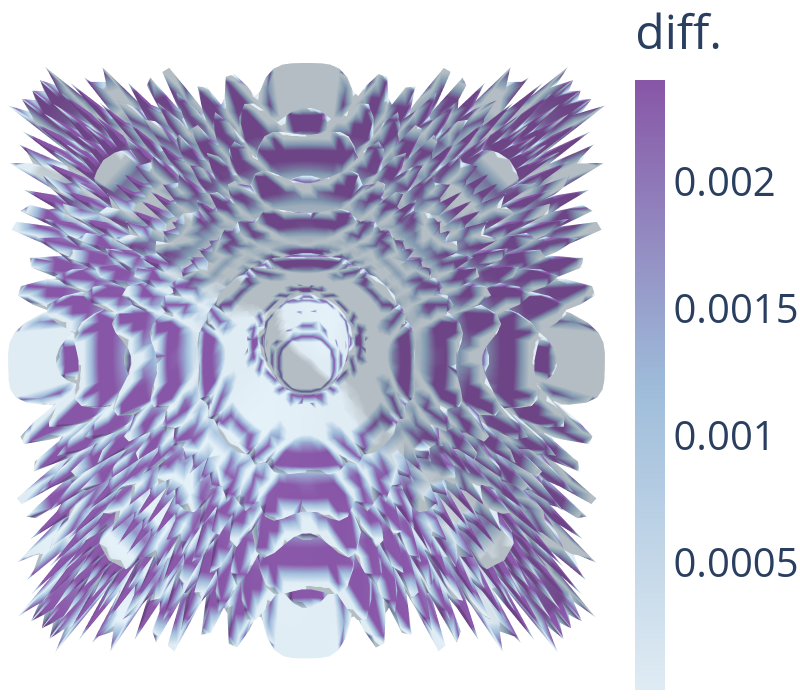}
        \caption{L vs C}
        \label{subfig:marschnerlobb_threshold_LC_64x64x64}
    \end{subfigure}
     %%%
    \begin{subfigure}[b]{0.16 \textwidth}
        \centering
        \includegraphics[width=0.99 \columnwidth]{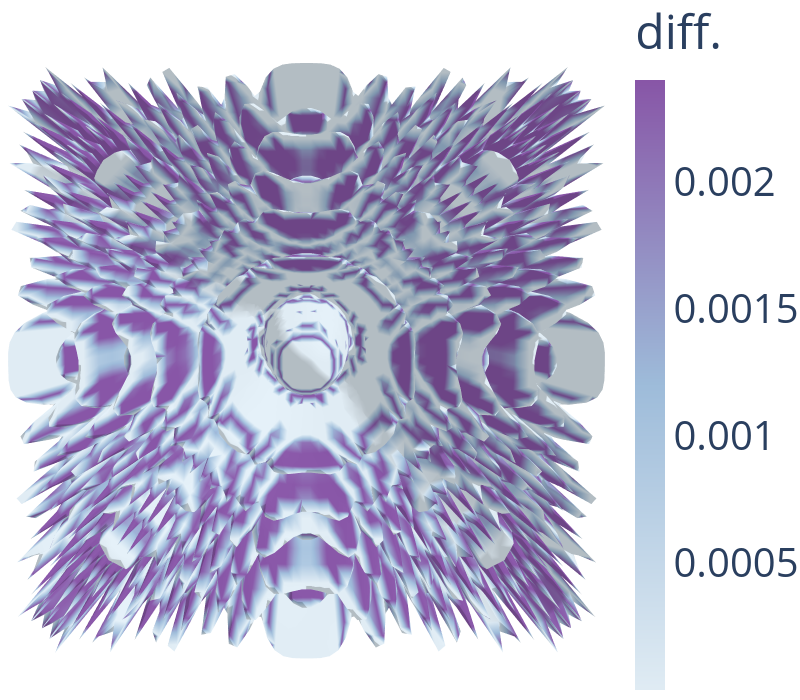}
        \caption{L vs W}
        \label{subfig:marschnerlobb_threshold_LW_64x64x64}
    \end{subfigure}
    %%%
     \begin{subfigure}[b]{0.16 \textwidth}
        \centering
        \includegraphics[width=0.99 \columnwidth]{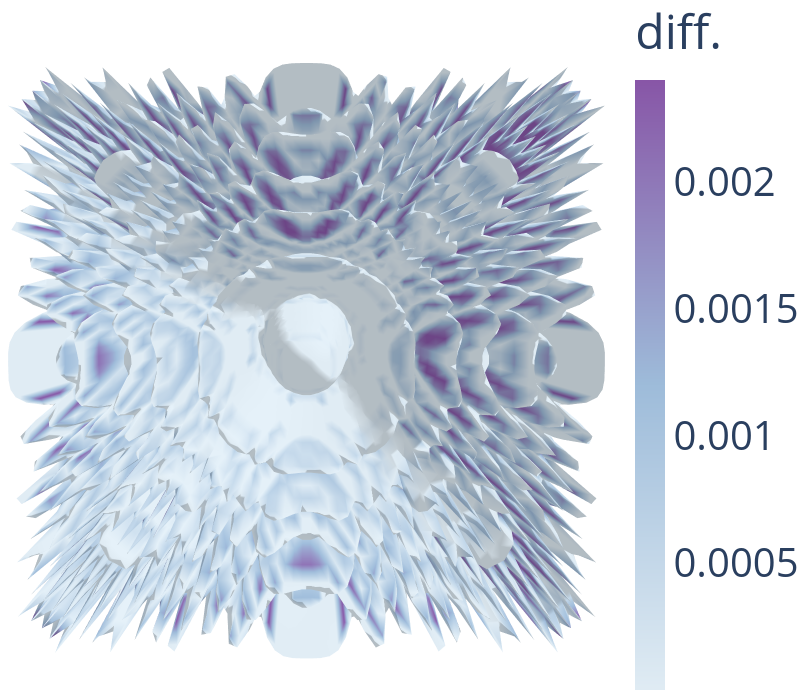}
        \caption{C vs W}
        \label{subfig:marschnerlobb_threshold_CW_64x64x64}
    \end{subfigure}
    % %%%%%%%%%%%%%%%%%%%%%%%%%%%%%%%%%%%%%%%%%%%%%%%%%%%%%%%%%%%%%%%%%%%%%%%%%%%%
    \vspace{-3mm}
    \caption{Isosurface comparison based on selected threshold value and interpolation methods (Components  $C_{1}$, $C_{3}$, $C_{4}$ and $C_{6}$). The first column shows the selected threshold using the plot of the cumulative percentage of vertices (y-axis) with respect to the approximated error (x-axis). The second column shows the binary colormap based on the selected threshold. The third column shows the maximum variation. The fourth, fifth, and sixth columns show a comparison of L vs. C, L vs. C, and C vs. W. 
    % These results show how our thresholded approximation errors in the second column from left provide a quick and reliable insight into positions of interpolation errors without needing to perform their actual computation.  
    }
    \label{fig:figs_threshol_global}
\end{figure*}

\begin{figure*}[!h]
    \begin{subfigure}[b]{0.19 \textwidth}
        \centering
        \includegraphics[width=0.99 \columnwidth]{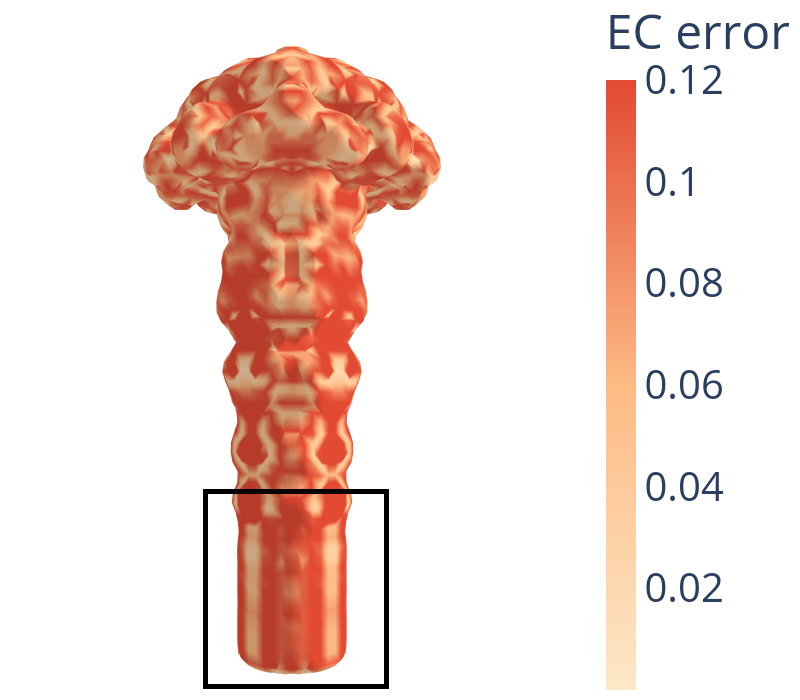}
        \caption{Fuel Approximated Error}
        \label{subfig:fuel_Approximated_Err}
    \end{subfigure}
    \begin{subfigure}[b]{0.19 \textwidth}
        \centering
        \includegraphics[width=0.99 \columnwidth]{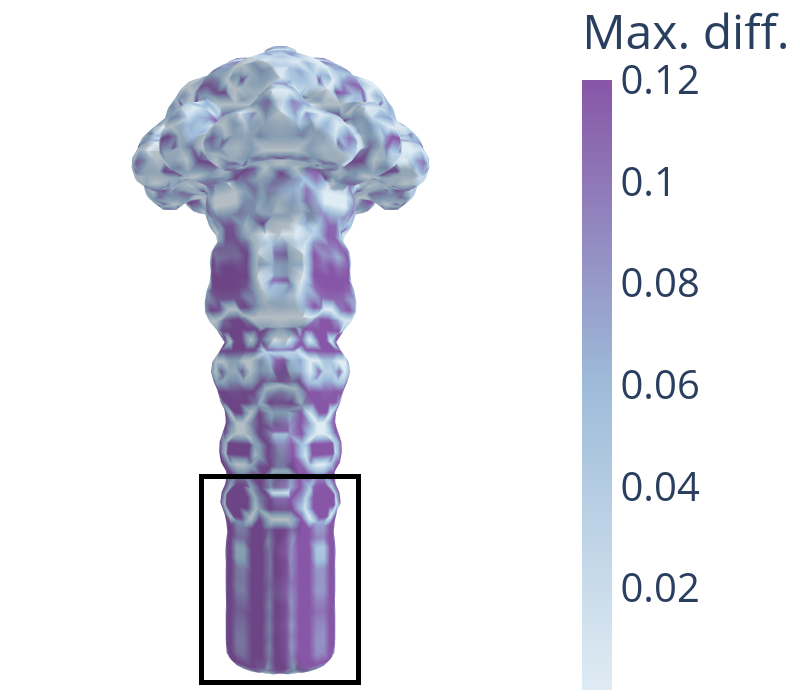}
        \caption{Max. Variation}
        \label{subfig:fuel_ALL}
    \end{subfigure}
    \begin{subfigure}[b]{0.19 \textwidth}
        \centering
        \includegraphics[width=0.99 \columnwidth]{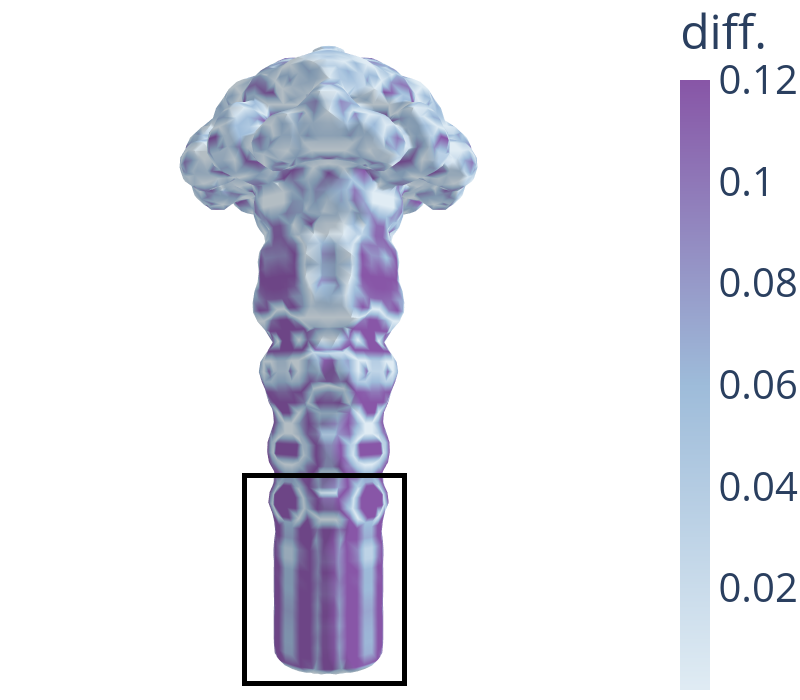}
        \caption{L vs C}
        \label{subfig:fuel_LC}
    \end{subfigure}
     % %
    \begin{subfigure}[b]{0.19 \textwidth}
        \centering
        \includegraphics[width=0.99 \columnwidth]{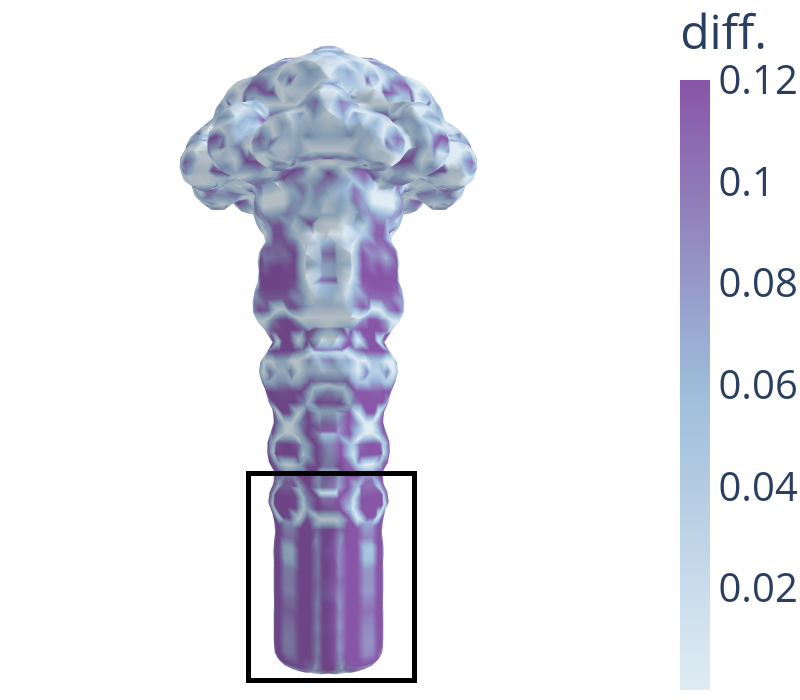}
        \caption{L v W}
        \label{subfig:fuel_LW}
    \end{subfigure}
     % %
    \begin{subfigure}[b]{0.19 \textwidth}
        \centering
        \includegraphics[width=0.99 \columnwidth]{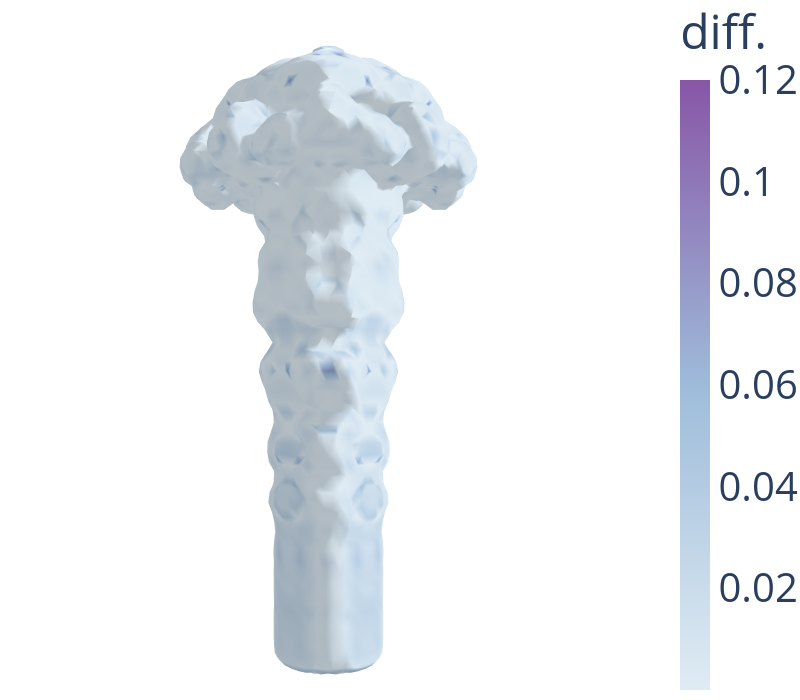}
        \caption{C vs W}
        \label{subfig:fuel_CW}
    \end{subfigure}
    %%%%%%%%%%%%%%%%%%%%%%%%%%%%%%%%%%%%%%%%%%%%%%%%%%%%%%%%%%%%%%%%%%%%%%%%%%%%
     \begin{subfigure}[b]{0.19 \textwidth}
        \centering
        \includegraphics[width=0.99 \columnwidth]{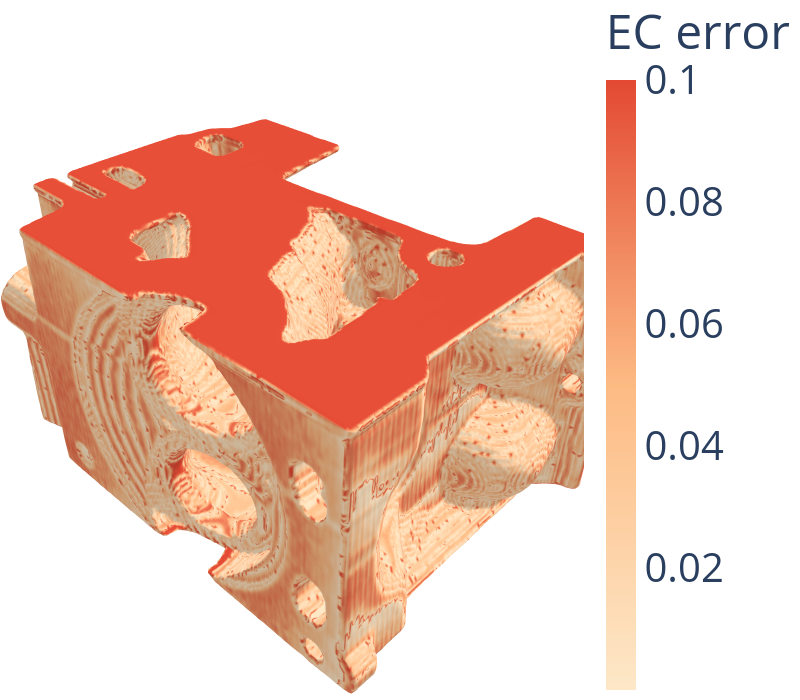}
        \caption{Fuel Approximated Error}
        \label{subfig:engine_Approximated_Err}
    \end{subfigure}
    \begin{subfigure}[b]{0.19 \textwidth}
        \centering
        \includegraphics[width=0.99 \columnwidth]{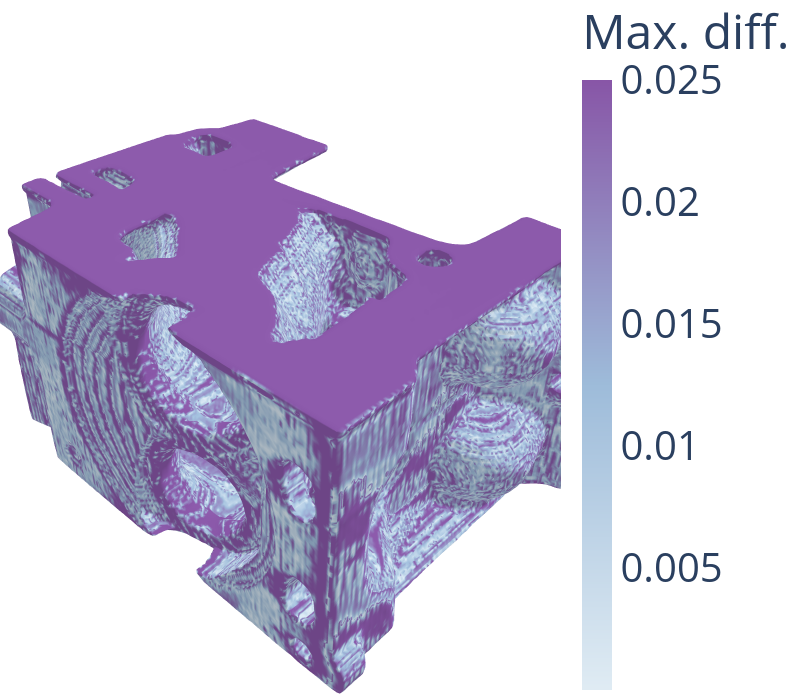}
        \caption{Max. Variation}
        \label{subfig:engine_ALL}
    \end{subfigure}
    \begin{subfigure}[b]{0.19 \textwidth}
        \centering
        \includegraphics[width=0.99 \columnwidth]{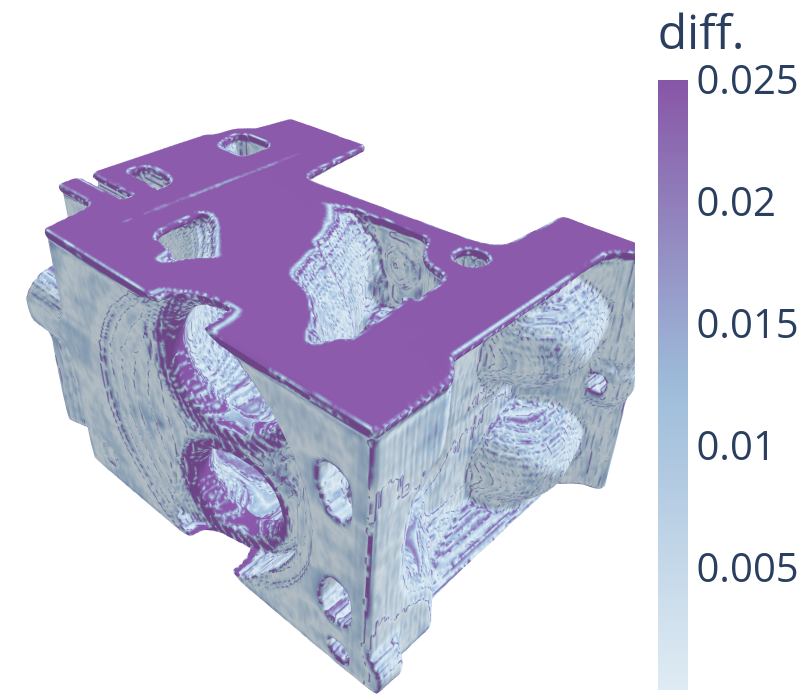}
        \caption{L vs C}
        \label{subfig:engine_LC}
    \end{subfigure}
     % %
    \begin{subfigure}[b]{0.19 \textwidth}
        \centering
        \includegraphics[width=0.99 \columnwidth]{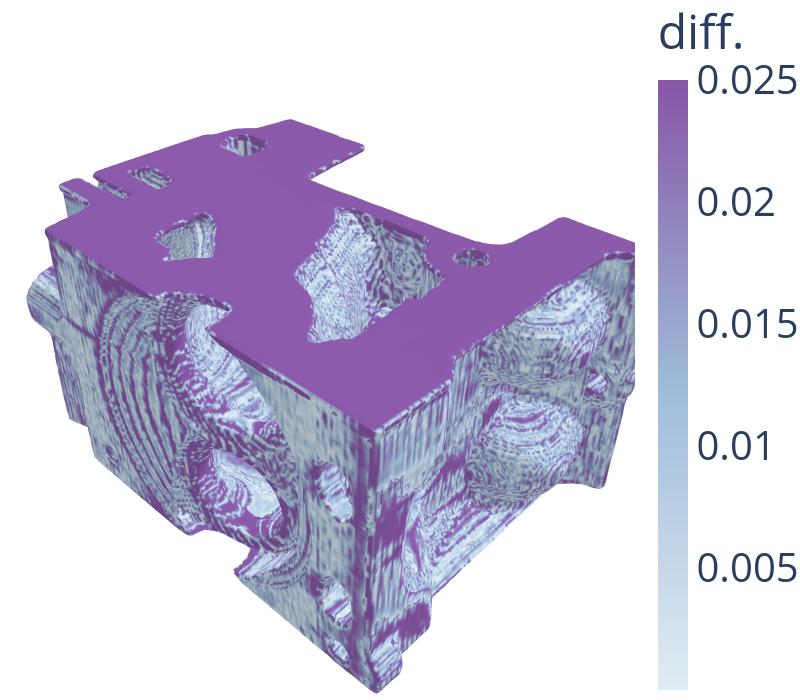}
        \caption{L vs C}
        \label{subfig:engine_LW}
    \end{subfigure}
     % %
    \begin{subfigure}[b]{0.19 \textwidth}
        \centering
        \includegraphics[width=0.99 \columnwidth]{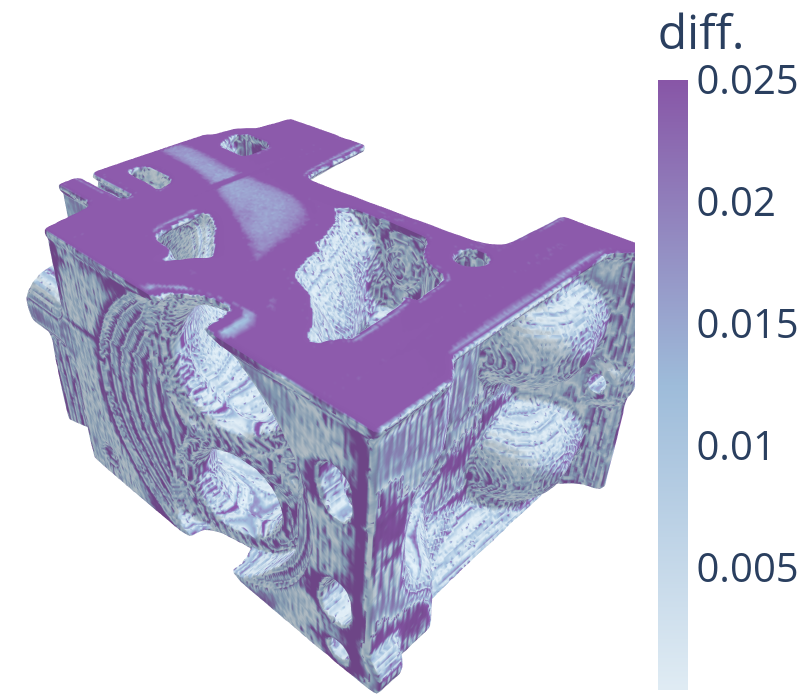}
        \caption{C vs W}
        \label{subfig:engine_CW}
    \end{subfigure}
    %%%%%%%%%%%%%%%%%%%%%%%%%%%%%%%%%%%%%%%%%%%%%%%%%%%%%%%%%%%%%%%%%%%%%%%%%%%%%%%
    \begin{subfigure}[b]{0.19 \textwidth}
        \centering
        \includegraphics[width=0.99 \columnwidth]{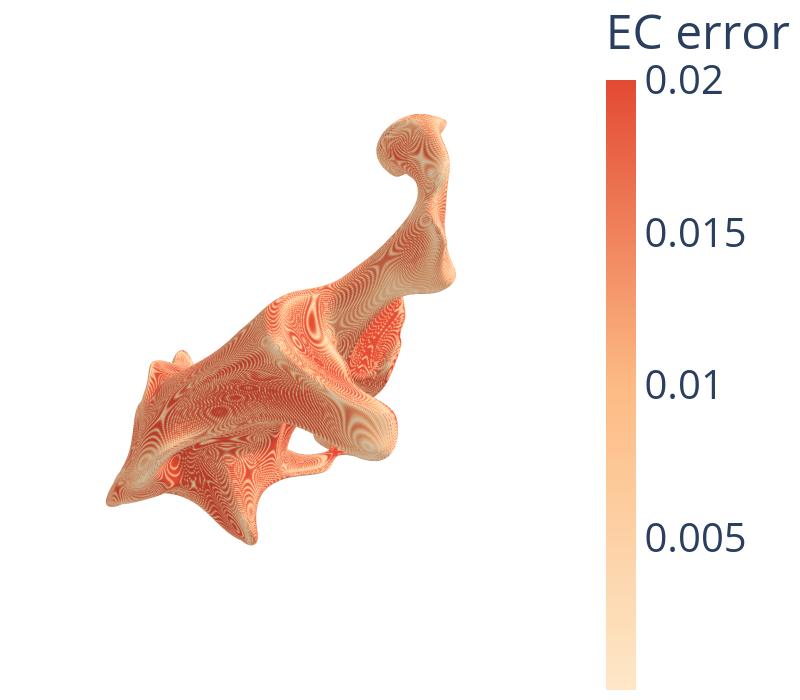}
        \caption{Fuel Approximated Error}
        \label{subfig:hcci_oh_Approximated_Err}
    \end{subfigure}
    \begin{subfigure}[b]{0.19 \textwidth}
        \centering
        \includegraphics[width=0.99 \columnwidth]{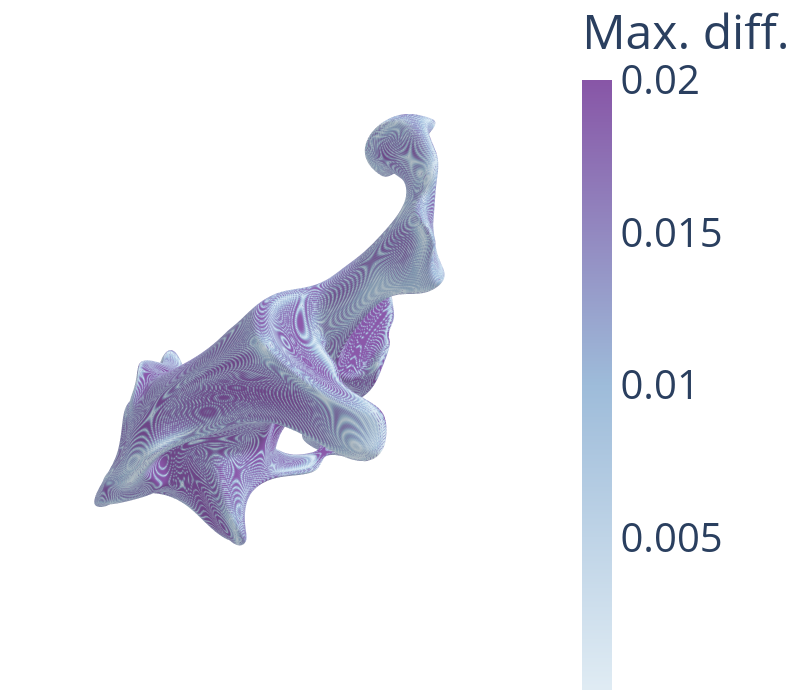}
        \caption{Max. Variation}
        \label{subfig:hcci_oh_ALL}
    \end{subfigure}
    \begin{subfigure}[b]{0.19 \textwidth}
        \centering
        \includegraphics[width=0.99 \columnwidth]{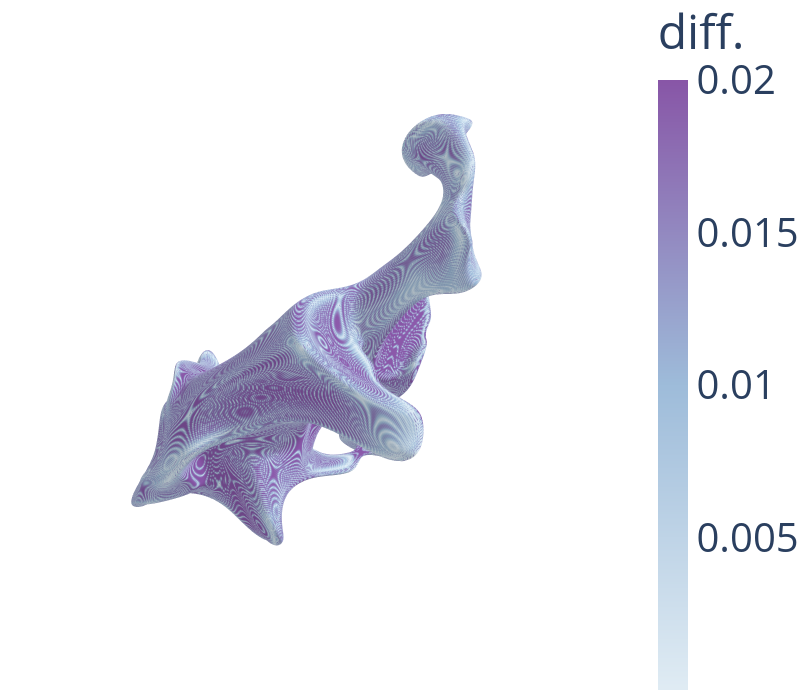}
        \caption{L vs C}
        \label{subfig:hcci_oh_LC}
    \end{subfigure}
     % %
    \begin{subfigure}[b]{0.19 \textwidth}
        \centering
        \includegraphics[width=0.99 \columnwidth]{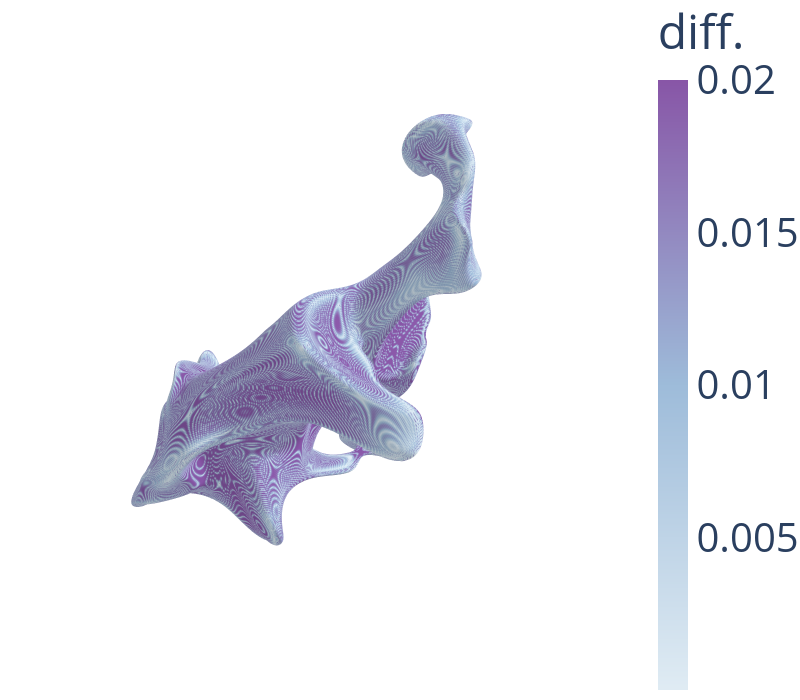}
        \caption{L vs W}
        \label{subfig:hcci_oh_LW}
    \end{subfigure}
     % %
    \begin{subfigure}[b]{0.19 \textwidth}
        \centering
        \includegraphics[width=0.99 \columnwidth]{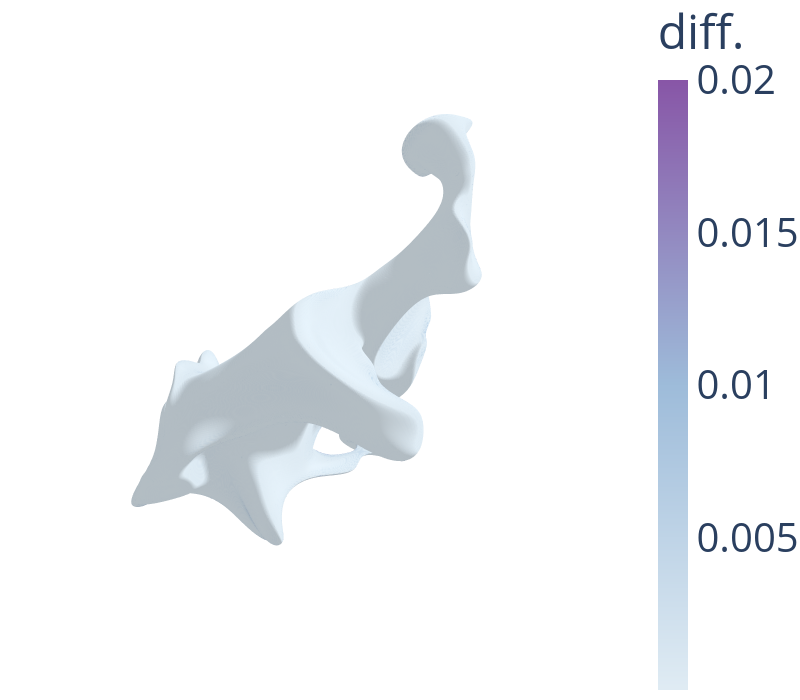}
        \caption{C vs W}
        \label{subfig:hcci_oh_CW}
    \end{subfigure}
     %%%%%%%%%%%%%%%%%%%%%%%%%%%%%%%%%%%%%%%%%%%%%%%%%%%%%%%%%%%%%%%%%%%%%%%%%%%%%%%
    \begin{subfigure}[b]{0.19 \textwidth}
        \centering
        \includegraphics[width=0.99 \columnwidth]{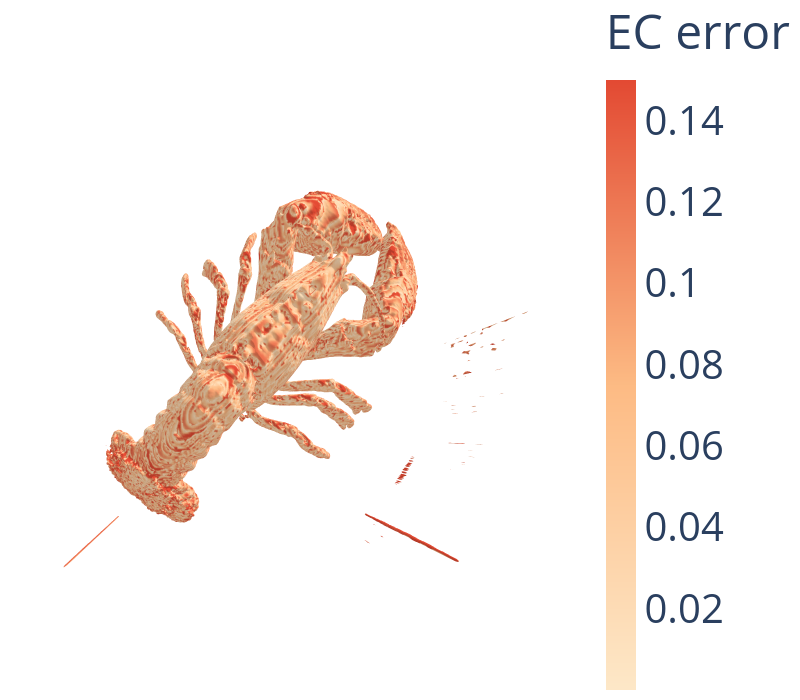}
        \caption{Fuel Approximated Error}
        \label{subfig:lobster_Approximated_Err}
    \end{subfigure}
    \begin{subfigure}[b]{0.19 \textwidth}
        \centering
        \includegraphics[width=0.99 \columnwidth]{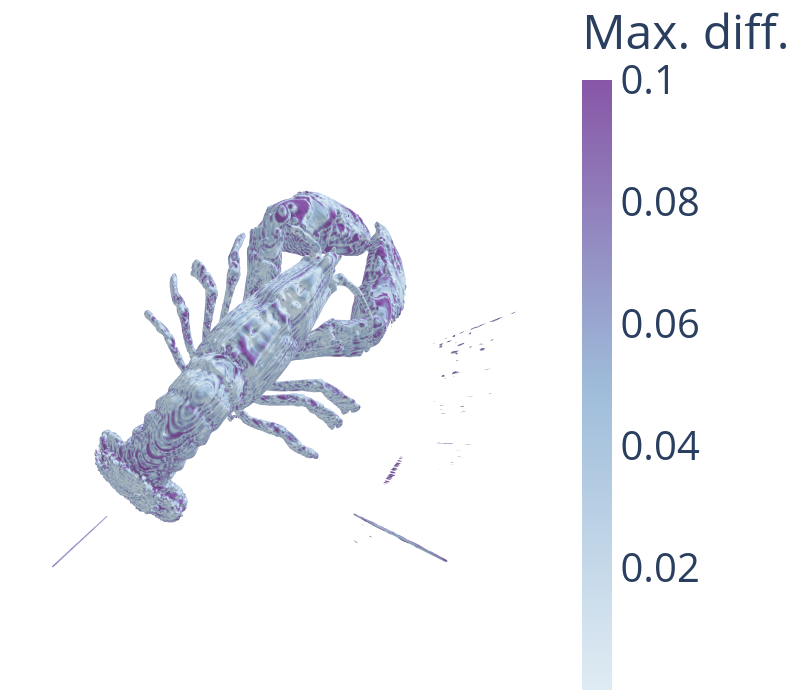}
        \caption{Max. Variation}
        \label{subfig:lobster_Approximated_Err}
    \end{subfigure}
    \begin{subfigure}[b]{0.19 \textwidth}
        \centering
        \includegraphics[width=0.99 \columnwidth]{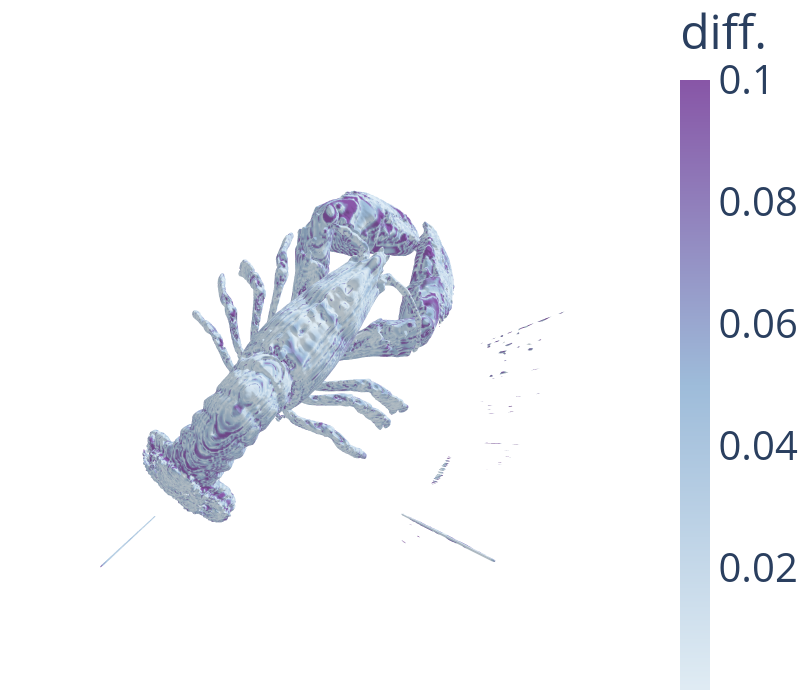}
        \caption{L vs C}
        \label{subfig:lobster_LC}
    \end{subfigure}
     % %
    \begin{subfigure}[b]{0.19 \textwidth}
        \centering
        \includegraphics[width=0.99 \columnwidth]{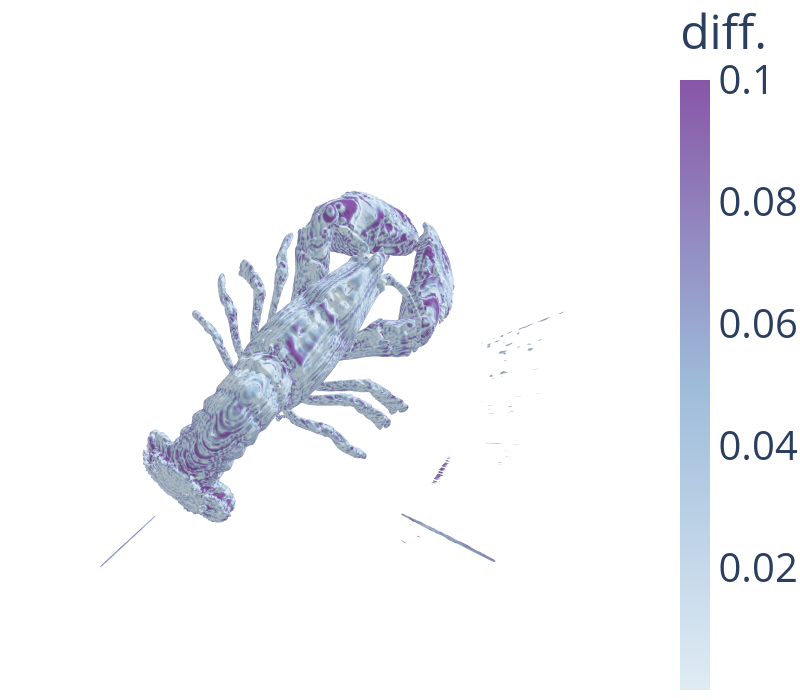}
        \caption{L vs W}
        \label{subfig:lobster_LW}
    \end{subfigure}
     % %
    \begin{subfigure}[b]{0.19 \textwidth}
        \centering
        \includegraphics[width=0.99 \columnwidth]{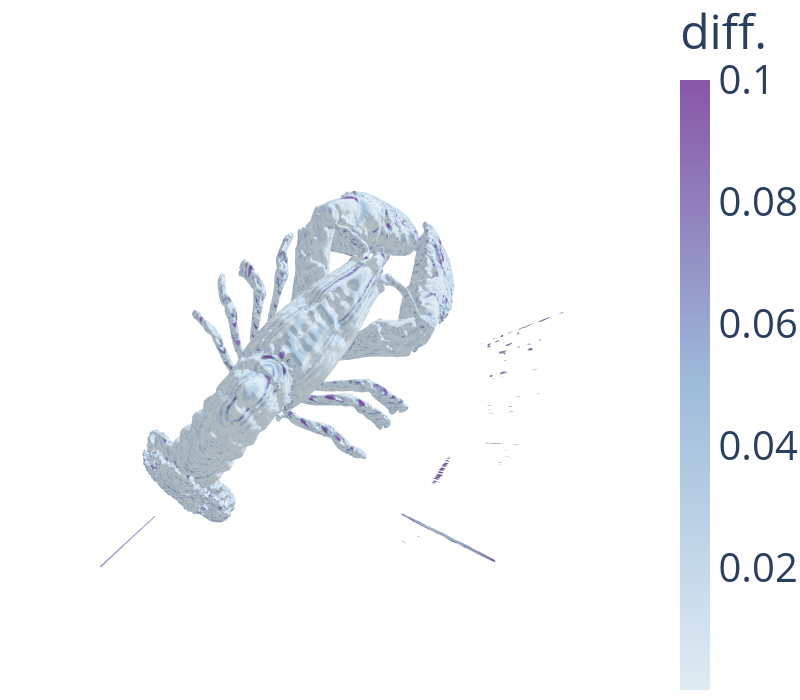}
        \caption{C vs W}
        \label{subfig:lobster_CW}
    \end{subfigure}
    \vspace{-3mm}
    \caption{Comparison of approximated error and different interpolation methods (Components $C_{1}$, $C_{3}$, $C_{4}$, and $C_{6}$). The first column 
    % (\cref{subfig:fuel_Approximated_Err}, \cref{subfig:engine_Approximated_Err}, \cref{subfig:hcci_oh_Approximated_Err}, \cref{subfig:lobster_Approximated_Err}) 
    corresponds to the approximated edge-crossing error. The second column shows the maximum variation. The remaining third, fourth, fifth, and sixth columns correspond to the comparison L vs. C, L vs. W, and C vs. W, respectively. These results show that using our methods in the first column from left effectively identifies regions with large errors and variations between linear and higher-order interpolation methods.}
    \label{fig:figs_real_world_examples_edge-crossing}
\end{figure*}

\section{Results}
\label{sec:results}
% Here, synthetic and real-world datasets are used to demonstrate and evaluate the uncertainty estimations and visualization methods introduced in \cref{sec:method}.

\subsection{Synthetic Examples}
\label{subsec:synthetic_examples}
The synthetic examples are based on the \textbf{Tangle}, \textbf{Torus}, \textbf{Marschner and Lobb}, \textbf{Teardrop},  and \textbf{Tubey}. \Cref{fig:figs_threshol_global} shows an exploration/analysis pipeline using the visualization framework to gain insight into the isosurface uncertainties from interpolation methods.
% using the components $C_{1}$, $C_{3}$, $C_{4}$ and $C_{6}$. 
The first column in \cref{fig:figs_threshol_global} shows a selected error threshold (vertical dashed line) and the corresponding percentage of vertices with errors larger than the specified error threshold is indicated with the red horizontal line. The results in the second column in \cref{fig:figs_threshol_global} are constructed from measured error obtained METRO~\cite{Cignoni1998}. These results show the isosurfaces with a binary colormap indicating the regions with isosurface errors that are above (in reg) and below (in light orange) the specified error threshold. The third column shows the maximum variation across all three interpolation methods. The fourth, fifth, and sixth columns in \cref{fig:figs_threshol_global} show a comparison between linear and cubic (L vs. C), linear and WENO (L vs. W), and cubic and WENO (C vs. W) in the regions with errors larger than the selected threshold. The similar patterns between the second, third, fourth, and fifth columns in \cref{fig:figs_threshol_global} indicate that the approximated error effectively identifies regions with large errors which corresponds to the regions with large variation between interpolation methods indicated in purple. 
% linear and high-order interpolation methods shown in yellow in the remaining third, fourth and fifth columns. 
% The validation of the error estimations demonstrated in the previous comparison is shown in \cref{fig:edge_crossing_error}. 
The fourth and fifth columns of \cref{fig:figs_threshol_global} show the possible accuracy improvement from linear to cubic and WENO.
% \color{blue}
The linear, cubic, WENO interpolation are $O(h^{2})$, $O(h^{4})$, and $O(h^{5})$ accurate. 
% \color{black}
The fifth column highlights the difference between cubic and WENO.  WENO further improves the cubic interpolation accuracy from $O(h^{4})$ to $O(h^{5})$. The improvement from cubic to WENO can be minor as shown in \cref{subfig:torus_threshold_CW_64x64x64}.

% The increase in order of accuracy from cubic to WENO is smaller than the increase between linear and cubic, or linear and WENO. 

The visualization tool in \cref{fig:frame-work} uses $C_{2}$ for local uncertainty exploration 
% by querying information about the locally selected region
, as shown in the first and second columns of the teaser image \cref{fig:figs_local_plus_hidden_futeres}.  The transparent boxes in \cref{subfig:teardrop_box_32x32x32} and \cref{subfig:tubey_box_32x32x32} indicate the selected local region of interest. These selections correspond to the regions with broken pieces and hidden features. \cref{subfig:teardrop_box_LC_LW_32x32x32} and \cref{subfig:tubey_box_LC_LW_32x32x32} show a local vertex-by-vertex comparison of the approximate edge-crossing error, L vs. C, and L vs. W. 
\cref{subfig:teardrop_LW_32x32x32} and \cref{subfig:tubey_LW_32x32x32} show a comparison of L vs. W. This local vertex-by-vertex comparison enables a detailed comparison of the magnitude of the approximated error and the difference between interpolation methods. The fourth and fifth columns show a comparison of the isosurface with (transparent orange) and without (opaque blue) hidden feature recovery. The zoomed-in versions in \cref{subfig:teardrop_zoom1_global_hidden_features_32x32x3} and \cref{subfig:tubey_zoom1_global_hidden_features_32x32x32} demonstrate that our proposed feature-reconstructions methods introduced in \cref{subsec:hidden_features} successfully constructs possible connection among the broken components (the transparent orange isosurface) and propose an alternate isosurface to be considered. 
% The MC algorithms with and without sharp feature recovery fail to recover the missing features in \cref{subfig:teardrop_global_hidden_features_32x32x32} and \cref{subfig:tubey_zoom1_global_hidden_features_32x32x32}. 
%Our hidden feature recovery method, therefore, presents a novel way of exploring uncertain feature topology.

\subsection{Real-World Examples}
\label{subsec:real_worl_examples}

The datasets obtained from  ~\cite{scivisdata} include a simulation of fuel injection into a combustion chamber ($64^{3}$), a CT scan of an engine ($256\times256\times128$), a CT scan of a lobster ($301 \times 324 \times 56$), a simulation of a homogeneous charge compression ignition ($560^{3}$), a rotational C-arm x-ray scan of the arteries of the right half of a human head ($256^{3}$), a CT scan of a Bonsai tree ($256^{3}$), and a CT scan of a carp fish ($256 \times 256 \times 512$).

\begin{figure*}[!ht]
    \centering
     % %
    \begin{subfigure}[b]{0.20 \textwidth}
        \centering
        \includegraphics[width=0.70 \columnwidth]{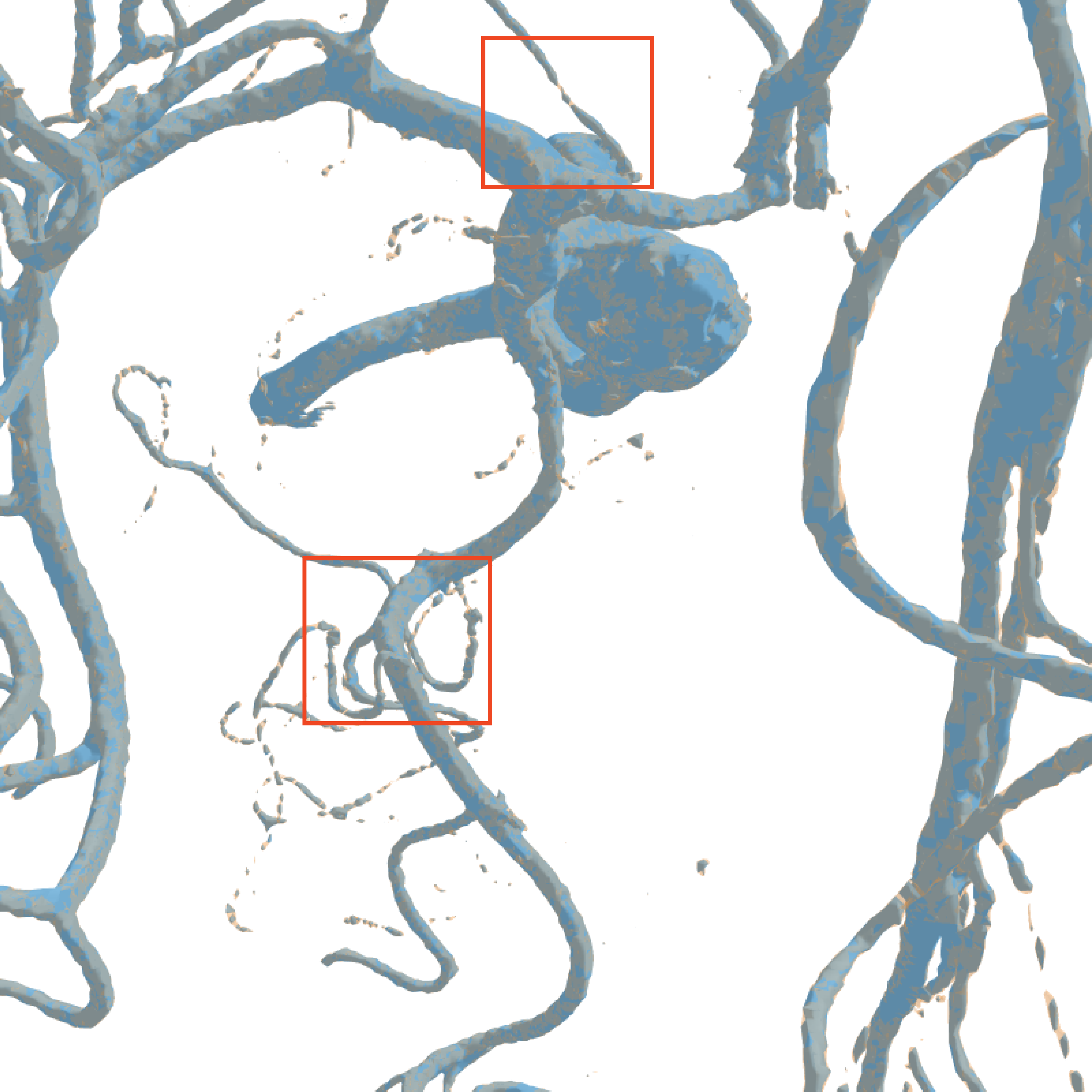}
        \caption{Aneurism  ($k=160.0$)}
        \label{subfig:aneurism}
    \end{subfigure}
    \begin{subfigure}[b]{0.19\textwidth}
        \centering
        \includegraphics[width=0.70 \columnwidth]{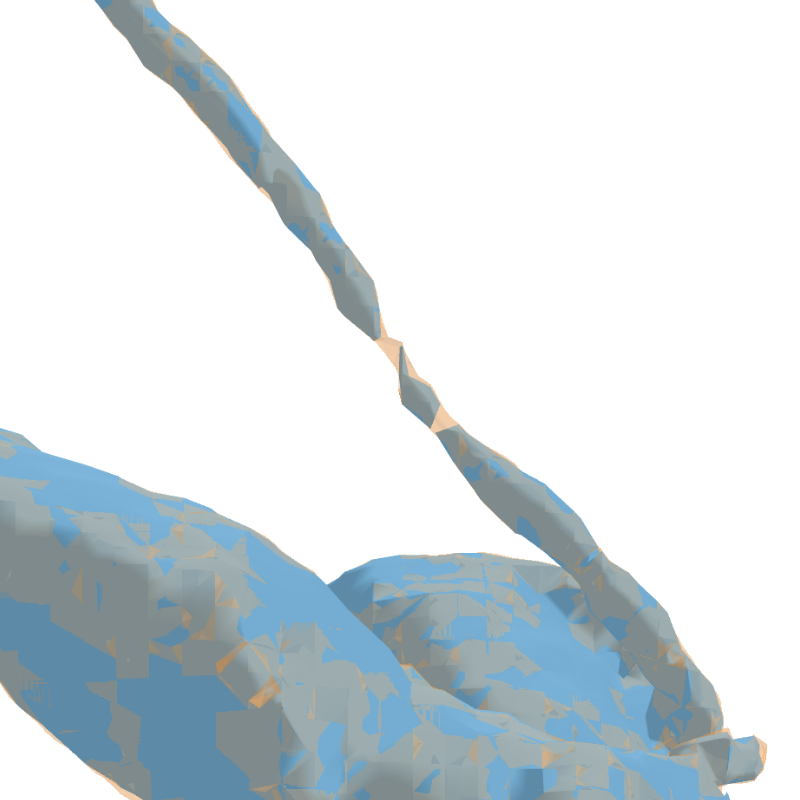}
        \caption{Hidden features (top box)}
        \label{subfig:aneurism_zoomed1}
    \end{subfigure}
    \begin{subfigure}[b]{0.19 \textwidth}
        \centering
        \includegraphics[width=0.70 \columnwidth]{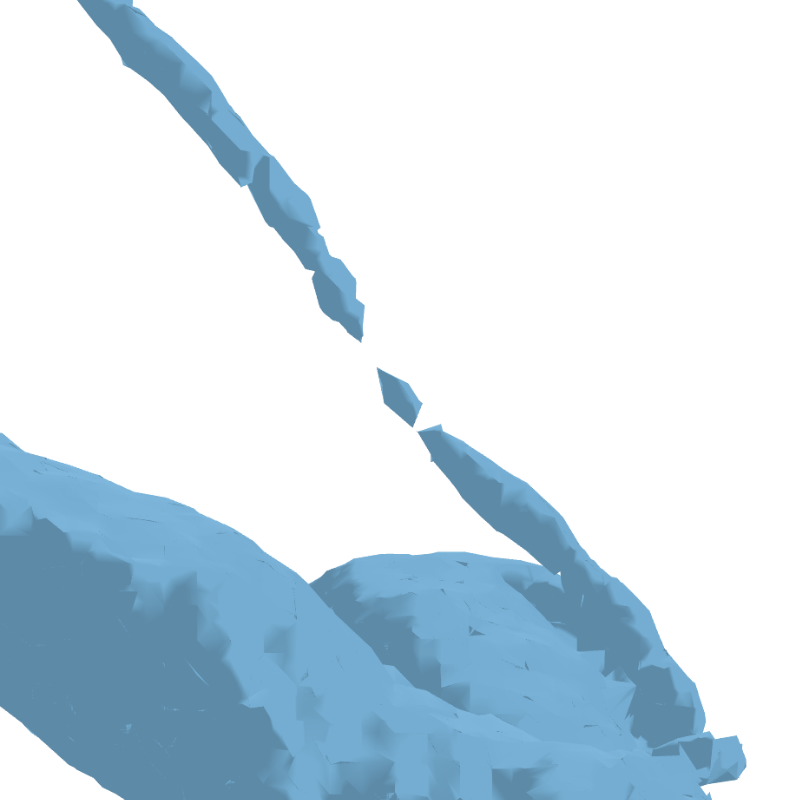}
        \caption{Dual Contouring (top box)}
        \label{subfig:aneurism_zoomed1_dc}
    \end{subfigure}
    \begin{subfigure}[b]{0.19 \textwidth}
        \centering
        \includegraphics[width=0.70 \columnwidth]{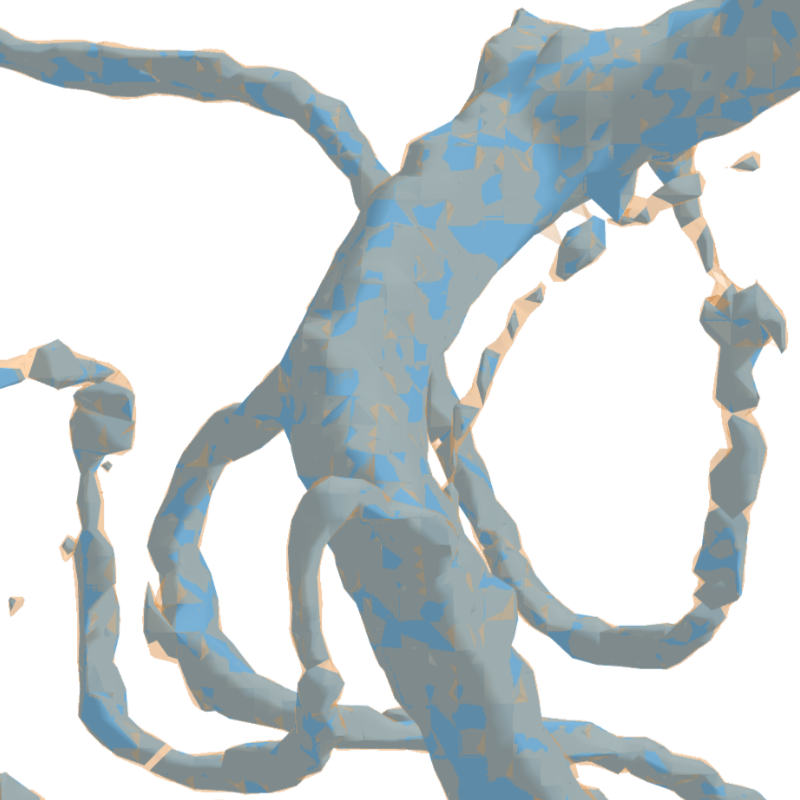}
        \caption{Hidden features (bottom box)}
        \label{subfig:aneurism_zoomed2}
    \end{subfigure}
    \begin{subfigure}[b]{0.19 \textwidth}
        \centering
        \includegraphics[width=0.70 \columnwidth]{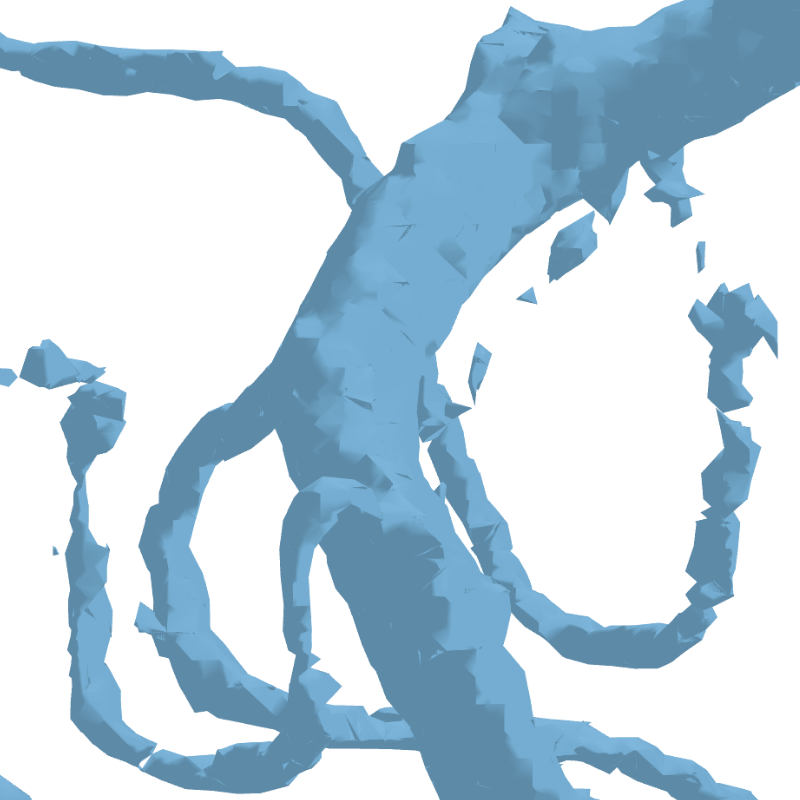}
        \caption{Dual contouring (bottom box)}
        \label{subfig:aneurism_zoomed2_dc}
    \end{subfigure}
     % %%%%%%%%%%%%%%%%%%%%%%%%%%%%%%%%%%%%%%%%%%%%%%%%%%%%%%%%%%%%%%%%%%%%%%%%%%%%%%%%%
    \begin{subfigure}[b]{0.20 \textwidth}
        \centering
        \includegraphics[width=0.70 \columnwidth]{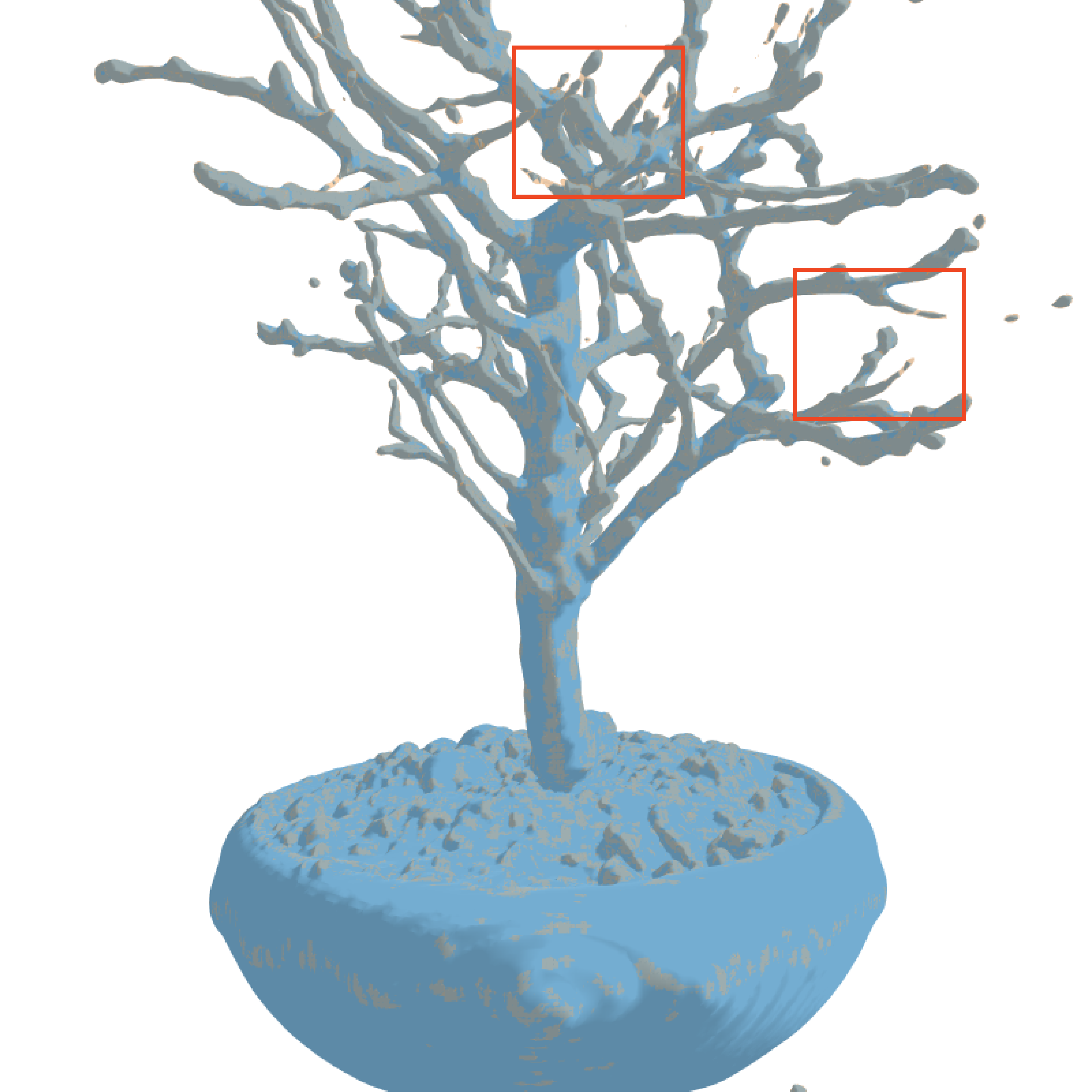}
        \caption{Bonsai ($k=75.0$)}
        \label{subfig:bonsai}
    \end{subfigure}
    \begin{subfigure}[b]{0.19 \textwidth}
        \centering
        \includegraphics[width=0.70 \columnwidth]{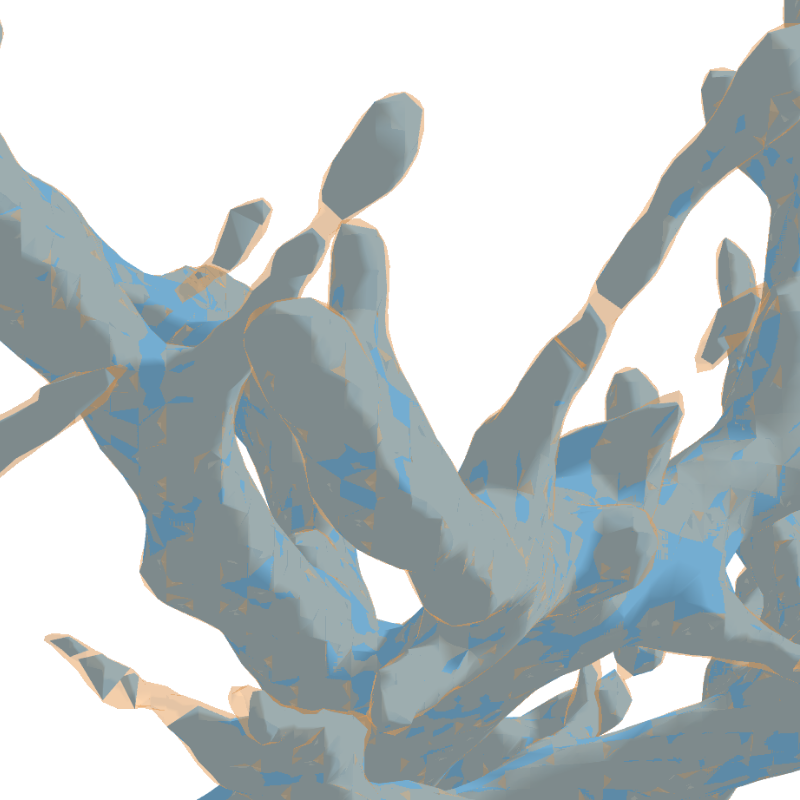}
        \caption{Bonsai zoomed in (top box)}
        \label{subfig:bonsai_zoomed1}
    \end{subfigure}
    %%%%
    \begin{subfigure}[b]{0.19 \textwidth}
        \centering
        \includegraphics[width=0.70 \columnwidth]{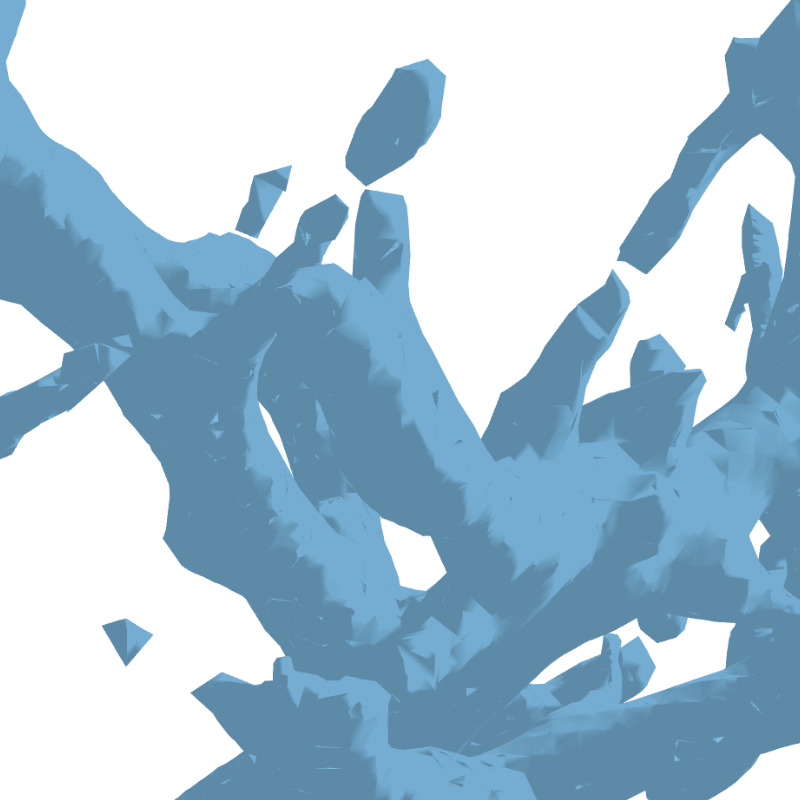}
        \caption{Dual contouring (top box)}
        \label{subfig:bonsai_zoomed1_dc}
    \end{subfigure}
    \begin{subfigure}[b]{0.19 \textwidth}
        \centering
        \includegraphics[width=0.70 \columnwidth]{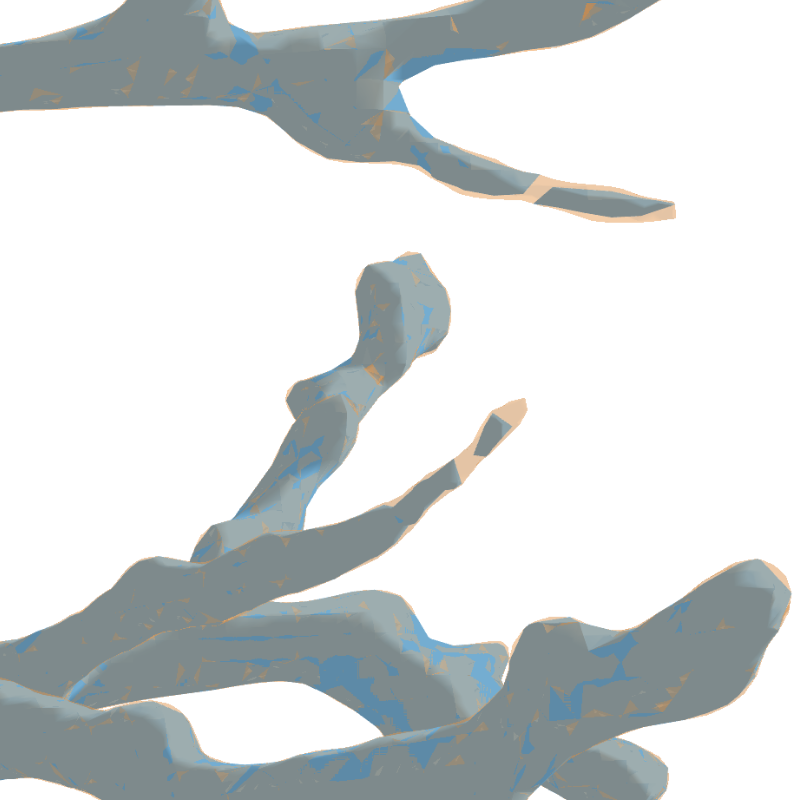}
        \caption{Hidden features (bottom box)}
        \label{subfig:bonsai_zoomed2}
    \end{subfigure}
    \begin{subfigure}[b]{0.19 \textwidth}
        \centering
        \includegraphics[width=0.70 \columnwidth]{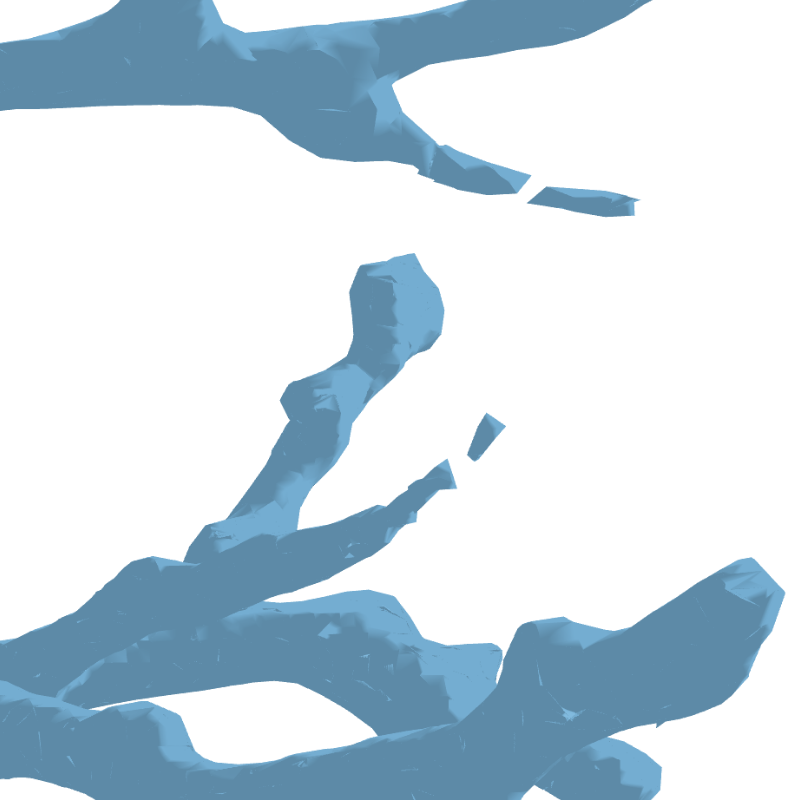}
        \caption{Dual contouring (bottom box)}
        \label{subfig:bonsai_zoomed2_dc}
    \end{subfigure}
      
    % %%%%%%%%%%%%%%%%%%%%%%%%%%%%%%%%%%%%%%%%%%%%%%%%%%%%%%%%%%%%%%%%%%%%%%%%%%%%%%%%%
    \begin{subfigure}[b]{0.20 \textwidth}
        \centering
        \includegraphics[width=0.70 \columnwidth]{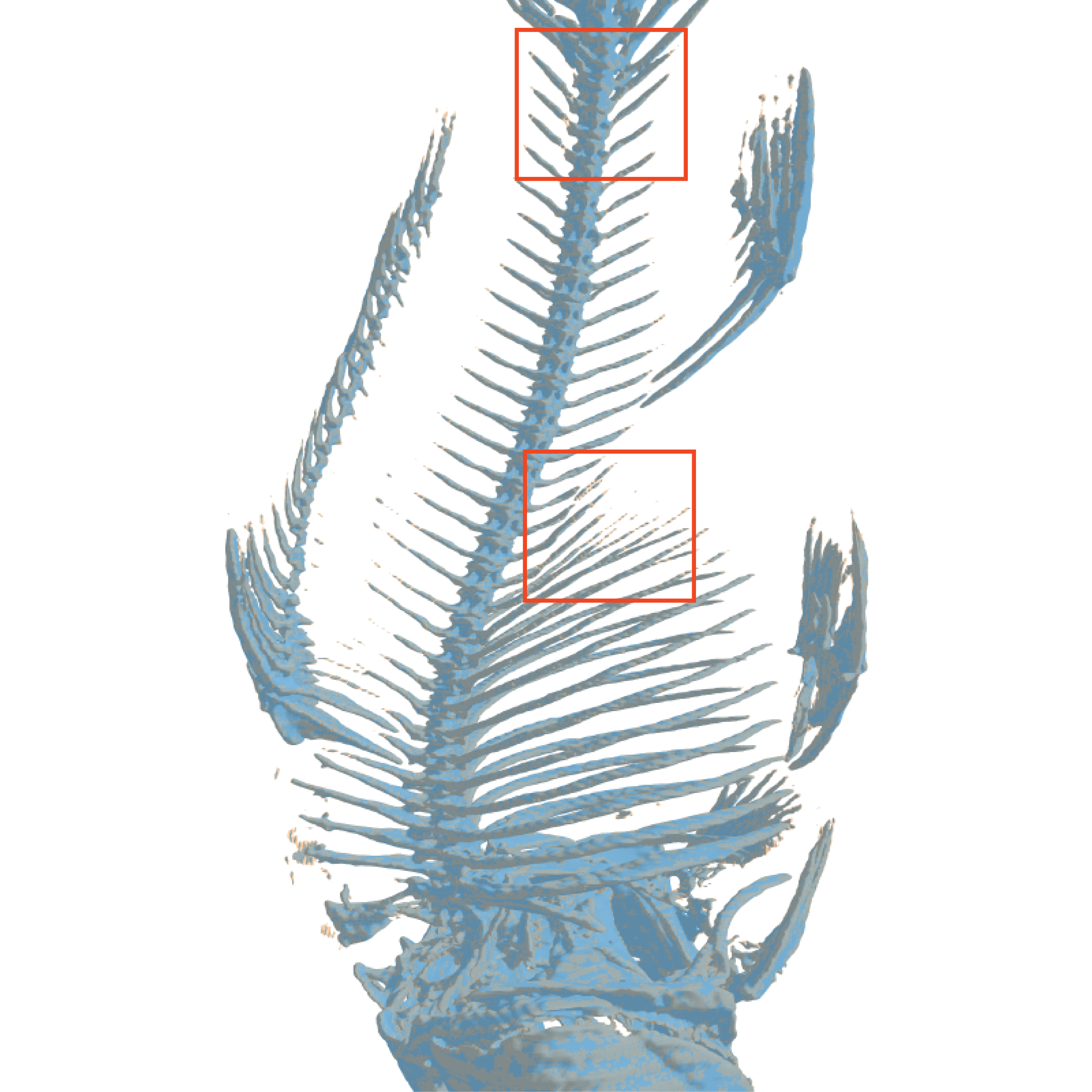}
        \caption{Carp ($k=1270.0$)}
        \label{subfig:carp}
    \end{subfigure}
    \begin{subfigure}[b]{0.19 \textwidth}
        \centering
        \includegraphics[width=0.70 \columnwidth]{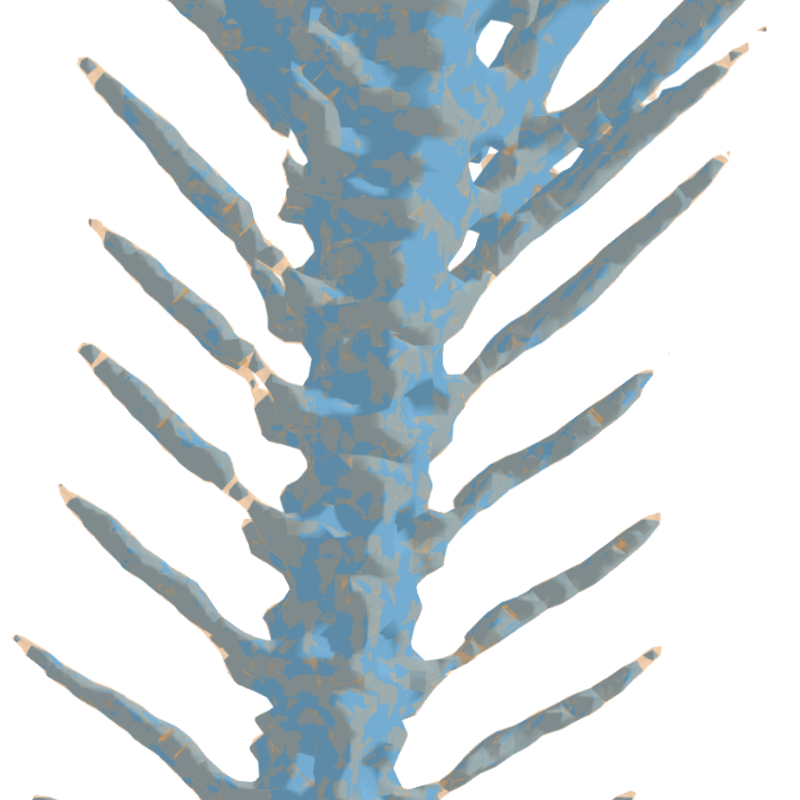}
        \caption{Hidden features (top box)}
        \label{subfig:carp_zoomed1}
    \end{subfigure}
    \begin{subfigure}[b]{0.19 \textwidth}
        \centering
        \includegraphics[width=0.70 \columnwidth]{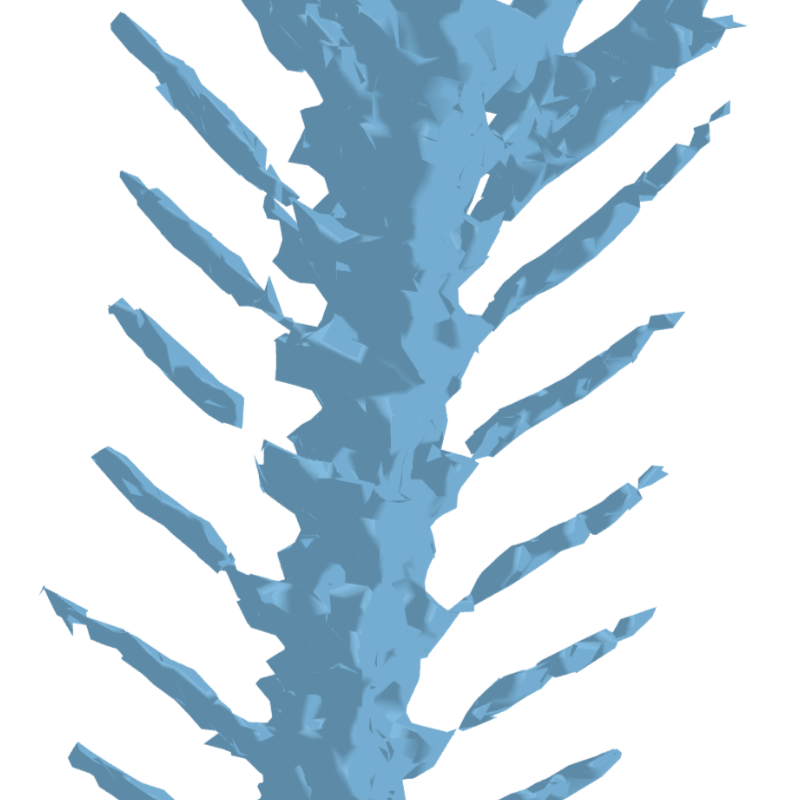}
        \caption{Dual contouring (top box)}
        \label{subfig:carp_zoomed1_dc}
    \end{subfigure}
    \begin{subfigure}[b]{0.19 \textwidth}
        \centering
        \includegraphics[width=0.70 \columnwidth]{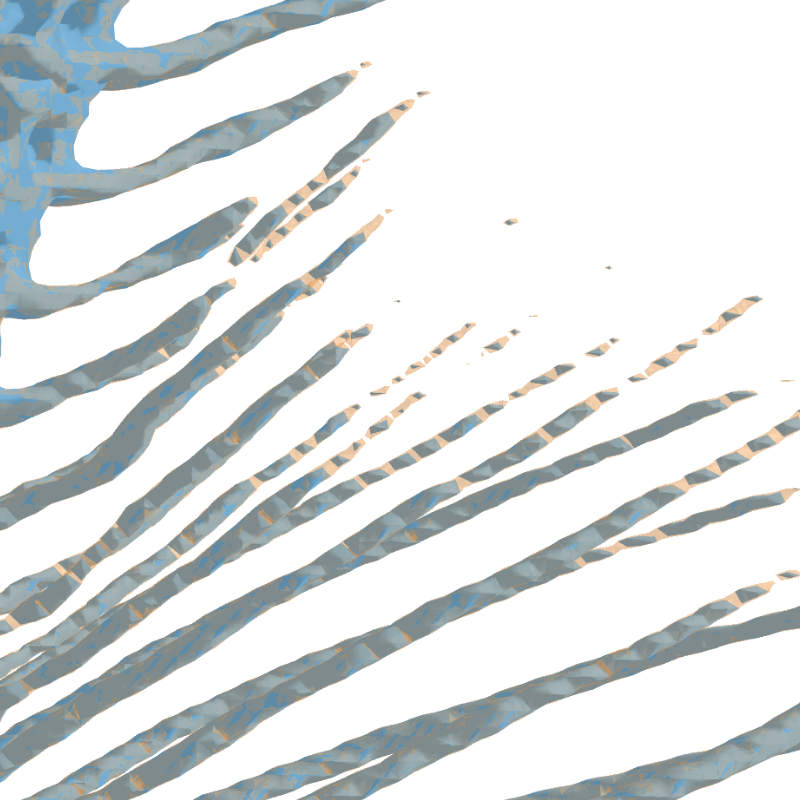}
        \caption{Hidden features (bottom box)}
        \label{subfig:carp_zoomed2}
    \end{subfigure}
    \begin{subfigure}[b]{0.19 \textwidth}
        \centering
        \includegraphics[width=0.70 \columnwidth]{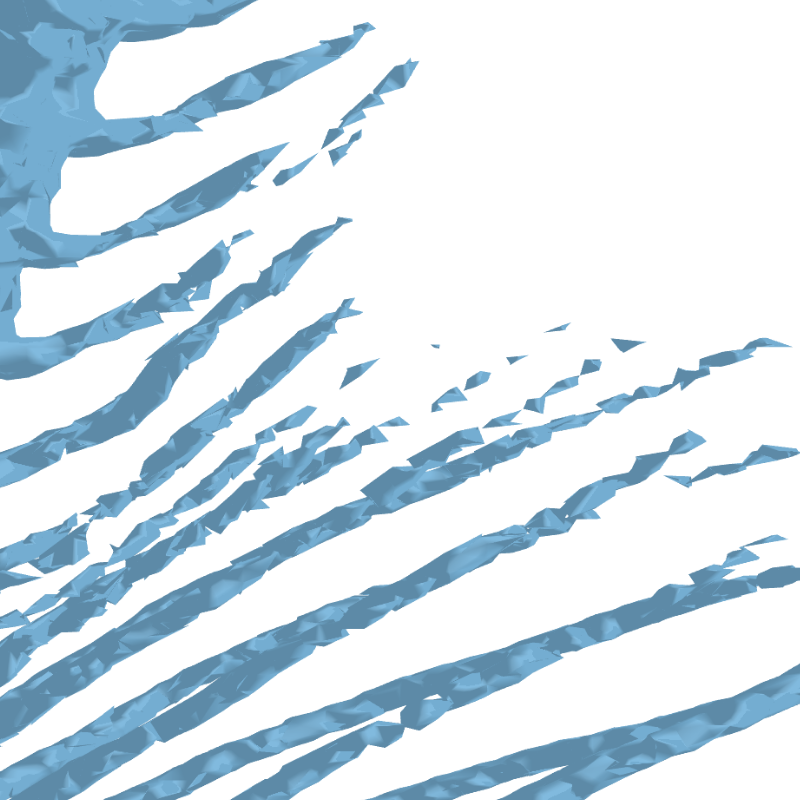}
        \caption{Dual contouring (bottom box)}
        \label{subfig:carp_zoomed2_dc}
    \end{subfigure}
    \vspace{-3mm}
    \caption{Comparison of isosurfaces from MC using our uncertain-feature recovery methods in $C_{6}$ (first, second, and fourth columns from left) vs. sharp feature recovery (third and fifth columns from left). The first column shows isosurfaces with selected regions indicated with red boxes. The second and fourth rows show a zoomed-in isosurface with (transparent orange) and without (blue) hidden feature recovery. The third and fifth columns show a zoomed-in isosurface from dual counting. The results indicate possible new interesting features and topological connections by semitransparent orange surfaces that are missed by the state-of-the-art MC feature extraction method.}
    \label{fig:figs_real_hidden_features}
\end{figure*}

\subsubsection{Edge-Crossing Uncertainty}
\label{subsubsec:edge-crossing}
The isosurface uncertainty shown in \cref{fig:figs_real_world_examples_edge-crossing} 
 is based on our edge-crossing error estimation introduced in \cref{subsec:error_approximation}, and the difference between interpolation methods. The approximated isosurface uncertainty shown in the first column (\cref{subfig:fuel_Approximated_Err}, \cref{subfig:engine_Approximated_Err}, \cref{subfig:hcci_oh_Approximated_Err}, \cref{subfig:lobster_Approximated_Err}) is larger than the 
 difference between interpolation methods shown in the remaining columns. The regions with high errors (red regions in the first column) correspond to the regions with large differences (uncertainty) between interpolation methods. 
For instance, in the fuel example, the black rectangle delineates a region of high error detected by our method in \cref{subfig:fuel_Approximated_Err}, corresponding to the same region with significant differences between linear and higher-order methods in \cref{subfig:fuel_LC} and \cref{subfig:fuel_LW}. These results show that the edge-crossing error estimation in \cref{subsec:error_approximation} identifies regions with large uncertainty in the context of practical datasets. In addition, the similarity observed demonstrates that our uncertainty estimation methods efficiently indicate the positions (red regions), where high-order interpolation can improve isosurface accuracy compared to linear interpolation. In the case of the Engine dataset, the improvements from linear in \cref{subfig:engine_LC}, \cref{subfig:engine_LW}, and \cref{subfig:engine_CW} are considerably smaller than the approximated errors in \cref{subfig:engine_Approximated_Err}. These results indicate that using higher-order interpolation methods in the case of the engine dataset 
% \color{blue} 
doesn't significantly improve the accuracy.
% \color{black} 
For the fuel, combustion simulation, and lobster examples, the differences between cubic and WENO are smaller compared to the differences between linear and high-order interpolation methods, as shown in \cref{subfig:fuel_CW}, \cref{subfig:hcci_oh_CW}, and \cref{subfig:lobster_CW}.
% The linear, cubic, WENO interpolation are $O(h^{2})$, $O(h^{4})$, and $O(h^{5})$ accurate. The increase in order of accuracy from cubic to WENO is smaller than the increase between linear and cubic, or linear and WENO. 

\subsubsection{Hidden Features Comparison}
\label{subsubsec:hidden-feature}
   The uncertain isosurface feature recovery using our proposed methods (\cref{subsec:hidden_features}) in the first, second, and fourth columns from the left of \cref{fig:figs_real_hidden_features} improves the reliability of results. State-of-the-art techniques miss these important features, as shown in the third and fifth columns from left in \cref{fig:figs_real_hidden_features}, which can lead to less reliable data analysis. The targeted regions of interest are shown with the rectangles in the first column. The zoomed-in regions shown in the remaining columns correspond to the red rectangles. The second and fourth columns show a comparison between the standard MC (opaque blue) and the hidden feature recovery method introduced in \cref{subsec:hidden_features} (transparent orange). Showing the standard MC and hidden feature recovery in the framework enables a visual comparison of isosurface features. The third and fifth columns show results for sharp feature recovery based on dual contouring \cite{Ju2002}.\\
   \indent The zoomed-in results show that many broken pieces in the isosurface without hidden feature recovery shown with the opaque blue in the second and fourth columns (\cref{subfig:aneurism_zoomed1}, \cref{subfig:aneurism_zoomed2}, \cref{subfig:bonsai_zoomed1}, \cref{subfig:bonsai_zoomed2}, \cref{subfig:carp_zoomed1}, and \cref{subfig:carp_zoomed2}) are connected in the isosurface with hidden feature recovery shown in transparent orange in the same figures. The dual contouring method connects the broken pieces but introduces sharp corners and edges.  

%% file: discussion_conclusion.tex
\section{Discussion}
\label{sec:discussion}

% We introduced an edge-crossing error approximation in \cref{subsec:error_approximation}, a hidden feature detection and \tajo{reconstruction} method in \cref{subsec:hidden_features}, and a visualization framework to explore and provide insights about isosurface uncertainty \cref{subsec:framework}. The framework shown in \cref{fig:frame-work} is suitable for global and local analysis of isosurface errors and feature differences. The visualization of approximated errors provides a global view of interpolation errors in components $C_{4}$. The box selection method (component $C_{2}$) enables exploration of local uncertainty ($C_{5}$, $C_{6}$) and understanding of hidden uncertain features ($C_{6}$). 
We presented an integrated interactive visualization system that provides valuable insights into the uncertainties inherent in the MC algorithm with linear interpolation. Our error estimation method introduced in \cref{subsec:error_approximation}, provides a more accurate approximation of the edge-crossing errors compared to standard error approximation, as demonstrated in \cref{fig:edge_crossing_error} and \cref{fig:figs_max_mean_rms}. The estimated error can be computed for any 3D volumetric data without the need for additional information, such as high-resolution data which underscores its broad applicability.  In addition, the error estimation is computationally efficient compared to measured error and other sampling methods because the errors are directly calculated using the local neighborhood cells.

% The results in \cref{fig:edge_crossing_error} and \cref{fig:figs_max_mean_rms} demonstrate that the edge-crossing error approximation described in \cref{subsec:error_approximation} is comparable to the measured error obtained by computing the difference between the extracted isosurface and a high-resolution target isosurface. The estimated error can be computed for any 3D volumetric data because it doesn't require any additional information, such as high-resolution data, to compare against. In addition, the error estimation is computationally efficient %\ta{I am quite sure that reviewers may ask to report performance results. You mentioned one example previously, but they might expect more. Though we may not have enought time to report some more results in terms of say speedup by xxx$\times$ factor} 
% compared to measured error and other sampling methods because the errors are directly calculated using the local neighborhood cells. Our proposed method effectively provides insight into edge-crossing errors without requiring high-resolution data. 

\Cref{fig:figs_threshol_global} and \cref{fig:figs_real_world_examples_edge-crossing} show examples of how the framework in \cref{fig:frame-work} can be used to visualize edge-crossing errors and compare different interpolation methods. The results in these figures further demonstrate that the isosurface uncertainty estimation, shown in the first column of both figures, successfully identifies regions with large errors (red) and variations between linear and higher-order interpolation methods. The orange regions in \cref{fig:figs_threshol_global} and \cref{fig:figs_real_world_examples_edge-crossing} show the accuracy improvement from linear to higher order methods.

Moreover, we introduced a hidden feature reconstruction method in \cref{subsec:hidden_features} that successfully identifies and reconstructs possible features that are ignored by the MC algorithms. The extraction of these features is based on fitting and refining the target cells using cubic interpolation. The polygon extraction at the boundary of the refined and unrefined cells causes cracks that are resolved using a similar approach in ~\cite{Kobbelt2001}, illustrated in \cref{fig:crack_patching}. we visualize the isosurfaces with (transparent orange) and without (opaque blue) to enable visual comparison of two possible isosurfaces from MC with different features and topological structures, as shown in \cref{fig:figs_local_plus_hidden_futeres}. \Cref{fig:figs_real_hidden_features} shows that our method for hidden feature reconstruction leads to a smoother connection between broken pieces compared to dual contouring which introduces sharp edges and corners. 

% The synthetic results in \cref{fig:figs_local_plus_hidden_futeres} show that the hidden feature reconstruction method introduced in \cref{subsec:hidden_features} identifies and reconstructs futures that are ignored by the MC algorithms. The extraction of these features is based on fitting and refining the target cells using cubic interpolation. The polygon extraction at the boundary of the refined and unrefined cells causes cracks that are resolved using a similar approach in ~\cite{Kobbelt2001}, illustrated in \cref{fig:crack_patching}. The visualization of the isosurfaces with (transparent orange) and without (opaque blue) enables visual comparison of two possible isosurfaces from MC with different features and topological structures. These differences provide the user with valuable insights into the potential features and limitations of the isosurface reconstructions, highlighting areas that may require more attention to improve the overall quality and accuracy of the isosurface. \Cref{fig:figs_real_hidden_features} shows that our method for hidden feature reconstruction leads to a smoother connection between broken pieces compared to dual contouring which introduces sharp edges and corners. 

% The work presented successfully employs various techniques and interactive visualization to provide valuable insights into the uncertainties inherent in the Marching Cubes algorithm with linear interpolation. 
It is important to note that the techniques introduced have some limitations that we plan to address as this work continues. Even though we didn't observe these issues in the datasets used, the cubic, WENO, and other high-order interpolation methods may introduce undesirable oscillations. These oscillations may be reduced without comprising the possible hidden features with bounded interpolation methods ~\cite{Ouermi2023, Ouermi2024}. In addition, non-polynomial-based methods could provide more accurate error approximation for data that are generated from processes that don't rely on polynomials.

\section{Conclusion}
\label{sec:conclusion}
In this paper, we presented an efficient method for estimating and visualizing isosurface uncertainty from Marching Cubes (MC) algorithms. We introduce a closed-form approximation of edge-crossing error using polynomial interpolation and develop a technique for detecting and reconstructing uncertain hidden features. These approaches provide valuable insights into isosurface uncertainty and highlight the limitations of linear interpolation. In addition, we developed an integrated visualization system for the exploration and analysis of these uncertainties. Our examples and results demonstrate the effectiveness of our methods in estimating and visualizing isosurface uncertainty associated with linear interpolation. 
% \color{blue}
This work focused on error estimations for linear, cubic, and WENO interpolation methods. Extending the error analysis to include higher-order polynomial and non-polynomial interpolation techniques in future work could further improve error characterization across a broader range of interpolation models. Additionally, the current framework visualizes errors and potential hidden features separately. Integrating these into a unified visualization could provide a more comprehensive analysis of isosurface uncertainty.
% \color{black}

%% file: appendix.tex
\section{Appendix}
    List of Synthetic examples used in this paper. 
    
  \textbf{Tangle}:
   \begin{equation}\label{eq:tangle}
        f_{1}(x,y,z) =  x^4 + y^4 + z^4 - (x^2 + y^2 + z^2 - 0.4), \textrm{ }x, y, z \in [-1,1].
    \end{equation}\label{eq:tangle}
    % \vspace{-0.9cm}
    %
    \textbf{Torus}:
    \begin{multline}\label{eq:torus}
        f_{2}(x,y,z) = r_{1} - \bigg(\sqrt{x^2 + y^2}\bigg)^{2} + z^{2} -r_{0}^{2}, x, y, z \in [-1,1],\\ 
                       r_{0}=0.1, \textrm{ } r_{1} = 0.3.
    \end{multline}
    % \vspace{-0.9cm}
    %
    \textbf{Marschner and Lobb}:
    \begin{multline}\label{eq:ML}
        f_{3}(x,y,z) = \frac{\bigg( 1-sin(\frac{\pi z}{2}) + \alpha (1 + \rho_{r}(\sqrt{x^2 + y^2 })\bigg)}{2(1+\alpha)}, \textrm{ }x, y,z \in [-1,1] \\
        \text{where: } \rho_{r} (r)= cos\big(2\pi f_{M} cos(\frac{\pi r}{2})\big), \textrm{ } f_{M} = 6 \textrm{ and } \alpha = 0.25.
    \end{multline}
   % \vspace{-0.7cm}
    %
   \textbf{Teardrop}:
    \begin{equation}\label{eq:teardrop}
        f_{4}(x,y,z) = 0.5x^{5} + 0.5x^{4} - y^{2} - z^{2}, \textrm{ } x, y, z \in [-1, 1]
    \end{equation}
    % \begin{align}\label{eq:hyperboloid}
    %     f_{4}(x,y,z) = \frac{x^2}{5} + \frac{y^2}{5} - \frac{z^2}{10} , \textrm{ } x,y,z \in [-1,1]
    % \end{align}
    %
    %
    \textbf{Tubey}:
    \begin{multline}\label{eq:tubey}
        f_{5}(x,y,z) = -3x^8 - 3y^8 - 2z^8 + 5x^4 y^2 z^2 + 3x^2 y^4 z^2 \\
        - 4(x^3 + y^3 + z^3 + 1) + (x + y + z + 1)^4 + 1, \textrm{ } x,y,z \in [-3, 3]    
    \end{multline}

%% file: main.bbl
\begin{thebibliography}{10}

\bibitem{Aspert2002}
N.~Aspert, D.~Santa-Cruz, and T.~Ebrahimi.
\newblock Mesh: measuring errors between surfaces using the hausdorff distance.
\newblock In {\em Proceedings. IEEE International Conference on Multimedia and Expo}, vol.~1, pp. 705--708 vol.1, 2002. doi: {{%
10\hspace{.1pt}\discretionary{.}{%
}{.}\hspace{.4pt}1109\discretionary{/}{%
}{/}ICME\hspace{.1pt}\discretionary{.}{%
}{.}\hspace{.4pt}2002\hspace{.1pt}\discretionary{.}{%
}{.}\hspace{.4pt}1035879}}


\bibitem{Athawale2013}
T.~Athawale and A.~Entezari.
\newblock Uncertainty quantification in linear interpolation for isosurface extraction.
\newblock {\em IEEE Transactions on Visualization and Computer Graphics}, 19(12):2723--2732, 2013. doi: {{%
10\hspace{.1pt}\discretionary{.}{%
}{.}\hspace{.4pt}1109\discretionary{/}{%
}{/}TVCG\hspace{.1pt}\discretionary{.}{%
}{.}\hspace{.4pt}2013\hspace{.1pt}\discretionary{.}{%
}{.}\hspace{.4pt}208}}


\bibitem{Athawale2016}
T.~Athawale, E.~Sakhaee, and A.~Entezari.
\newblock Isosurface visualization of data with nonparametric models for uncertainty.
\newblock {\em IEEE Transactions on Visualization and Computer Graphics}, 22(1):777--786, 2016. doi: {{%
10\hspace{.1pt}\discretionary{.}{%
}{.}\hspace{.4pt}1109\discretionary{/}{%
}{/}TVCG\hspace{.1pt}\discretionary{.}{%
}{.}\hspace{.4pt}2015\hspace{.1pt}\discretionary{.}{%
}{.}\hspace{.4pt}2467958}}


\bibitem{Athawale2021}
T.~M. Athawale, S.~Sane, and C.~R. Johnson.
\newblock Uncertainty visualization of the marching squares and marching cubes topology cases.
\newblock In {\em 2021 IEEE Visualization Conference (VIS)}, pp. 106--110, 2021. doi: {{%
10\hspace{.1pt}\discretionary{.}{%
}{.}\hspace{.4pt}1109\discretionary{/}{%
}{/}VIS49827\hspace{.1pt}\discretionary{.}{%
}{.}\hspace{.4pt}2021\hspace{.1pt}\discretionary{.}{%
}{.}\hspace{.4pt}9623267}}


\bibitem{BAGLEY2016}
B.~Bagley, S.~P. Sastry, and R.~T. Whitaker.
\newblock A marching-tetrahedra algorithm for feature-preserving meshing of piecewise-smooth implicit surfaces.
\newblock {\em Procedia Engineering}, 163:162--174, 2016.
\newblock 25th International Meshing Roundtable. doi: {{%
10\hspace{.1pt}\discretionary{.}{%
}{.}\hspace{.4pt}1016\discretionary{/}{%
}{/}j\hspace{.1pt}\discretionary{.}{%
}{.}\hspace{.4pt}proeng\hspace{.1pt}\discretionary{.}{%
}{.}\hspace{.4pt}2016\hspace{.1pt}\discretionary{.}{%
}{.}\hspace{.4pt}11\hspace{.1pt}\discretionary{.}{%
}{.}\hspace{.4pt}042}}


\bibitem{Bourke2003}
P.~Bourke.
\newblock Tubey, 2003.
\newblock https://paulbourke.net/geometry/tubey/.

\bibitem{Brodlie2012}
K.~Brodlie, R.~Allendes~Osorio, and A.~Lopes.
\newblock {\em A Review of Uncertainty in Data Visualization}, pp. 81--109.
\newblock Springer London, London, 2012. doi: {{%
10\hspace{.1pt}\discretionary{.}{%
}{.}\hspace{.4pt}1007\discretionary{/}{%
}{/}978\discretionary{%
}{-}{-}1\discretionary{%
}{-}{-}4471\discretionary{%
}{-}{-}2804\discretionary{%
}{-}{-}5\_6}}


\bibitem{CATMULL1974}
E.~Catmull and R.~Rom.
\newblock A class of local interpolating splines.
\newblock In R.~E. BARNHILL and R.~F. RIESENFELD, eds., {\em Computer Aided Geometric Design}, pp. 317--326. Academic Press, 1974. doi: {{%
10\hspace{.1pt}\discretionary{.}{%
}{.}\hspace{.4pt}1016\discretionary{/}{%
}{/}B978\discretionary{%
}{-}{-}0\discretionary{%
}{-}{-}12\discretionary{%
}{-}{-}079050\discretionary{%
}{-}{-}0\hspace{.1pt}\discretionary{.}{%
}{.}\hspace{.4pt}50020\discretionary{%
}{-}{-}5}}


\bibitem{Chen2022}
Z.~Chen, A.~Tagliasacchi, T.~Funkhouser, and H.~Zhang.
\newblock Neural dual contouring.
\newblock {\em ACM Trans. Graph.}, 41(4), jul 2022. doi: {{%
10\hspace{.1pt}\discretionary{.}{%
}{.}\hspace{.4pt}1145\discretionary{/}{%
}{/}3528223\hspace{.1pt}\discretionary{.}{%
}{.}\hspace{.4pt}3530108}}


\bibitem{Chen2021}
Z.~Chen and H.~Zhang.
\newblock Neural marching cubes.
\newblock {\em ACM Trans. Graph.}, 40(6), dec 2021. doi: {{%
10\hspace{.1pt}\discretionary{.}{%
}{.}\hspace{.4pt}1145\discretionary{/}{%
}{/}3478513\hspace{.1pt}\discretionary{.}{%
}{.}\hspace{.4pt}3480518}}


\bibitem{Cignoni1998}
P.~Cignoni, C.~Rocchini, and R.~Scopigno.
\newblock {Metro: Measuring Error on Simplified Surfaces}.
\newblock {\em Computer Graphics Forum}, 1998. doi: {{%
10\hspace{.1pt}\discretionary{.}{%
}{.}\hspace{.4pt}1111\discretionary{/}{%
}{/}1467\discretionary{%
}{-}{-}8659\hspace{.1pt}\discretionary{.}{%
}{.}\hspace{.4pt}00236}}


\bibitem{Cohen-Or2000}
D.~Cohen-Or, A.~Kadosh, D.~Levin, and R.~Yagel.
\newblock Smooth boundary surfaces from binary {3D} datasets.
\newblock In M.~Chen, A.~E. Kaufman, and R.~Yagel, eds., {\em Volume Graphics}, pp. 71--78. Springer London, London, 2000. doi: {{%
10\hspace{.1pt}\discretionary{.}{%
}{.}\hspace{.4pt}1007\discretionary{/}{%
}{/}978\discretionary{%
}{-}{-}1\discretionary{%
}{-}{-}4471\discretionary{%
}{-}{-}0737\discretionary{%
}{-}{-}8\_4}}


\bibitem{Balazs2019}
B.~Cs\'{e}bfalvi.
\newblock Beyond trilinear interpolation: Higher quality for free.
\newblock {\em ACM Trans. Graph.}, 38(4), jul 2019. doi: {{%
10\hspace{.1pt}\discretionary{.}{%
}{.}\hspace{.4pt}1145\discretionary{/}{%
}{/}3306346\hspace{.1pt}\discretionary{.}{%
}{.}\hspace{.4pt}3323032}}


\bibitem{Dietrich2009}
C.~A. Dietrich, C.~E. Scheidegger, J.~Schreiner, J.~L. Comba, L.~P. Nedel, and C.~T. Silva.
\newblock Edge transformations for improving mesh quality of marching cubes.
\newblock {\em IEEE Transactions on Visualization and Computer Graphics}, 15(1):150--159, 2009. doi: {{%
10\hspace{.1pt}\discretionary{.}{%
}{.}\hspace{.4pt}1109\discretionary{/}{%
}{/}TVCG\hspace{.1pt}\discretionary{.}{%
}{.}\hspace{.4pt}2008\hspace{.1pt}\discretionary{.}{%
}{.}\hspace{.4pt}60}}


\bibitem{EDMUNDS2012}
M.~Edmunds, R.~S. Laramee, G.~Chen, N.~Max, E.~Zhang, and C.~Ware.
\newblock Surface-based flow visualization.
\newblock {\em Computers and Graphics}, 36(8):974--990, 2012.
\newblock Graphics Interaction Virtual Environments and Applications 2012. doi: {{%
10\hspace{.1pt}\discretionary{.}{%
}{.}\hspace{.4pt}1016\discretionary{/}{%
}{/}j\hspace{.1pt}\discretionary{.}{%
}{.}\hspace{.4pt}cag\hspace{.1pt}\discretionary{.}{%
}{.}\hspace{.4pt}2012\hspace{.1pt}\discretionary{.}{%
}{.}\hspace{.4pt}07\hspace{.1pt}\discretionary{.}{%
}{.}\hspace{.4pt}006}}


\bibitem{FENG2023}
Y.-F. Feng, L.-Y. Shen, C.-M. Yuan, and X.~Li.
\newblock Deep shape representation with sharp feature preservation.
\newblock {\em Computer-Aided Design}, 157:103468, 2023. doi: {{%
10\hspace{.1pt}\discretionary{.}{%
}{.}\hspace{.4pt}1016\discretionary{/}{%
}{/}j\hspace{.1pt}\discretionary{.}{%
}{.}\hspace{.4pt}cad\hspace{.1pt}\discretionary{.}{%
}{.}\hspace{.4pt}2022\hspace{.1pt}\discretionary{.}{%
}{.}\hspace{.4pt}103468}}


\bibitem{Fuhrmann2015}
S.~Fuhrmann, M.~Kazhdan, and M.~Goesele.
\newblock Accurate isosurface interpolation with hermite data.
\newblock In {\em 2015 International Conference on 3D Vision}, pp. 256--263, 2015. doi: {{%
10\hspace{.1pt}\discretionary{.}{%
}{.}\hspace{.4pt}1109\discretionary{/}{%
}{/}3DV\hspace{.1pt}\discretionary{.}{%
}{.}\hspace{.4pt}2015\hspace{.1pt}\discretionary{.}{%
}{.}\hspace{.4pt}36}}


\bibitem{Gao2020}
J.~Gao, W.~Chen, T.~Xiang, A.~Jacobson, M.~McGuire, and S.~Fidler.
\newblock Learning deformable tetrahedral meshes for {3D} reconstruction.
\newblock In {\em Proceedings of the 34th International Conference on Neural Information Processing Systems}, NIPS'20. Curran Associates Inc., Red Hook, NY, USA, 2020.

\bibitem{Gillmann2021}
C.~Gillmann, D.~Saur, T.~Wischgoll, and G.~Scheuermann.
\newblock Uncertainty-aware visualization in medical imaging - a survey.
\newblock {\em Computer Graphics Forum}, 40(3):665--689, 2021. doi: {{%
10\hspace{.1pt}\discretionary{.}{%
}{.}\hspace{.4pt}1111\discretionary{/}{%
}{/}cgf\hspace{.1pt}\discretionary{.}{%
}{.}\hspace{.4pt}14333}}


\bibitem{Han2022}
M.~Han, T.~M. Athawale, D.~Pugmire, and C.~R. Johnson.
\newblock Accelerated probabilistic marching cubes by deep learning for time-varying scalar ensembles.
\newblock In {\em 2022 IEEE Visualization and Visual Analytics (VIS)}, pp. 155--159, 2022. doi: {{%
10\hspace{.1pt}\discretionary{.}{%
}{.}\hspace{.4pt}1109\discretionary{/}{%
}{/}VIS54862\hspace{.1pt}\discretionary{.}{%
}{.}\hspace{.4pt}2022\hspace{.1pt}\discretionary{.}{%
}{.}\hspace{.4pt}00040}}


\bibitem{hildebrand1987introduction}
F.~B. Hildebrand.
\newblock {\em Introduction to numerical analysis}.
\newblock Courier Corporation, 1987.

\bibitem{Ho2004}
C.-C. Ho, Y.-H. Lu, H.-T. Lin, S.-H. Guan, S.-Y. Cho, R.-H. Liang, B.-Y. Chen, and M.~Ouhyoung.
\newblock Feature refinement strategy for extended marching cubes: Handling on dynamic nature of real-time sculpting application.
\newblock In {\em 2004 IEEE International Conference on Multimedia and Expo (ICME) (IEEE Cat. No.04TH8763)}, vol.~2, pp. 855--858 Vol.2, 2004. doi: {{%
10\hspace{.1pt}\discretionary{.}{%
}{.}\hspace{.4pt}1109\discretionary{/}{%
}{/}ICME\hspace{.1pt}\discretionary{.}{%
}{.}\hspace{.4pt}2004\hspace{.1pt}\discretionary{.}{%
}{.}\hspace{.4pt}1394335}}


\bibitem{Ho2005}
C.-C. Ho, F.-C. Wu, B.-Y. Chen, Y.-Y. Chuang, and M.~Ouhyoung.
\newblock {Cubical Marching Squares: Adaptive Feature Preserving Surface Extraction from Volume Data}.
\newblock {\em Computer Graphics Forum}, 2005. doi: {{%
10\hspace{.1pt}\discretionary{.}{%
}{.}\hspace{.4pt}1111\discretionary{/}{%
}{/}j\hspace{.1pt}\discretionary{.}{%
}{.}\hspace{.4pt}1467\discretionary{%
}{-}{-}8659\hspace{.1pt}\discretionary{.}{%
}{.}\hspace{.4pt}2005\hspace{.1pt}\discretionary{.}{%
}{.}\hspace{.4pt}00879\hspace{.1pt}\discretionary{.}{%
}{.}\hspace{.4pt}x}}


\bibitem{Jiang1996}
G.-S. Jiang and C.-W. Shu.
\newblock Efficient implementation of weighted {ENO} schemes.
\newblock {\em Journal of Computational Physics}, 126(1):202--228, 1996. doi: {{%
10\hspace{.1pt}\discretionary{.}{%
}{.}\hspace{.4pt}1006\discretionary{/}{%
}{/}jcph\hspace{.1pt}\discretionary{.}{%
}{.}\hspace{.4pt}1996\hspace{.1pt}\discretionary{.}{%
}{.}\hspace{.4pt}0130}}


\bibitem{Johnson2004}
C.~Johnson.
\newblock Top scientific visualization research problems.
\newblock {\em IEEE Computer Graphics and Applications}, 24(4):13--17, 2004. doi: {{%
10\hspace{.1pt}\discretionary{.}{%
}{.}\hspace{.4pt}1109\discretionary{/}{%
}{/}MCG\hspace{.1pt}\discretionary{.}{%
}{.}\hspace{.4pt}2004\hspace{.1pt}\discretionary{.}{%
}{.}\hspace{.4pt}20}}


\bibitem{Ju2002}
T.~Ju, F.~Losasso, S.~Schaefer, and J.~Warren.
\newblock Dual contouring of hermite data.
\newblock {\em ACM Trans. Graph.}, 21(3):339–346, jul 2002. doi: {{%
10\hspace{.1pt}\discretionary{.}{%
}{.}\hspace{.4pt}1145\discretionary{/}{%
}{/}566654\hspace{.1pt}\discretionary{.}{%
}{.}\hspace{.4pt}566586}}


\bibitem{scivisdata}
P.~Klacansky.
\newblock Open scivis datasets, December 2017.
\newblock \small \texttt{https://klacansky.com/open-scivis-datasets/}.

\bibitem{Knoll2007}
A.~Knoll, Y.~Hijazi, C.~Hansen, I.~Wald, and H.~Hagen.
\newblock Interactive ray tracing of arbitrary implicits with simd interval arithmetic.
\newblock In {\em 2007 IEEE Symposium on Interactive Ray Tracing}, pp. 11--18, 2007. doi: {{%
10\hspace{.1pt}\discretionary{.}{%
}{.}\hspace{.4pt}1109\discretionary{/}{%
}{/}RT\hspace{.1pt}\discretionary{.}{%
}{.}\hspace{.4pt}2007\hspace{.1pt}\discretionary{.}{%
}{.}\hspace{.4pt}4342585}}


\bibitem{Kobbelt2001}
L.~P. Kobbelt, M.~Botsch, U.~Schwanecke, and H.-P. Seidel.
\newblock Feature sensitive surface extraction from volume data.
\newblock In {\em Proceedings of the 28th Annual Conference on Computer Graphics and Interactive Techniques}, SIGGRAPH '01, p. 57–66. Association for Computing Machinery, New York, NY, USA, 2001. doi: {{%
10\hspace{.1pt}\discretionary{.}{%
}{.}\hspace{.4pt}1145\discretionary{/}{%
}{/}383259\hspace{.1pt}\discretionary{.}{%
}{.}\hspace{.4pt}383265}}


\bibitem{LEE2001}
T.-Y. Lee and C.-H. Lin.
\newblock Growing-cube isosurface extraction algorithm for medical volume data.
\newblock {\em Computerized Medical Imaging and Graphics}, 25(5):405--415, 2001. doi: {{%
10\hspace{.1pt}\discretionary{.}{%
}{.}\hspace{.4pt}1016\discretionary{/}{%
}{/}S0895\discretionary{%
}{-}{-}6111\discretionary{%
}{(}{(}00\discretionary{)}{%
}{)}00084\discretionary{%
}{-}{-}7}}


\bibitem{Willam1987}
W.~E. Lorensen and H.~E. Cline.
\newblock Marching cubes: A high resolution {3D} surface construction algorithm.
\newblock {\em SIGGRAPH Comput. Graph.}, 21(4):163–169, aug 1987. doi: {{%
10\hspace{.1pt}\discretionary{.}{%
}{.}\hspace{.4pt}1145\discretionary{/}{%
}{/}37402\hspace{.1pt}\discretionary{.}{%
}{.}\hspace{.4pt}37422}}


\bibitem{Marschner1994}
S.~Marschner and R.~Lobb.
\newblock An evaluation of reconstruction filters for volume rendering.
\newblock In {\em Proceedings Visualization '94}, pp. 100--107, 1994. doi: {{%
10\hspace{.1pt}\discretionary{.}{%
}{.}\hspace{.4pt}1109\discretionary{/}{%
}{/}VISUAL\hspace{.1pt}\discretionary{.}{%
}{.}\hspace{.4pt}1994\hspace{.1pt}\discretionary{.}{%
}{.}\hspace{.4pt}346331}}


\bibitem{Mitchell1988}
D.~P. Mitchell and A.~N. Netravali.
\newblock Reconstruction filters in computer-graphics.
\newblock {\em SIGGRAPH Comput. Graph.}, 22(4):221–228, jun 1988. doi: {{%
10\hspace{.1pt}\discretionary{.}{%
}{.}\hspace{.4pt}1145\discretionary{/}{%
}{/}378456\hspace{.1pt}\discretionary{.}{%
}{.}\hspace{.4pt}378514}}


\bibitem{Moller1996}
T.~Moller, R.~Machiraju, K.~Mueller, and R.~Yagel.
\newblock Classification and local error estimation of interpolation and derivative filters for volume rendering.
\newblock In {\em Proceedings of 1996 Symposium on Volume Visualization}, pp. 71--78, 1996. doi: {{%
10\hspace{.1pt}\discretionary{.}{%
}{.}\hspace{.4pt}1109\discretionary{/}{%
}{/}SVV\hspace{.1pt}\discretionary{.}{%
}{.}\hspace{.4pt}1996\hspace{.1pt}\discretionary{.}{%
}{.}\hspace{.4pt}558045}}


\bibitem{Moller1997}
T.~Moller, R.~Machiraju, K.~Mueller, and R.~Yagel.
\newblock Evaluation and design of filters using a taylor series expansion.
\newblock {\em IEEE Transactions on Visualization and Computer Graphics}, 3(2):184--199, 1997. doi: {{%
10\hspace{.1pt}\discretionary{.}{%
}{.}\hspace{.4pt}1109\discretionary{/}{%
}{/}2945\hspace{.1pt}\discretionary{.}{%
}{.}\hspace{.4pt}597800}}


\bibitem{Moller1998}
T.~Moller, K.~Mueller, Y.~Kurzion, R.~Machiraju, and R.~Yagel.
\newblock Design of accurate and smooth filters for function and derivative reconstruction.
\newblock In {\em IEEE Symposium on Volume Visualization (Cat. No.989EX300)}, pp. 143--151, 1998. doi: {{%
10\hspace{.1pt}\discretionary{.}{%
}{.}\hspace{.4pt}1109\discretionary{/}{%
}{/}SVV\hspace{.1pt}\discretionary{.}{%
}{.}\hspace{.4pt}1998\hspace{.1pt}\discretionary{.}{%
}{.}\hspace{.4pt}729596}}


\bibitem{Muller2009}
M.~M\"{u}ller.
\newblock Fast and robust tracking of fluid surfaces.
\newblock In {\em Proceedings of the 2009 ACM SIGGRAPH/Eurographics Symposium on Computer Animation}, SCA '09, p. 237–245. Association for Computing Machinery, New York, NY, USA, 2009. doi: {{%
10\hspace{.1pt}\discretionary{.}{%
}{.}\hspace{.4pt}1145\discretionary{/}{%
}{/}1599470\hspace{.1pt}\discretionary{.}{%
}{.}\hspace{.4pt}1599501}}


\bibitem{NEWMAN2006}
T.~S. Newman and H.~Yi.
\newblock A survey of the marching cubes algorithm.
\newblock {\em Computers and Graphics}, 30(5):854--879, 2006. doi: {{%
10\hspace{.1pt}\discretionary{.}{%
}{.}\hspace{.4pt}1016\discretionary{/}{%
}{/}j\hspace{.1pt}\discretionary{.}{%
}{.}\hspace{.4pt}cag\hspace{.1pt}\discretionary{.}{%
}{.}\hspace{.4pt}2006\hspace{.1pt}\discretionary{.}{%
}{.}\hspace{.4pt}07\hspace{.1pt}\discretionary{.}{%
}{.}\hspace{.4pt}021}}


\bibitem{Ouermi2023}
T.~A. Ouermi, R.~M. Kirby, and M.~Berzins.
\newblock Eno-based high-order data-bounded and constrained positivity-preserving interpolation.
\newblock {\em Numerical Algorithms}, 92(3):1517--1551, 2023.

\bibitem{Ouermi2024}
T.~A.~J. Ouermi, R.~M. Kirby, and M.~Berzins.
\newblock Algorithm 1041: Hippis -- a high-order positivity-preserving mapping software for structured meshes.
\newblock {\em ACM Trans. Math. Softw.}, 50(1), mar 2024. doi: {{%
10\hspace{.1pt}\discretionary{.}{%
}{.}\hspace{.4pt}1145\discretionary{/}{%
}{/}3632291}}


\bibitem{Pothkow2013}
K.~P\"{o}thkow and H.-C. Hege.
\newblock Nonparametric models for uncertainty visualization.
\newblock {\em Computer Graphics Forum}, 32(3pt2):131--140, 2013. doi: {{%
10\hspace{.1pt}\discretionary{.}{%
}{.}\hspace{.4pt}1111\discretionary{/}{%
}{/}cgf\hspace{.1pt}\discretionary{.}{%
}{.}\hspace{.4pt}12100}}


\bibitem{Remelli2020}
E.~Remelli, A.~Lukoianov, S.~R. Richter, B.~Guillard, T.~Bagautdinov, P.~Baque, and P.~Fua.
\newblock {MeshSDF}: Differentiable iso-surface extraction.
\newblock In {\em Proceedings of the 34th International Conference on Neural Information Processing Systems}, NIPS'20. Curran Associates Inc., Red Hook, NY, USA, 2020.

\bibitem{RISTOVSKI2014}
G.~Ristovski, T.~Preusser, H.~K. Hahn, and L.~Linsen.
\newblock Uncertainty in medical visualization: Towards a taxonomy.
\newblock {\em Computers and Graphics}, 39:60--73, 2014. doi: {{%
10\hspace{.1pt}\discretionary{.}{%
}{.}\hspace{.4pt}1016\discretionary{/}{%
}{/}j\hspace{.1pt}\discretionary{.}{%
}{.}\hspace{.4pt}cag\hspace{.1pt}\discretionary{.}{%
}{.}\hspace{.4pt}2013\hspace{.1pt}\discretionary{.}{%
}{.}\hspace{.4pt}10\hspace{.1pt}\discretionary{.}{%
}{.}\hspace{.4pt}015}}


\bibitem{Schlegel2012}
S.~Schlegel, N.~Korn, and G.~Scheuermann.
\newblock On the interpolation of data with normally distributed uncertainty for visualization.
\newblock {\em IEEE Transactions on Visualization and Computer Graphics}, 18(12):2305--2314, 2012. doi: {{%
10\hspace{.1pt}\discretionary{.}{%
}{.}\hspace{.4pt}1109\discretionary{/}{%
}{/}TVCG\hspace{.1pt}\discretionary{.}{%
}{.}\hspace{.4pt}2012\hspace{.1pt}\discretionary{.}{%
}{.}\hspace{.4pt}249}}


\bibitem{Spackman2021}
P.~R. Spackman, M.~J. Turner, J.~J. McKinnon, S.~K. Wolff, D.~J. Grimwood, D.~Jayatilaka, and M.~A. Spackman.
\newblock {{\it CrystalExplorer}: a program for Hirshfeld surface analysis, visualization and quantitative analysis of molecular crystals}.
\newblock {\em Journal of Applied Crystallography}, 54(3):1006--1011, Jun 2021. doi: {{%
10\hspace{.1pt}\discretionary{.}{%
}{.}\hspace{.4pt}1107\discretionary{/}{%
}{/}S1600576721002910}}


\bibitem{Theisel2002}
H.~Theisel.
\newblock Exact isosurfaces for marching cubes.
\newblock {\em Computer Graphics Forum}, 21(1):19--32, 2002. doi: {{%
10\hspace{.1pt}\discretionary{.}{%
}{.}\hspace{.4pt}1111\discretionary{/}{%
}{/}1467\discretionary{%
}{-}{-}8659\hspace{.1pt}\discretionary{.}{%
}{.}\hspace{.4pt}00563}}


\bibitem{Wald2021}
I.~Wald, S.~Zellmann, W.~Usher, N.~Morrical, U.~Lang, and V.~Pascucci.
\newblock Ray tracing structured amr data using exabricks.
\newblock {\em IEEE Transactions on Visualization and Computer Graphics}, 27(2):625--634, 2021. doi: {{%
10\hspace{.1pt}\discretionary{.}{%
}{.}\hspace{.4pt}1109\discretionary{/}{%
}{/}TVCG\hspace{.1pt}\discretionary{.}{%
}{.}\hspace{.4pt}2020\hspace{.1pt}\discretionary{.}{%
}{.}\hspace{.4pt}3030470}}


\bibitem{Wang2023}
Z.~Wang, T.~M. Athawale, K.~Moreland, J.~Chen, C.~R. Johnson, and D.~Pugmire.
\newblock {$FunMC^2$: A Filter for Uncertainty Visualization of Marching Cubes on Multi-Core Devices}.
\newblock In R.~Bujack, D.~Pugmire, and G.~Reina, eds., {\em Eurographics Symposium on Parallel Graphics and Visualization}. The Eurographics Association, 2023. doi: {{%
10\hspace{.1pt}\discretionary{.}{%
}{.}\hspace{.4pt}2312\discretionary{/}{%
}{/}pgv\hspace{.1pt}\discretionary{.}{%
}{.}\hspace{.4pt}20231081}}


\bibitem{weber2012}
G.~H. Weber, H.~Childs, and J.~S. Meredith.
\newblock Efficient parallel extraction of crack-free isosurfaces from adaptive mesh refinement (amr) data.
\newblock In {\em IEEE Symposium on Large Data Analysis and Visualization (LDAV)}, pp. 31--38, 2012. doi: {{%
10\hspace{.1pt}\discretionary{.}{%
}{.}\hspace{.4pt}1109\discretionary{/}{%
}{/}LDAV\hspace{.1pt}\discretionary{.}{%
}{.}\hspace{.4pt}2012\hspace{.1pt}\discretionary{.}{%
}{.}\hspace{.4pt}6378973}}


\bibitem{weber2001}
G.~H. Weber, O.~Kreylos, T.~J. Ligocki, J.~M. Shalf, H.~Hagen, B.~Hamann, and K.~I. Joy.
\newblock Extraction of crack-free isosurfaces from adaptive mesh refinement data.
\newblock In D.~S. Ebert, J.~M. Favre, and R.~Peikert, eds., {\em Data Visualization 2001}, pp. 25--34. Springer Vienna, Vienna, 2001.

\bibitem{weisstein2002}
E.~W. Weisstein.
\newblock Cubic formula.
\newblock {\em https://mathworld. wolfram. com/}, 2002.

\bibitem{WOLF2005}
I.~Wolf, M.~Vetter, I.~Wegner, T.~Böttger, M.~Nolden, M.~Schöbinger, M.~Hastenteufel, T.~Kunert, and H.-P. Meinzer.
\newblock The medical imaging interaction toolkit.
\newblock {\em Medical Image Analysis}, 9(6):594--604, 2005.
\newblock ITK. doi: {{%
10\hspace{.1pt}\discretionary{.}{%
}{.}\hspace{.4pt}1016\discretionary{/}{%
}{/}j\hspace{.1pt}\discretionary{.}{%
}{.}\hspace{.4pt}media\hspace{.1pt}\discretionary{.}{%
}{.}\hspace{.4pt}2005\hspace{.1pt}\discretionary{.}{%
}{.}\hspace{.4pt}04\hspace{.1pt}\discretionary{.}{%
}{.}\hspace{.4pt}005}}


\bibitem{YU2008}
Z.~Yu, M.~J. Holst, Y.~Cheng, and J.~McCammon.
\newblock Feature-preserving adaptive mesh generation for molecular shape modeling and simulation.
\newblock {\em Journal of Molecular Graphics and Modelling}, 26(8):1370--1380, 2008. doi: {{%
10\hspace{.1pt}\discretionary{.}{%
}{.}\hspace{.4pt}1016\discretionary{/}{%
}{/}j\hspace{.1pt}\discretionary{.}{%
}{.}\hspace{.4pt}jmgm\hspace{.1pt}\discretionary{.}{%
}{.}\hspace{.4pt}2008\hspace{.1pt}\discretionary{.}{%
}{.}\hspace{.4pt}01\hspace{.1pt}\discretionary{.}{%
}{.}\hspace{.4pt}007}}


\bibitem{ZHANG2016103}
Y.-T. Zhang and C.-W. Shu.
\newblock Chapter 5 - {ENO} and {WENO} schemes.
\newblock In R.~Abgrall and C.-W. Shu, eds., {\em Handbook of Numerical Methods for Hyperbolic Problems}, vol.~17 of {\em Handbook of Numerical Analysis}, pp. 103--122. Elsevier, 2016. doi: {{%
10\hspace{.1pt}\discretionary{.}{%
}{.}\hspace{.4pt}1016\discretionary{/}{%
}{/}bs\hspace{.1pt}\discretionary{.}{%
}{.}\hspace{.4pt}hna\hspace{.1pt}\discretionary{.}{%
}{.}\hspace{.4pt}2016\hspace{.1pt}\discretionary{.}{%
}{.}\hspace{.4pt}09\hspace{.1pt}\discretionary{.}{%
}{.}\hspace{.4pt}009}}


\end{thebibliography}
